\documentclass[11pt]{article}
\usepackage{amssymb,amsmath,amsfonts,mathtools}
\usepackage{commath}
\usepackage{subfigure}
\usepackage{enumerate}
\usepackage{bbm}
\usepackage{epsfig}
\usepackage{wrapfig}
\usepackage{yfonts}
\usepackage{geometry}                		
\usepackage{graphicx}				
\usepackage{color}
\definecolor{hyperref}{RGB}{026,028,185}
\usepackage[bookmarks=true,colorlinks=true,linkcolor=hyperref,citecolor=hyperref,urlcolor=hyperref,bookmarksnumbered]{hyperref}
\usepackage{enumerate}

\usepackage{physics} 

\usepackage[sort,compress]{cite}
\numberwithin{equation}{section}
\def\[{\begin{equation}}
\def\]{\end{equation}}
\newcommand{\be}{\begin{eqnarray}}
\newcommand{\ee}{\end{eqnarray}}
\newcommand{\nn}{\nonumber}

\def\Z{ {\mathbb Z} }
\def\RR{ {\mathbb R} }

\def\G{\Gamma}

\def\half{{1\over 2}}

\def\D{\mathcal{D}}
\def\M{\mathcal{M}}
\def\PP{\mathcal{P}}

\def\tr{\mathrm{Tr}}
\newcommand{\p}{\partial}

\usepackage{color}

\newcommand{\stirling}{\genfrac{[}{]}{0pt}{}} 
\newcommand{\sumprime}{\sideset{}{'}\sum}

\begin{document}
\renewcommand{\thefootnote}{\arabic{footnote}}
 
\overfullrule=0pt
\parskip=2pt
\parindent=12pt
\headheight=0in \headsep=0in \topmargin=0in \oddsidemargin=0in

\vspace{ -3cm} \thispagestyle{empty} \vspace{-1cm}
\begin{flushright} 
\footnotesize
\end{flushright}%

\begin{center}
\vspace{1.2cm}
{\Large\bf \mathversion{bold}
Entanglement Entropy in Generalised Quantum Lifshitz Models}

\vspace{0.8cm} {J.~Angel-Ramelli $^{a,}$\footnote{{\tt jfa1@hi.is}}, 
V.~Giangreco~M. Puletti $^{a,}$\footnote{{\tt vgmp@hi.is}}, and L.~Thorlacius $^{a,b,}$\footnote{{\tt lth@hi.is}}}

\vskip  0.5cm

\small
{\em
(a) University of Iceland,
Science Institute,
Dunhaga 3,  107 Reykjav\'ik, Iceland
\vskip 0.5cm
(b) The Oskar Klein Centre for Cosmoparticle Physics \& Department of Physics, \\ Stockholm University, 
AlbaNova, 106 91 Stockholm, Sweden.
}
\normalsize

 \end{center}

\vspace{0.3cm}
\begin{abstract}
We compute universal finite corrections to entanglement entropy for generalised quantum Lifshitz models 
in arbitrary odd spacetime dimensions. These are generalised free field theories with Lifshitz scaling 
symmetry, where the dynamical critical exponent $z$ equals the number of spatial dimensions $d$, and which 
generalise the 2+1-dimensional quantum Lifshitz model to higher dimensions. We analyse two cases: 
one where the spatial manifold is a $d$-dimensional sphere and the entanglement entropy is evaluated for 
a hemisphere, and another where a $d$-dimensional flat torus is divided into two cylinders. In both examples
the finite universal terms in the entanglement entropy are scale invariant and depend on the compactification 
radius of the scalar field.  
\end{abstract}
\newpage

\tableofcontents
\pagenumbering{arabic}

\setcounter{footnote}{0}
\newpage

\section{Introduction}
\label{sec:intro}

Quantum entanglement refers to a correlation of a purely quantum mechanical 
nature between degrees of freedom in a physical system. Consider a quantum
system that can be divided into two subsystems $A$ and $B$, such that the 
Hilbert space can be written as a direct product of the Hilbert spaces of the 
subsystems, $\mathcal H= \mathcal H_A \otimes \mathcal H_B$, and take the 
full system to be in a state described by a density matrix $\rho$.
The entanglement entropy of subsystem $A$ is then defined as the von 
Neumann entropy of the reduced density matrix, obtained by taking a trace 
over the degrees of freedom in subsystem $B$, i.e.
\be
\label{eq:von-neumann-entropy}
S[A] = - \mathrm{\Tr}\left(\rho_A \log\rho_A\right)\,,
\ee
with $\rho_A= \Tr_B \rho$. We will take $\rho$ to be the ground state density 
matrix of the full system but our results can be extended to more general states.

Entanglement entropy is a useful theoretical probe
that encodes certain universal properties of field theories describing critical systems, 
see {\it e.g.} \cite{Hsu:2008af, Calabrese:2005zw, Amico:2008aa, Eisert:2010aa, 
LAFLORENCIE20161}. 
A well known example in this respect is the entanglement entropy of a two-dimensional 
conformal field theory (CFT) \cite{Callan:1994aa, Holzhey:1994we, Calabrese:2004eu}, 
which has a {\it universal} logarithmic term, 
\be
S[A] = {c\over 3} \log\left({L_A\over \varepsilon}\right)\,,
\ee
where $c$ is the central charge of the CFT in question, $L_A$ is the spatial size of the 
subsystem $A$, and $\varepsilon$ is a UV-cutoff. Here we are assuming that $A$
is connected and that its size is small compared to the full system size $L_A\ll L$.
The logarithmic UV behaviour of the entanglement entropy tells us that the system 
has long-range entangled degrees of freedom (in contrast to an area-law where short-range 
entanglement would mainly contribute).

In  recent years considerable effort has been devoted to investigating such universal terms 
in the entanglement entropy of CFTs in arbitrary dimensions 
(see \cite{Rangamani:2016dms, Nishioka:2018khk} for reviews). 
What will be important for us is the following general UV behavior of entanglement entropy
in even $D$-dimensional CFTs,
\be
S[A]= c_{D-2} {\Sigma_{D-2}\over \varepsilon^{D-2}} + \dots 
+ c_0 \log\left({L_A\over \varepsilon}\right)+\dots \,,
\ee
where $\Sigma_{D-2}$ is the area of the $(D{-}2)$-dimensional entangling surface $\p A$
and $L_A$ is a characteristic length associated with $\p A$.
Only $c_0$ is universal in this expression. The other coefficients depend on the regularisation 
scheme used. One also finds a universal sub-leading term in odd-dimensional 
CFTs but in this case it is finite rather than logarithmic, see {\it e.g.} \cite{Nishioka:2018khk} 
and references therein. The coefficient of the universal term is a function of topological and 
geometric invariants, such as the Euler density and Weyl invariants constructed from the 
entangling (hyper)-surface \cite{Solodukhin:2008dh, Fursaev:2013fta}. This reflects the fact 
that the logarithmic term in $S[A]$ is related to the conformal anomaly of the stress-energy 
tensor of the corresponding CFT.

In the present work we will study entanglement entropy, including universal finite terms
({\it i.e.} of order $\mathcal O(1)$ with respect to $L_A$), in a family of $d{+}1$-dimensional 
scale invariant quantum field theories introduced in \cite{Keranen:2016ija}. The scale symmetry 
is a non-relativistic Lifshitz symmetry that acts asymmetrically on the time and spatial coordinates,
\be\label{L-scaling}
\vec x \to \lambda\, \vec x, \qquad \tau \to \lambda^z \tau\,,
\ee
with a dynamical critical exponent equal to the number of spatial dimensions $z=d$.
These theories are closely related to the well known quantum Lifshitz model (QLM), first studied 
in the seminal work \cite{Ardonne:2003wa}. This is a scale invariant free field theory in 
$2+1$-dimensional spacetime with a $z=2$ dynamical critical exponent. It is an effective field 
theory for certain quantum dimer models (and their universality class) in square lattices at a 
critical point and involves a {\it compactified\/} free massless scalar field \cite{Ardonne:2003wa}.
Non-trivial Lifshitz scaling is achieved via a kinetic term that is asymmetric between time 
and space (with higher derivatives acting in the spatial directions). 
The higher-derivative construction can easily be extended to free scalar field theories in any 
number of spatial dimensions $d$ with $z=d$ Lifshitz scaling. In \cite{Keranen:2016ija} such 
theories were dubbed generalised quantum Lifshitz models (GQLMs) and several interesting
symmetry properties were revealed in correlation functions of scaling operators. The periodic
identification of the scalar field did not figure in that work but turns out be important when one 
considers entanglement entropy in a GQLM defined on a topologically nontrivial geometry. 

A key property of the QLM and GQLM theories is that the ground state wave functional is invariant 
under conformal transformations involving {\it only} the spatial dimensions \cite{Ardonne:2003wa, Keranen:2016ija}. 
The spatial conformal symmetry is a rather special feature (the corresponding 
critical points are called conformal quantum critical points \cite{Ardonne:2003wa}) 
and it manifests in the scaling properties of entanglement entropy. 
In essence, the symmetry allows us to map a $(d{+}1)$-dimensional Lifshitz field theory with $z=d$ 
to a $d$-dimensional Euclidean CFT. In the $d=2$ QLM the CFT is the standard free boson CFT 
but for $d>2$ GQLM's the spatial CFT is a higher-derivative generalised free field theory.
Such higher-derivative CFTs have been discussed in a number of contexts, for instance 
in relation to higher spin theories, {\it e.g.}~\cite{Brust:2016gjy, Beccaria:2017aqc}, as  
models in elastic theory {\it e.g.}~\cite{Nakayama:2016dby}, in a high-energy physics setting in 
connection with the naturalness problem {\it e.g.} \cite{Griffin:2014bta, Griffin:2015hxa}, and in
the context of dS/CFT duality {\it e.g.} \cite{Strominger:2001pn, Anninos:2011ui}. 
These theories are not unitary but their $n$-point correlation functions are well defined in 
Euclidean spacetime and being free field theories they have no interactions that trigger 
instability. 

Entanglement entropy and its scaling properties have been extensively studied in the QLM
\cite{Fradkin:2006mb, Hsu:2008af, Hsu:2010ag, Stephan:2009aa, Oshikawa:2010kv, 
Zaletel:2011ir, Zhou:2016ykv}.%
\footnote{See also \cite{Stephan:2011gy} for related study of 
R\'enyi entropies in quantum dimer models.}
The replica method in the QLM was first developed in~\cite{Fradkin:2006mb}, where it was 
found that for a smooth entanglement boundary, the scaling behaviour of the entanglement entropy in the QLM 
(and more generally for conformal quantum critical points in $(2{+}1)$-dimensions) is of the form 
\be
S[A] = c_1 {L_A\over \varepsilon}+ {c\over 6} \Delta \chi \log {L_A\over \varepsilon}+\dots\,,
\ee 
where $c_1$ depends on the regulator and $\Delta \chi$ is the change in the Euler characteristic 
(upon dividing the the system in two), which in turns depends on the topology of the system and 
on the entangling surface. The above behavior follows from general expectations for the free energy 
of a two-dimensional CFT with boundary \cite{Cardy:1988tk}. 
This result is for the entanglement entropy of a non-relativistic $(2{+}1)$-dimensional theory, however 
its computation starts from a ground state which is a ``time-independent'' conformal invariant. 
As the time coordinate only appears as a spectator, the final result displays features of a 
two-dimensional CFT.  This is in line with the results of \cite{Ardonne:2003wa, Keranen:2016ija}, 
where it was shown that equal-time correlation functions of local scaling operators in the QLM and
GQLM can be expressed in terms of correlation functions of a $d$-dimensional Euclidean CFT. 

Furthermore, by choosing a smooth partition, such that we have no contribution from the logarithm ({\it i.e.} $\Delta \chi=0$),
a further sub-leading (of order one in $L_A$) universal term appears in the entanglement entropy for the QLM \cite{Hsu:2008af}, 
that is 
\be
S_{EE}= c_1 {L_A \over \varepsilon}+ \gamma_{QCP}+\dots\,.
\ee
The universal term $\gamma_{QCP}$, where QCP stands for quantum critical point, depends both on the geometry and 
topology of the manifolds \cite{Hsu:2010ag, Oshikawa:2010kv, Zaletel:2011ir, Zhou:2016ykv}. 
It depends on the geometry in the sense that it includes a scaling function written in terms of aspect 
ratios typical of the given subsystem, and on the topology through a contribution from zero modes and 
non-trivial winding modes. 
In this sense, the entanglement entropy of the QLM is able to capture long-range (non-local) properties of the system. 
In particular, $\gamma_{QCP}$ was computed by various methods for a spatial manifold in the form of a cylinder 
in \cite{Stephan:2009aa, Oshikawa:2010kv, Zaletel:2011ir, Zhou:2016ykv, Stephan:2009dy}, for a sphere in \cite{Zhou:2016ykv}, 
and the toroidal case was treated in \cite{Oshikawa:2010kv} by means of a boundary field theory approach. 
The toroidal case was further investigated in \cite{Stephan:2012du}, where analytic results were obtained for R\'enyi entropies.%
\footnote{For related studies of the scaling properties of entanglement entropy for a toroidal manifold in scale 
invariant $(2{+}1)$-dimensional systems, see also \cite{Chen:2014zea, Chen:2016bk}.}

Our aim is to extend the study of these finite universal terms in the entanglement entropy to generalised quantum Lifshitz models. 
In particular, we analyse their scaling properties in full generality, in any number of spatial dimensions $d$ with Lifshitz exponent 
$z=d$. For technical reasons (which we explain below) we restrict attention to the case of even integer $d$. 
More concretely, we obtain universal terms in the entanglement entropy in GQLM on two classes of manifolds. 
On the one hand, we divide a $d$-dimensional sphere into two $d$-dimensional hemispheres, on the other hand we consider
a $d$-dimensional flat torus, sliced into two $d$-dimensional cylinders.%
\footnote{Here a $d$-dimensional cylinder refers to the product of an interval and a $(d{-}1)$-dimensional flat torus.}
Our computations are purely field theoretical. The theories we consider represent rare examples of 
non-relativistic critical theories, for which entanglement entropy can be obtained analytically. 
We view them as toy-models where we can explore quantum entanglement for different values of the
dynamical critical exponent $z$. 
Further motivation comes from a puzzling aspect of Lifshitz holography, where one considers gravitational solutions that 
realise the Lifshitz scaling \eqref{L-scaling}, see {\it e.g.} \cite{Kachru:2008yh,Taylor:2008tg}. In AdS/CFT the 
entanglement entropy is computed by means of the Ryu-Takayanagi formula \cite{Ryu:2006ef, Ryu:2006bv}. 
The usual working assumption is to apply the RT prescription also in Lifshitz holography, but in a static Lifshitz spacetime
the holographic entanglement entropy does not depend on the critical exponent at all (see {\it e.g.} \cite{Keranen:2013vla}).
From a field theory point of view, however, one expects the higher-derivative terms to dominate at short distances, 
and thus the UV behaviour of entanglement should reflect the value of the dynamical critical exponent. 
The absence of $z$ from the holographic entanglement entropy is puzzling if Lifshitz spacetime is the
gravitational dual of a strongly coupled field theory with Lifshitz symmetry. Similar considerations motivated the work in
\cite{Gentle:2017ywk, He:2017wla, MohammadiMozaffar:2017nri}. While we clearly see a dependence on $z$, 
it is difficult to compare our results directly to those of these authors as we are not working within the same class of 
field theories and we focus on universal sub-leading terms rather than the leading area terms.

The paper is arranged as follows. 
In Section \ref{sec:gen-qlm} we briefly review the construction of generalised quantum Lifshitz models and extend 
their definition to two specific compact manifolds. These are higher-derivative field theories so we must ensure that 
the variational problem is well posed. This amounts to imposing $z$ conditions on the variations \eqref{cond-variation}, 
\eqref{bc-sphere-fluctuations} and including a boundary action \eqref{boundary-action-torus} for the $d$-torus or 
\eqref{s-boundary-sphere} for the $d$-sphere.
In Section \ref{sec:replica} we discuss the replica method, which we use to compute the entanglement entropy. 
In essence, this approach maps the problem of computing the $n$th-power of the reduced density matrix to a 
density matrix of an $n$ times replicated field theory. 
The goal is to produce a result that can be analytically continued in $n$, in order to calculate the entanglement 
entropy according to \eqref{EE-replicamethod}. 
As we explain in Section \ref{sec:replica}, the replica method forces all the replicated fields to be equal at the cut,
since the cut is not physical and our original field theory only has one field. There is an additional subtlety in 
implementing this condition due to the periodic identification of the scalar field in the GQLM and in order to 
ensure the correct counting of degrees of freedom we separate the replicated fields into classical and fluctuating parts.
The fluctuating fields obey Dirichlet boundary conditions as well as the vanishing of the conformal Laplacian 
and its integer powers at the cut. 
Their contribution is encoded in partition functions computed via functional determinants defined on the sphere and 
torus respectively. 
The classical fields give rise to winding sectors that are encoded in the function $W(n)$ described in Section \ref{sec:replica}. 
For the spherical case this contribution is simple and only amounts to a multiplicative factor $\sqrt{n}$. 
For the toroidal case, the contribution from the classical fields is less trivial, and requires summing over  
classical vacua of the action. 
For higher-derivative theories some further conditions have to be implemented in the classical sector, 
and we argue that a compatible prescription is to use Neumann boundary conditions for these fields.
At this point no freedom and/or redundancy is left, and it is straightforward to compute the sum over  
winding modes. We collect the contributions from the classical and fluctuating fields to the universal 
finite term of the entanglement entropy for a $d$-sphere and a $d$-torus in Sections \ref{sec:sphere} 
and \ref{sec:torus} respectively. We conclude with some open questions in Section \ref{sec:conclusions}. 

Most of the technical details are relegated to appendices. 
In Appendix \ref{sec:dowker} we review the computation of the functional determinant contribution for 
the spherical case, which was originally worked out in \cite{Dowker2011}.
In Appendix \ref{sec:alternative-dowker} we develop an alternative expression for the formulae presented 
in Appendix \ref{sec:dowker}, which we find more transparent and better suitable for numerical evaluation. 
In Appendix \ref{sec:Torusdet} we compute the functional determinant contribution for the toroidal case. 
Finally we compute the winding sector contribution for the $d$-torus in Appendix \ref{sec:winding-torus}. 

\section{Generalised quantum Lifshitz models in (d+1)-dimensions}
\label{sec:gen-qlm}

The $2{+}1$-dimensional quantum Lifshitz model \cite{Ardonne:2003wa} can be generalised to 
$d{+}1$-dimensions~\cite{Keranen:2016ija}. Whenever the dynamical critical exponent $z$ is equal 
to the  number of spatial dimensions, the ground state wave-functional is invariant under 
$d$-dimensional conformal transformations acting in the spatial directions, extending the connection 
between the quantum Lifshitz model and a free conformal field theory in one less dimension to any $d$. 
We recall that in the $2{+}1$-dimensional case the scalar field is compactified \cite{Ardonne:2003wa}, and 
below we will also compactify the scalar field in the GQLM at general $d$ on a circle of radius $R_c$, 
that is identify $\phi \sim \phi +2\pi R_c$.

The ground state wave functional of the GQLM is \cite{Keranen:2016ija}
\be\label{def-gs-wf}
\left| \psi_0\rangle\right. = {1\over \sqrt Z}\, \int [\D\phi] e^{-\half S[\phi]} |\phi\rangle \,,
\ee
where $\left\{ | \phi\rangle\right\}$ is an orthonormal basis of states in the Hilbert space made up of
eigenstates of the field operator, and the partition function $Z$ is given by
\be
Z=  \int \left[\D \phi\right] e^{- S[\phi]}\,. 
\ee
We are interested in computing the sub-leading universal finite term of the entanglement entropy 
\eqref{eq:von-neumann-entropy} in the ground state, {\it i.e.} with $\rho=| \psi_0\rangle\langle\psi_0 |$, 
when the manifold $\M$ is a $d$-sphere or a $d$-torus. The subsystem $A$ will consist of  
field degrees of freedom on a submanifold of $\M$.
For technical reasons we restrict attention to the case where $d$ is an even (positive) integer. 

We follow the normalisation convention of \cite{Zhou:2016ykv} and write the action as
\be
\label{def-action-general}
S[\phi]=S_0[\phi]+S_{\p \M}[\phi]= {\kappa\over 4\pi} 
\int\limits_\M \dif^{\,d}x \sqrt{G}\, \phi \,\mathcal P_{z,\,{\tiny \M}}\, \phi\, +S_{\partial \M}[\phi]\,,
\ee
where $G=\det G_{ab}$ ($a, b=1,\dots, d$) is the determinant of the Euclidean metric on the manifold $\M$, 
and $\mathcal P_{z,\,{\tiny \M}}$ is a proper conformal differential operator of degree $z$ in $d$ dimensions. 
The specific form of $\mathcal P_{z,\,{\tiny \M}}$ depends on $\M$, as we will discuss at the end of this section.
In order to have a well-defined variational problem, the action has to include a suitable boundary term 
$S_{\partial \M}$ whose specific form is also given below. 
Note that the scalar field $\phi$ has dimension zero under Lifshitz scaling in the GQLM
at general $d$. 
We find it convenient to use the shorthand $g={\kappa \over 4\pi}$.%
\footnote{The normalisation of the action in 
\eqref{def-action-general} and the compactification radius $R_c$ are not independent. A rescaling of the scalar 
field will affect both $g$ and $R_c$, while physical quantities that are independent of rescaling are expressed 
in terms of $2\pi R_c \sqrt{g}$~\cite{DiFrancesco1997, Ginsparg:1988ui}.}
We note that for a flat manifold, the compactification of the field implies a global shift 
symmetry compatible with conformal symmetry~\cite{DiFrancesco1997, Ginsparg:1988ui}.
This is also true for the $z=d$ theory on the $d$-sphere (and more generally on any Einstein manifold)
provided the action includes appropriate terms that generalise the notion of conformal coupling to a 
higher-derivative setting. 

Let us consider how the action in \eqref{def-action-general} is constructed concretely for the two cases, 
mentioned above. To keep the discussion somewhat general, we assume that both $z$ and $d$ are even 
positive integers and do not insist on $z=d$ for the time being. The case when $d$ is an odd integer 
is also interesting but raises a number of technical issues that we do not address in this work.
The boundary terms in the action will be important once we divide the system into subsystems and
apply the replica method (cf. Section \ref{sec:replica}). 
\vskip 0.5 cm
\paragraph{$d$-torus.}
In section \ref{sec:torus}, we consider a torus obtained as the quotient space of $\RR^d$ and a $d$-dimensional lattice. 
The manifold is flat, and in this case the operator appearing in the action $S_0[\phi]$ in \eqref{def-action-general} is 
simply the $z/2$ power of the Laplace-Beltrami operator,  
\be
\label{operator-torus}
\mathcal{P}_{z, T^d} = (-1)^{z/2+1} \Delta^{z/2}= (-1)^{z/2+1} \left(-\p_a \p^a\right)^{z/2}\,,
\ee
that is 
\be
\label{bulk-action-torus}
S_0[\phi] =(-1)^{z/2+1} g \int\limits_M d^{d}  x\, \phi \Delta^{z/2} \phi\,.
\ee
For $d=z=2$ this reduces to $S=g\int \left(\nabla \phi\right)^2$ 
as in \cite{Ardonne:2003wa} (after integrating by parts). Varying $S_0[\phi]$ we obtain
\begin{align}
\begin{split}
\delta S_0[\phi]
&=2(-1)^{z/2+1}\,g \int\limits_M d^{d} x\ (\Delta^{z/2}\phi)\delta\phi +(-1)^{z/2+1}\, g \int\limits_{\partial M} d^{d-1}x\ \sum_{k=0}^{z-1}(-1)^k(\partial_n^k\phi)(\partial_n^{z-1-k}\delta\phi)
\end{split}
\end{align}
where the partial derivatives should be understood as follows 
\be
\p_n^{2\ell}= (\p_a \p^a)^\ell \,, \qquad \p_n^{2\ell+1}= n_a\p^a (\p_b\p^b)^\ell\,,
\ee
with $n_a$ an oriented unit vector normal to the boundary. 
We need to choose appropriate boundary conditions for the variations. One possibility is to demand that
\be\label{cond-variation}
\p_n^{2\ell}\delta\phi\big\vert_{\p\M}=0\,, \qquad \ell=0, \dots, {z\over 2}-1\,.
\ee
The reason behind this choice is that we will be interested in the eigenvalue problem of  $\Delta^{z/2}$, 
for which we require the operator to be self-adjoint and to have a complete set of consistent boundary conditions. 
The replica method forces us to choose Dirichlet conditions on the field at the boundary (cf. Section \ref{sec:replica}) 
and the remaining conditions are chosen to be consistent with the self-adjointness of the operator.
Equipped with \eqref{cond-variation}, the variation of the Lagrangian reduces to 
\begin{align}
\begin{split}
\delta S_0[\phi]
&=2(-1)^{z/2+1}\,g \int\limits_M d^dx\ (\Delta^{z/2}\phi)\delta\phi 
+(-1)^{z/2+1}\, g \int\limits_{\partial M} d^{d-1}x\ 
\sum_{\ell=0}^{{z\over 2}-1}(\partial_n^{2\ell}\phi)(\partial_n^{z-1-2\ell}\delta\phi)\,. 
\end{split}
\end{align}
Hence, defining the following boundary action
\begin{equation}
\label{boundary-action-torus}
S_\partial[\phi]=(-1)^{z/2+1}g\, \int\limits_{\partial M}d^{d-1}x\ 
\sum\limits_{k=0}^{\frac{z}{2}-1}(\p_n^{2k} \phi)(\partial_n^{z-2k-1}\phi),
\end{equation}
with variation given by
\begin{align}
\delta S_\partial[\phi]
&= (-1)^{z/2+1} g\,\int\limits_{\partial M} d^{d-1}x \sum\limits_{k=0}^{\frac{z}{2}-1}(\p_n^{2k} \phi)(\partial_n^{z-2k-1}\delta\phi),
\end{align}
clearly gives a well-defined variation for the total action $S[\phi]=S_0[\phi]+S_\p[\phi]$ and leads to the 
following equations of motion
\begin{gather}\label{eom-torus}
\Delta^{z/2}\phi=0,\quad\text{and}\quad \p_n^{2k}\delta\phi\big\vert_{\partial M}=0,\quad \text{for}\ k=0,\ldots,\frac{z}{2}-1\,.
\end{gather}

\vskip 0.5 cm
\paragraph{$d$-sphere.}
When the manifold $\M$ is a unit $d$-sphere, the operator in \eqref{def-action-general} is the so-called GJMS 
operator on a $d$-sphere~\cite{GJMS}. In essence, GJMS operators generalise the conformal Laplacian to higher 
derivatives and $d$-dimensional curved manifolds~\cite{GJMS} (see 
\cite{Branson:aa, Gover_2004, Gover:aa, Graham_2007, Fefferman:2012aa, Paneitz_2008, Juhl:2009aa, Juhl:2011aa, bookGJMS} 
for more references on the subject). This means that a GJMS operator of degree $2k$ in $d$-dimensions
(where $k$ is a positive integer) is constructed so that it transforms in a simple way under a Weyl transformation 
of the metric, $G_{ab} \to e^{2\omega} G_{ab}$, 
\be
\label{weyl-transf-GJMS}
\mathcal{P}_{2k} (e^{2\omega} G)=e^{-(d/2+k)\omega} \mathcal{P}_{2k} (G)\, e^{(d/2-k)\omega}\,.
\ee
In general, the operator $\mathcal{P}_{2k}$ is well defined for $k=1, \dots\, d/2$ for even $d$, in the sense that it 
reduces to the standard Laplacian of degree $k$ in flat space \cite{Gover_2004}. For odd $d$-dimensional manifolds 
operators satisfying \eqref{weyl-transf-GJMS} exist for all $k\ge 1$ \cite{Gover_2004,Juhl:2009aa, Juhl:2011aa}. 

On a unit $d$-sphere the GJMS operator of degree $2k$ explicitly reads as
\begin{equation}
\label{GJMS_op_sphere-1}
\mathcal{P}_{2k, S^d}  = \prod_{j=1}^{k} \left[ {\Delta}_{S^d} +\left(\frac{d}{2}-j\right)\left( \frac{d}{2}+j-1\right)\right]\,,
\end{equation}
where $\Delta_{S^d}= -{1\over \sqrt{G}} \p_a \left(\sqrt{G} \,G^{ab} \p_b\right)$, with $a, b=1, \dots d$,
is the Laplace-Beltrami operator.
The case of most interest to us is to consider a GJMS operator of degree $2k=d$. This is known in the 
literature as the critical case, while $k<{d/2}$ is referred to as the subcritical case. 
It is straightforward to check, using \eqref{weyl-transf-GJMS}, that the final action $S_0$ \eqref{def-action-general} 
is invariant under Weyl transformations. 
The factorisation in \eqref{GJMS_op_sphere-1} is a general characteristic of Einstein manifolds.
Since the eigenfunctions of the Laplace-Beltrami operator on a compact Riemannian manifold form an orthonormal basis, 
one can easily obtain the spectrum of the GJMS operator on the sphere from the factorisation above.
This will play an important role later in the computation of partition functions in Section \ref{sec:sphere}.

When $S^d$ is divided into hemispheres, the action $S_0[\phi]$ has to be complemented by boundary terms.
As before, in order to have a well-defined variational problem, we compute the variation of the action $\delta S_0$, 
impose $z$ boundary conditions on $\delta \phi$ and its derivatives, and then cancel any remaining terms against 
appropriate boundary terms. 
We impose the following boundary conditions 
\be\label{bc-sphere-fluctuations}
\delta\phi\big\vert_{\p\mathcal M}=0 \,, \quad \Delta^k \delta\phi\big\vert_{\p\mathcal M}=0 \qquad k=1, \dots {z\over 2}-1\,,
\ee
that is Dirichlet boundary conditions on the variation $\delta \phi$ and vanishing of its Laplacian and its powers 
at the boundary. This is analogous to \eqref{cond-variation} for a curved manifold. 
The explicit expression for $S_\p[\phi]$ is
\be
\label{s-boundary-sphere}
S_\p[\phi] = g \sum_{\ell=1}^{d/2}\ \int\limits_{\M} d^{d}x\ \p_a &&\left\{\sqrt{G}G^{ab}\left[\prod_{j=1}^{d/2-\ell} 
\left(\Delta +\left(\frac{d}{2}-j\right)\left( \frac{d}{2}+j-1\right)\right)\phi\right]\right.
\\ \nn
&& \left.\times 
\p_b \left[\prod_{k=d/2-\ell+2}^{d/2} \left(\Delta +\left(\frac{d}{2}-k\right)\left( \frac{d}{2}+k-1\right)\right)\phi\right]
\right\}\,,
\ee
where the products in the above expression are taken to be empty when the upper extreme is less than the lower extreme. 
Finally, the classical equation of motion is 
\be\label{eom-classical-sphere}
\mathcal{P}_{d, S^d} \,\phi=  \prod_{j=1}^{d/2} \left[ {\Delta}_{S^d} +\left(\frac{d}{2}-j\right)\left( \frac{d}{2}+j-1\right)\right]\, \phi =0\,.
\ee

\section{Replica method}
\label{sec:replica}

The entanglement entropy of subsystem $A$ can be defined as 
\be\label{EE-replicamethod}
S[A]= -\lim_{n\to 1} \p_n \mathrm{Tr}(\rho_A^n) \,, 
\ee
where an analytic continuation of the index $n$ is assumed. 
This definition is equivalent to the von Neumann entropy of $\rho_A$ \eqref{eq:von-neumann-entropy}.
Following \cite{Calabrese:2004eu, Calabrese:2009qy}, we will use the replica approach to evaluate \eqref{EE-replicamethod}. 
At the heart of this method is to view each appearance of the density matrix $\rho$ in $\Tr(\rho_A^n)$ as coming
from an independent copy of the original theory, so that one ends up working with $n$ replicated scalar fields. 
The process of taking partial traces and multiplying the replicas of $\rho$ then induces a specific set of boundary 
conditions at the entanglement cuts on the replica fields.  

\begin{figure}
\centering\includegraphics[scale=0.351]{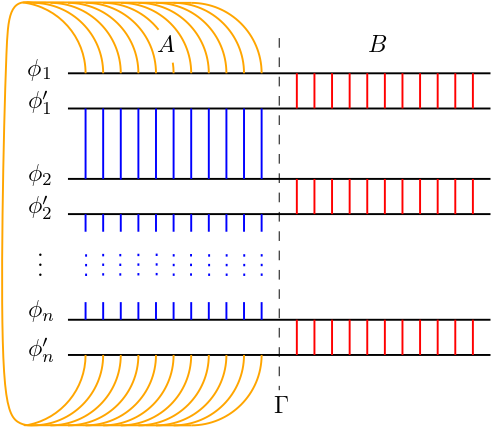}
\caption{The gluing procedure due to the replica trick. Gluing due to the partial trace over $B$ is represented in red, due to multiplication of the reduced density matrices $\rho_A$ in blue, and due to the final total trace in yellow.}
\label{fig:replica}
\end{figure}

In this section we adapt the replica trick to generalised quantum Lifshitz theories.
For the QLM the replica method was reviewed in \cite{Zaletel:2011ir, Zhou:2016ykv}.
Our starting point is the ground state density matrix $\rho=\left | \psi_0 \rangle \langle \psi_0\right|$, 
with $| \psi_0 \rangle$ as in \eqref{def-gs-wf}. 
Now divide the manifold into two regions $A$ and $B$ and assume that the Hilbert space splits as 
$\mathcal{H}=\mathcal{H}_A\otimes\mathcal{H}_B$. This allows us to write the density matrix as
\begin{equation}
\label{eq:compl-dens-matrix}
\rho=\frac{1}{Z}\int [\D\phi^A][\D\phi^B][\D\phi'^A][\D\phi'^B]e^{-\frac{1}{2}S[\phi^A]-\frac{1}{2}S[\phi^B]-\frac{1}{2}S[\phi'^A]-\frac{1}{2}S[\phi'^B]}|\phi^A\rangle \otimes  | \phi^B \rangle \langle  \phi'^A | \otimes \langle \phi'^B |,
\end{equation}
where the field eigenstates $\{|\phi^A\rangle\}$ and $\{|\phi^B\rangle\}$ provide orthonormal bases in 
$\mathcal{H}_A$ and $\mathcal{H}_B$, respectively, and we have used that $\phi^{A}$ and $\phi^{B}$ 
are free fields with support on non-overlapping subsets of $\M$ to write $S[\phi]=S[\phi^A]+S[\phi^B]$. 
We then construct $\rho_A^n$ by the gluing procedure represented in Figure~\ref{fig:replica}.
Each copy of $\rho$ is represented by a path integral as in \eqref{eq:compl-dens-matrix} with fields 
labelled by a replica index $i=1,\ldots,n$. The partial trace over field degrees of freedom with support in $B$ 
gives the following reduced density matrix for the $i$-th replica,
\begin{align}
\label{eq:part-trace}
\begin{split}
\rho_A &= \Tr_B\left(\rho\right) \\
&=\frac{1}{Z}\int [\D\phi^A_i][\D\phi'^A_i][\D\phi_i^B]e^{-\frac{1}{2}S[\phi_i^A]-\frac{1}{2}S[\phi_i'^A]-S[\phi_i^B]} 
| \phi_i^A \rangle \langle \phi_i'^A |.
\end{split}
\end{align}
Multiplying together two adjacent copies of the reduced density matrix gives
\begin{align}
\begin{split}
\rho_A^2 &= \frac{1}{Z^2}\int [\D\phi^A_i][\D\phi'^A_{i+1}] [\D\phi^A_{i+1}] [\D\phi^B_i][\D\phi^B_{i+1}]
e^{-\frac{1}{2}S[\phi_i^A]-\frac{1}{2}S[\phi_{i+1}'^A]-S[\phi_{i+1}^A]-S[\phi_i^B]-S[\phi_{i+1}^B]} | \phi_i^A \rangle  \langle \phi_{i+1}'^A |
\end{split}.
\end{align}
The $\delta$-function coming from $\langle  \phi_{i}'^A | \phi_{i+1}^A \rangle$ forces the identification
$\phi_i'^A=\phi_{i+1}^A$, effectively gluing together the primed field from replica $i$ and the unprimed 
field from replica $i+1$, as indicated in Figure~\ref{fig:replica}. 
It follows that multiplying $n$ copies of the reduced density matrix gives rise to pairwise gluing conditions 
$\phi_i'^A=\phi_{i+1}^A$ for $i=1,\ldots,n-1$, and when we take the trace of the complete expression we get 
a gluing condition between the first and last replicas, $\phi_1^A=\phi_n'^A$. 
Combining this with the gluing condition $\phi_i^B=\phi_i'^B$ for $i=1,\ldots,n$ from the partial trace in 
\eqref{eq:part-trace}, and the boundary condition $\phi_i^A\big\vert_\Gamma=\phi_i^B\big\vert_\Gamma$, 
where $\Gamma$ denotes the entangling surface separating $A$ and $B$, we see that \emph{all} the replica 
fields are forced to agree on $\Gamma$. The final result for $\rho_A^n$ is then
\be
\label{tr-nrhoa-v1}
\tr_A\left(\rho_A^n\right)= {1\over Z_{A\cup B}^n}   \int_{\rm bc}\prod_{i=1}^n [\D\phi_i^A]  e^{-\sum_{i=1}^n 
S[\phi_i^A] } \int_{\rm bc}\prod_{i=1}^n [\D\phi_i^B] \, e^{ -\sum_{i=1}^n S[\phi_i^B]} 
\ee
with
\be\label{bc-at-the-cut-v1}
{\rm bc}~~: \quad {\phi_i^A}\big\vert_{\G}(x)={\phi_j^B}\big\vert_{\G}(x)\equiv\text{cut}(x)\,, \quad i, j=1, \dots, n\,, 
\ee
where $\text{cut}(x)$ is some function of the boundary coordinates.
We write the denominator in \eqref{tr-nrhoa-v1} as  $Z_{A\cup B}^n$ to emphasise that it contains $n$ 
copies of the partition function of the original system before any subdivision into fields on $A$ and $B$.
In the numerator, however, the field configurations of the different replicas are integrated over independently,
except that the replicated fields are subject to the boundary conditions \eqref{bc-at-the-cut-v1} (up to the periodic 
identification $\phi \sim \phi+2\pi R_c$). 

In order to take the periodic identification into account when applying boundary conditions, we separate each 
replicated field into a classical mode and a 
fluctuation, following a long tradition, see {\it e.g.} \cite{DiFrancesco1997, Ginsparg:1988ui},
\be
\phi^{A(B)}_i= \phi^{A(B)}_{i, \rm cl} + \varphi^{A(B)}_i\,, \qquad i=1, \dots, n\,. 
\ee
The modes $\phi_{i, \rm cl} $ satisfy the following classical equations of motion and boundary conditions,
\be\label{eom-classical}
\mathcal P_z \phi^{A(B)}_{i, \rm cl} (x) =0\,, \qquad {\phi^{A(B)}_{i, \rm cl}}(x)\big\vert_{\G}
= \text{cut}(x)+ 2\pi R_c w_i^{A(B)}\,, \qquad i=1, \dots, n\,,
\ee
where $w_i^{A(B)}$ are integers indicating the winding sector. The classical field determines the total field value 
at the entanglement cut, while the fluctuating field $\varphi$ satisfies Dirichlet boundary conditions,
\be\label{dbc}
{\varphi^{A(B)}_i}(x)\big\vert_{\G}=0  \qquad i=1, \dots, n\,. 
\ee
In two dimensions this condition, along with the equation of motion of the classical fields, ensures that the 
action factorises~\cite{Zhou:2016ykv}, %
\footnote{Derivatives of the classical fields are in general discontinuous at the cut, but the left- and right-derivatives 
remain bounded.}
\be\label{split-s}
S[\phi^{A(B)}_i]= S[\phi^{A(B)}_{i, \rm cl} ]+S[ \varphi^{A(B)}_i]\,, \qquad i=1, \dots, n\,. 
\ee
The decomposition of the action is less trivial in higher dimensions but it can be achieved if the Dirichlet 
boundary condition  
on the fluctuating field at the entanglement cut is augmented by further conditions. 
It is straightforward to check that imposing \eqref{dbc} along with \eqref{cond-variation}-\eqref{bc-sphere-fluctuations} 
on the fluctuating fields at the cut leads to a well-posed variational problem as well as self-adjointness of the 
operator $\mathcal{P}_{z,\M}$ on $\M$. As was discussed earlier, this combination of conditions amounts to the vanishing 
of the Laplace operator and its integer powers acting on the fluctuating fields at the boundary.
This turns out to be enough to ensure that the {\it total} action splits according to \eqref{split-s} (once again the equations 
of motion for the classical fields have to be used to achieve factorization). 
We note, that with this prescription and using the classical equations of motion, the boundary terms in the action 
can be written in a form that only depends on the classical part of the field,
\be\label{split-s-specific}
S_\p[\phi^{A(B)}_i]= S_\p[\phi^{A(B)}_{i, \rm cl}]\,.
\ee
In the presence of winding modes, there remains some redundancy in the classical part of the action, as further discussed
in Appendix~\ref{sec:winding-torus} where we compute the contribution from the classical winding sector for the $d$-torus. 

As a consequence of \eqref{split-s}, the fluctuating modes $\varphi_i$ simply contribute as $n$ independent fields 
obeying Dirichlet boundary conditions \eqref{dbc} at the entanglement cut. 
For the classical modes, on the other hand, we can solve for the $A$ and $B$ sectors simultaneously, as the boundary value 
problem~\eqref{eom-classical} has a unique solution in $A\cup B\smallsetminus\G$, up to winding numbers.
At the entanglement cut only relative winding numbers matter and we can choose to write the boundary 
conditions for the classical fields as~\cite{Zaletel:2011ir}
\be\label{eom-classical-final}
{\phi_{i, \rm cl}}\big\vert_{\G}(x)= \text{cut}(x)+ 2\pi R_c w_i\,, \qquad i=1, \dots, n-1 \,,  \qquad 
{\phi_{n, \rm cl}}\big\vert_{\G}(x)= \text{cut}(x)\,. 
\ee
Thus, the trace of the $n$-th power of the reduced density matrix reads
\be
\label{tr-nrhoa-v2}
\tr_A\left(\rho_A^n\right)= {1\over Z_{A\cup B}^n}  \, \prod_{i=1}^n \int_{\rm D} [\D\varphi_i^A]  
e^{-S[\varphi_i^A] }\, \prod_{i=1}^n \int_{\rm D} [\D\varphi_i^B] \, e^{ -S[\varphi_i^B]} \, W(n)\,,
\ee
where $W(n)$ is the contribution coming from summing over all classical field configurations satisfying the 
boundary conditions \eqref{eom-classical-final}. 
The subscript $D$ on the integral sign is a reminder that the the fluctuating fields obey Dirichlet boundary 
conditions. 

At this point we need to distinguish the spherical case from the toroidal one. We start by analysing the 
problem on the $d$-sphere, which turns out to be particularly simple. 

\paragraph{$d$-sphere.}

We closely follow the treatment of the two-dimensional case in \cite{Zhou:2016ykv}.
The crucial observation here is that the winding mode can be reabsorbed by the global shift symmetry 
of the action, $S[\phi_i]=S[\phi_i+\phi_i^0]$ with constant $\phi_i^0$, as mentioned in Section \ref{sec:gen-qlm}.
Indeed, since the fields satisfying the classical equation of motion include any constant part of the total field, 
we can use the symmetry to rewrite their boundary conditions as
\be
\phi^{\rm cl}_{i|\Gamma} (x) = \text{cut}(x) +2\pi\, R_c \,\omega_i+\phi_{i}^0\,.
\ee 
We then choose $\phi_i^0=-2\pi\, R_c \,\omega_i$ to cancel out all winding numbers. 
The boundary conditions then become
\be\label{classical-phi-sphere}
\phi^{\rm cl}_{i|\Gamma} (x) =\text{cut}(x)\,, \qquad i=1, \dots, n\,,
\ee  
and the sum over classical configurations can be written as
\be
W(n)= \sum_{\phi_{i, \rm cl}} e^{-\sum_i S[\phi_{i, \rm cl}]}=
\sum_{\phi_{n, \rm cl}} e^{- n \, S[\phi_{n, \rm cl}]}=
\sum_{\phi_{n, \rm cl}} e^{- S[\sqrt{n} \, \phi_{n, \rm cl}]}\,.
\ee
Consequently, we have for the $d$-sphere
\be\label{rho_intermediate_expression}
\tr_A\left(\rho_A^n\right)= {1\over Z_{A\cup B}^n}  \prod_{i=1}^n \int_{\rm D} [\D\varphi_i^A]  e^{-S[\varphi_i^A] } \prod_{i=1}^n \int_{\rm D} [\D\varphi_i^B] \, e^{ -S[\varphi_i^B]} \, \sum_{\phi_{n, \rm cl}} e^{- S[\sqrt{n} \, \phi_{n, \rm cl}]}\,.
\ee
We can now combine the $n$-th fluctuating fields with support on $A$ and $B$ and the $n$-th classical field to define
\be
\Phi_n= \varphi_n+ \sqrt{n}\, \phi_{n, \rm cl}\,,
\ee
with $\varphi_n=\varphi_n^{A(B)}$ in $A(B)$. 
Notice that the effective compactification radius of $\Phi_n$ is now  $\sqrt{n} R_c$~\cite{Zaletel:2011ir, Zhou:2016ykv}. 
The path integral over $\varphi_n^A$ and $\varphi_n^B$ along with the contribution $W(n)$ from the rescaled classical field
amounts to the partition function on the whole $d$-sphere for the combined field $\Phi_n$, which is equal to the partition function
of the original field up to a factor of $\sqrt{n}$ due to the different compactification radius, and it therefore almost exactly cancels
one power of the original partition function in the denominator in \eqref{rho_intermediate_expression},
\be
\tr_A\left(\rho_A^n\right) &=& {1\over Z_{A\cup B}^n}  \prod_{i=1}^{n-1} \int_{\rm D} [\D\varphi_i^A]  e^{-S[\varphi_i^A] } \prod_{i=1}^{n-1} \int_{\rm D} [\D\varphi_i^B] \, e^{ -S[\varphi_i^B]} \, \sqrt{n} \, Z_{A\cup B}
\\ \nn
&=&
\sqrt{n}\, \left({Z_{D, A} Z_{D, B}\over Z_{A\cup B}}\right)^{n-1}\,,
\ee
where $Z_{D, A(B)}\equiv\int_{\rm D} [\D\varphi_i^{A(B)}]  e^{-S[\varphi_i^{A(B)}] }$ denotes the Dirichlet partition function on $A(B)$.
Hence, the entanglement entropy is given by
\be
\label{EE-sphere}
S_{EE}= - \lim_{n\to 1} \p_n \tr_A\left(\rho_A^n\right)= -\log{\left({Z_{D, A} Z_{D, B}\over Z_{A\cup B}}\right)} -{1\over 2}\,,
\ee
and the original problem has been reduced to the computation of partition functions with appropriate boundary conditions 
on the regions $A$ and $B$ and $A\cup B$. 

We will consider the case where the $d$-sphere is divided into two hemispheres. Then we only have to compute 
the partition function on the full sphere and a Dirichlet partition function on a hemisphere. These are in turn 
given by determinants of the appropriate GJMS operators. The detailed computation is described in 
Section~\ref{sec:sphere} and Appendices~\ref{sec:dowker} and~\ref{sec:alternative-dowker}.

\paragraph{$d$-torus.}

We now apply the replica method in the case of a $d$-torus. 
We cut the torus into two parts, thus introducing two boundaries which are given by two disjoint periodically identified $(d-1)$-intervals (in $d=2$ this is simply an $S^1$). 
As explained before, all the fields have to agree at the cuts $\Gamma_a$ (where now the index $a=1,2$ labels each cut). For the quantum fields this simply implies that they need to satisfy Dirichlet boundary conditions, that is
\be\label{dirichlet-conds}
\varphi_{i}^{A(B)}\big\vert_{\Gamma_a} = 0\,, \quad\quad i=1, \dots, n, \qquad a=1,2\, .
\ee 
As explained earlier, further conditions are necessary in dimensions $d>2$, and we demand that the conditions \eqref{cond-variation} hold at the cut for the fluctuating fields.

Now consider the classical fields on the torus. 
We can use the global shift symmetry discussed in Section \ref{sec:gen-qlm}
to write the boundary conditions for the classical fields as
\be
{\phi_{i, \rm cl}}\big\vert_{\G_a}(x) = \text{cut}_a(x)+2\pi R_c \,\omega_i^a +\phi_{i}^0\,
\ee
with $\phi_i^0$ constant on the whole torus. As in the spherical case, we can choose the $\phi_i^0$ 
so that they absorb the winding numbers from one of the cuts,
\be
&&{\phi_{i, \rm cl}}\big\vert_{\G_1}(x) = \text{cut}_1(x)\,,\qquad\qquad i=1, \dots, n\,,
\\
&& {\phi_{i, \rm cl}}\big\vert_{\G_2}(x) = \text{cut}_2(x)+2\pi R_c \,\omega_i\,, \qquad i=1, \dots, n\,,
\ee
where $\omega_i:\, =\left(\omega_i^2 - \omega_i^1\right)$. 
We are effectively left to deal with winding sectors at a single entanglement cut and since only the relative 
winding number between adjacent replicas matters we can eliminate one more winding number to obtain 
\be
&&{\phi_{i, \rm cl}}\big\vert_{\G_1}(x) = \text{cut}_1(x)\,,\qquad i=1, \dots, n\,
\\
&& {\phi_{i, \rm cl}}\big\vert_{\G_2}(x) = \text{cut}_2(x)+2\pi R_c \,\omega_i\,, \quad i=1, \dots, n-1, \qquad ~~
{\phi_{n, \rm cl}}\big\vert_{\G_2}(x) = \text{cut}_2(x)\,.~~~~~
\ee
At this point, we can use the same unitary rotation $U_n$ as in \cite{Zhou:2016ykv, Zaletel:2011ir}, 
to bring the classical fields ${\phi_{i, \rm cl}}$ $i=1, \dots, n$ into a canonical form constructed to separate 
the contribution from the winding modes from the contribution from modes subject to boundary conditions 
given by the functions $\text{cut}_{1,2}(x)$. 
Concretely, we define the matrices
\be\label{def-matrix-U}
U_n= 
\begin{bmatrix}
{1\over \sqrt 2} ~&~ -{1\over \sqrt{2}}~ &~0 & ~&\dots\\
{1\over \sqrt 6} ~& ~-{1\over \sqrt{6}} ~ & ~ - {2\over \sqrt{6}} & ~~0& \dots \\
\vdots 
\\
{1\over \sqrt{n(n-1)}} ~& ~{1\over \sqrt{n(n-1)}} & \dots~ & ~\dots & -\sqrt{1-{1\over n}}\\
{1\over \sqrt{n}} ~& ~{1\over \sqrt{n}} & \dots~ & ~\dots & {1\over \sqrt{n}}\\
\end{bmatrix}\,,
\ee
and 
\be
M_{n-1} =\text{diag}\left(1, \dots, 1, {1\over n} \right) U_{n-1}\,,
\ee
so that we have
\be
\label{initial-bc-torus2-v2}
&& \bar\phi^{\rm cl}_i|_{\G_1}(x)=0 \quad\quad\quad\quad\quad\quad\quad\quad\quad\quad\quad\quad\quad  i=1, \dots, n-1\,,\nonumber\\
&&\bar\phi^{\rm cl}_n|_{\G_1} (x)=  \sqrt{n}\, \text{cut}_1(x)\,, \\ \nn
&& \bar\phi^{\rm cl}_i|_{\G_2}(x)= 2\pi R_c\,  \left(M_{n-1}\right)_{ij}\omega_j\,, \quad\quad\quad\quad\quad i=1,\dots, n-1\,, \nonumber\\
&& \bar\phi^{\rm cl}_n|_{\G_2} (x)=  \sqrt{n}\, \text{cut}_2(x) + {2\pi\, R_c\over \sqrt n} \sum_{i=1}^{n-1} \omega_i \,.
\ee
Hence, the sum over all the classical configurations reduces to a sum over the vector 
$\mathbf w=\left(\omega_1, \dots, \omega_{n-1}\right) \in\Z^{n-1}$ and an integral over the $n$-th classical mode. 
Notice that, as for the spherical case,  the $n$-th classical mode $\bar\phi^{\rm cl}_n$ has a compactification 
radius amplified by $\sqrt{n}$, due to the rotation \eqref{def-matrix-U}. 
We want to use this mode to reconstruct a full partition function on the torus, that is define
\be
\label{Phi_field}
\Phi_n= \varphi_n+\bar \phi_n^{\rm cl}\,,
\ee
so that 
\be\label{int-n-field-torus}
\int_{\rm D} [d\varphi_n^A] e^{-S[\varphi_n^A]}\int_{\rm D} [d\varphi_n^B] e^{-S[\varphi_n^B]} 
\int  [d\bar{\phi}_n^{\rm cl}] e^{-S[ \bar{\phi}_n^{\rm cl}]} = \sqrt{n} \, Z_{A\cup B}\,,
\ee
where the $\sqrt{n}$ factor on the right-hand-side of \eqref{int-n-field-torus} accounts for the different 
compactification radius. 
Thus, the replica method finally gives
\be\label{tr-rho-and-wind-sect-torus}
\tr_A\left(\rho_A^n\right) = \sqrt{n} \left({Z_{D, A} Z_{D, B} \over Z_{\rm free, A\cup B}}\right)^{n-1} \overline{W}(n)\,,
\qquad
\ee
where $\overline{W}(n)$ contains the contributions from the first $n-1$ classical configurations 
satisfying the boundary conditions in \eqref{initial-bc-torus2-v2} at $\Gamma_{1 (2)}$. 

In two dimensions the classical fields are uniquely determined by the equations of motion and 
the boundary conditions \eqref{initial-bc-torus2-v2}, and thus the classical action has only one vacuum. 
The contribution from the winding sector is then simply given by the sum over the corresponding 
winding modes \cite{Zhou:2016ykv, Zaletel:2011ir},
\be
\overline{W}(n)= \sum_{\mathbf{w} \in \Z^{n-1}} e^{-\sum_{i=1}^{n-1}S[\bar{\phi}^{\rm cl}_i]}\,.
\ee
However, in higher dimensions ($d>2$) the conditions \eqref{initial-bc-torus2-v2} do not uniquely specify the 
vacua of the classical action. In other words, our construction is consistent for more than one set of boundary 
conditions applied on derivatives of the classical fields and the value of the boundary action depends on the 
boundary conditions.  This is the redundancy mentioned in Section \ref{sec:gen-qlm}. 

The classical field satisfies a higher-derivative equation of motion, whose general solution
is parametrised by $z/2$ constants. The boundary condition imposed on $\Phi_n$ will fix one
of these constants but we need to add $z/2-1$ further boundary conditions for the classical field to fix 
the rest. The value of the boundary terms in the action will in general depend on the choice of 
boundary conditions.

In the present work we impose a generalised form of Neumann boundary conditions on 
derivatives of the classical fields,
\begin{equation}
\label{eq:ext-boundary-conditions-appendix}
\partial_n\Delta^{k}\bar{\phi}^{\rm cl}_i\big\vert_{\partial_r}=0\quad k=0,\ldots,\frac{z}{2}-2.
\end{equation} 
This prescription is compatible with the conditions imposed on the fluctuations,
\be
\Delta^{k}\phi_i\big\vert_{\partial M}&=\Delta^{k}(\bar{\phi}^{\rm cl}_i+\varphi_i)\big\vert_{\partial M}
&=\Delta^{k}\bar{\phi}^{\rm cl}_i\big\vert_{\partial M},\quad \text{for}\ k=1,\ldots,\frac{z}{2}-1\,, 
\ee
and, at the same time, it gives a non-vanishing classical boundary action, which is important in 
order for the sum over winding modes to converge.
The contribution from winding modes is then accounted for in any number 
dimentions by computing 
\be\label{Wn-torus}
\overline{W}(n)= \sum_{\mathbf{w} \in \Z^{n-1}} e^{-\sum_{i=1}^{n-1}S[\bar{\phi}^{\rm cl}_i]}\,,
\ee
where the classical fields $\bar{\phi}^{\rm cl}_i$ satisfy the boundary conditions \eqref{initial-bc-torus2-v2} and 
the Neumann conditions in \eqref{eq:ext-boundary-conditions-appendix}. 
The details of the computation are reported in Appendix \ref{sec:winding-torus}. 

Finally, the entanglement entropy  for the $d$-torus is given by
\be
\label{EE-torus-done}
S_{EE} = -\log{\left({ Z_{D, A} Z_{D, B} \over Z_{ A\cup B}}\right)} -{1\over 2} -\overline{W}'(1)\,,
\ee
since the winding sector is normalised such that $\overline{W}(1)=1$.
The computation of the partition functions for the $d$-torus is presented in Section \ref{sec:torus} below. 

We close this section by noting that even though winding numbers come into play across entanglement 
cuts in our computation, we are restricting our attention to a single topological sector of the original theory 
on the $d$-torus.
Indeed, since we periodically identify the field, we could 
consider winding sectors on the $d$-torus itself,
\be
\phi(x_1+ L_1, \dots, x_d+L_d) = \phi(x_1, \dots, x_d) +  2\pi L^I m_I, \quad I=1, \dots, d\,, ~~ m_I\in \Z\,.
\ee
We have set $m_I=0$ in our calculations in the present paper but a more general study can be carried out, 
evaluating the contribution from winding sectors $W(n,m_L)$ associated with an entanglement cut for each 
topological configuration, and then summing over the $m_I$. The corresponding topological contributions to 
entanglement entropy in a scalar field theory on a two-dimensional cylinder, are obtained in \cite{Zaletel:2011ir}.


\section{Entanglement entropy on a hemisphere}
\label{sec:sphere}

\begin{figure}
\centering\includegraphics[scale=0.351]{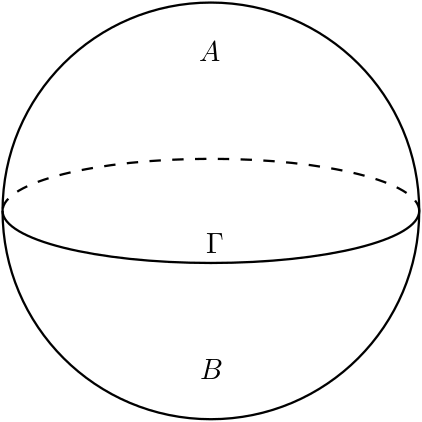}
\caption{The sphere is cut into hemispheres $A$ and $B$ by an entangling cut at the equator.}
\label{fig:sphere}
\end{figure}

In this section we calculate the universal finite terms of entanglement entropy in GQLM 
resulting from the division of a $d$-sphere into two $d$-hemispheres $A$ and $B$ by an 
entanglement cut at the equator as shown in Figure \ref{fig:sphere}. 
According to the replica calculation in Section \ref{sec:replica}, we have to compute \eqref{EE-sphere}, 
where now $A$ and $B$ are the two $d$-hemispheres, and the bulk action contains the 
GJMS operator \eqref{GJMS_op_sphere-1} with $2k=d$. 
The partition function on the whole manifold ($Z_{A\cup B}$ in \eqref{EE-sphere}) 
contains a zero mode, which should be treated separately \cite{DiFrancesco1997, Ginsparg:1988ui},
\be\label{eq-partition-sphere}
Z_{A\cup B} = 2\pi R_c \sqrt{{g \over \pi} \mathcal{A}_d} \, \left({\det}'{\left(\frac{g}{\pi}\mathcal{P}_{d, S^d}\right)}\right)^{-\half}
\ee
where $\mathcal{A}_d$ is the area of the $d$-sphere and the $\sqrt{{g \over \pi} \mathcal{A}_d}$ factor 
comes from the normalisation of the eigenfunction corresponding to the zero eigenvalue.  
The functional determinant ${\det}'$  is the (regularised) product of the non-zero eigenvalues.
The operator \eqref{GJMS_op_sphere-1} on the unit $d$-hemisphere with Dirichlet boundary conditions 
does not have a zero eigenvalue, so the partition functions on the subsystems $A$ and $B$ can be directly 
computed via regularised functional determinants. 
Hence we can write \eqref{EE-sphere} as
\begin{equation}
\label{eq:entropy-GJMS-sphere}
S[A]=\frac{1}{2}\log \left(\frac{\det \mathcal{P}_{d, H^d_D} \det \mathcal{P}_{d, H^d_D} }{{\det}' {\mathcal{P}_{d, S^d}}} \right)+\log\left(\sqrt{4\pi g \mathcal{A}_d}R_c\right)-\frac{1}{2},
\end{equation}
where the $D$ subscript on $H^d_D$ indicates Dirichlet boundary conditions on the fields at the boundary 
of the $d$-hemisphere $H^d$. At the end of the day, the entanglement entropy can only depend on 
the combination $g \mathcal{A}_d$. All factors of $g/\pi$ inside functional 
determinants must therefore cancel out in the final result and going forward we simply leave them out of our formulas.

We now turn to the explicit computation of the functional determinants appearing in \eqref{eq:entropy-GJMS-sphere}. 
In a series of papers \cite{Chang1993, Dowker1994, Dowker2011, Dowker2013}, Dowker calculates determinants of 
GJMS operators on spheres in any even dimension $d$ and for any degree $k\le {d/2}$ via $\zeta$-function methods. 
We give a self-contained review of these calculations in Appendix~\ref{sec:dowker}, partly to adapt them to our notation
and partly to have all the results we want to use in one place. Determinants of critical GJMS operators (where the degree 
$2k$ of the operator matches $d$) on spheres and hemispheres are expressed in terms of multiple $\Gamma$-functions
in~\cite{Dowker2011}. A simplified version of these results, expressing them in terms of the more familiar 
Riemann $\zeta$-function, is presented in Appendix~\ref{sec:alternative-dowker}.

The starting point of Dowker's computation is the observation that the determinant of the GJMS operator on a 
$d$-sphere, given in terms of the spectral $\zeta$-function, can be obtained as a sum of the corresponding 
determinant on a $d$-hemisphere with Dirichlet and Neumann boundary conditions~\cite{Dowker:1993jv, Dowker2011} 
(again expressed in terms of spectral $\zeta$-functions). 
On the hemisphere with Dirichlet boundary conditions the log-determinant of the GJMS operator is given by
\begin{equation}
\label{eq:GJMS-logdet-crit-hem-D}
\log \det \mathcal{P}_{d,H^d_{D}}=-Z'_d(0,a_D, {d/2})=-\sum_{n=0}^{d}h^{D}_n(d)\zeta'(-n)- f^D(d),
\end{equation}
where $Z_d(s,a_D, {d/2})$ is the spectral $\zeta$-function corresponding to the GJMS operator of degree $2k=d$,
cf.~\eqref{eq:spectralzeta} and \eqref{eq:logdetp2k}. 
Here $\zeta$ is the Riemann $\zeta$-function, and $h^D_n$, $f^D$ are given by
\begin{align}
\label{eq:det-hem-dir-coeff-h-crit}
h^D_n(d)&=-\frac{1}{(d-1)!}\stirling{d}{n+1}+\frac{1}{d!}\stirling{d+1}{n+1}-
\sum_{j=0}^{d-n}\frac{(-1)^j}{(d-j)!}\binom{d+1}{j}\stirling{d-j+1}{n+1},\\
\label{eq:det-hem-dir-coeff-f-crit}
f^D(d)&= -\frac{1}{d!}\sum_{l=1}^{d-1}\log(l)(l-d)_{d+1}+M(d).
\end{align}
The $\stirling{d}{k}$ are Stirling numbers of the first kind, $(z)_k$ is a Pochhammer symbol, 
and $M(d)$ is a sum of harmonic numbers and generalised Bernoulli polynomials whose explicit 
form is not important to us, as it cancels in the final expression for the entanglement entropy. 
The derivations of $h^D_n$ and $f^D$ can be found in Appendix \ref{sec:det-hem-alt}, 
while the derivation of $M(d)$ can be found in \ref{sec:GJMS-det}, its explicit form is given in equation \eqref{eq:m-explicit}.   
These functions may seem quite complicated at first sight, but they all consist of well understood algebraic functions 
that can easily be evaluated using a computer. 
For the determinant of a critical GJMS operator on a hemisphere with Neumann boundary conditions 
we find a similar result
\begin{equation}
\label{eq:GJMS-logdet-crit-hem-N}
\log \det \PP_{d,H^d_{N}} =-Z'_d(0,a_N, {d/2})=-\sum_{n=0}^{d}h^{N}_n(d)\zeta'(-n)- f^N(d),
\end{equation}
with $h^N_n$ and $f^N$ given by
\begin{align}
\label{eq:det-hem-dir-coeff-h-crit}
h^N_n(d)&=\frac{1}{d!}\stirling{d+1}{n+1}-\sum_{j=0}^{d-n}\frac{(-1)^j}{(d-j)!}\binom{d}{j}\stirling{d-j+1}{n+1},\\
\label{eq:det-hem-dir-coeff-f-crt}
f^N(d) &=\log (d-1)!+f^D(d).
\end{align}
We note that our result in \eqref{eq:GJMS-logdet-crit-hem-N} differs from~\cite{Dowker2011} by a sign in the 
term $\log(d-1)!$. This is because we treat the zero mode separately as is apparent in \eqref{eq-partition-sphere} 
and \eqref{eq:entropy-GJMS-sphere}. 

\begin{figure}
\begin{center}
\includegraphics[scale=0.5]{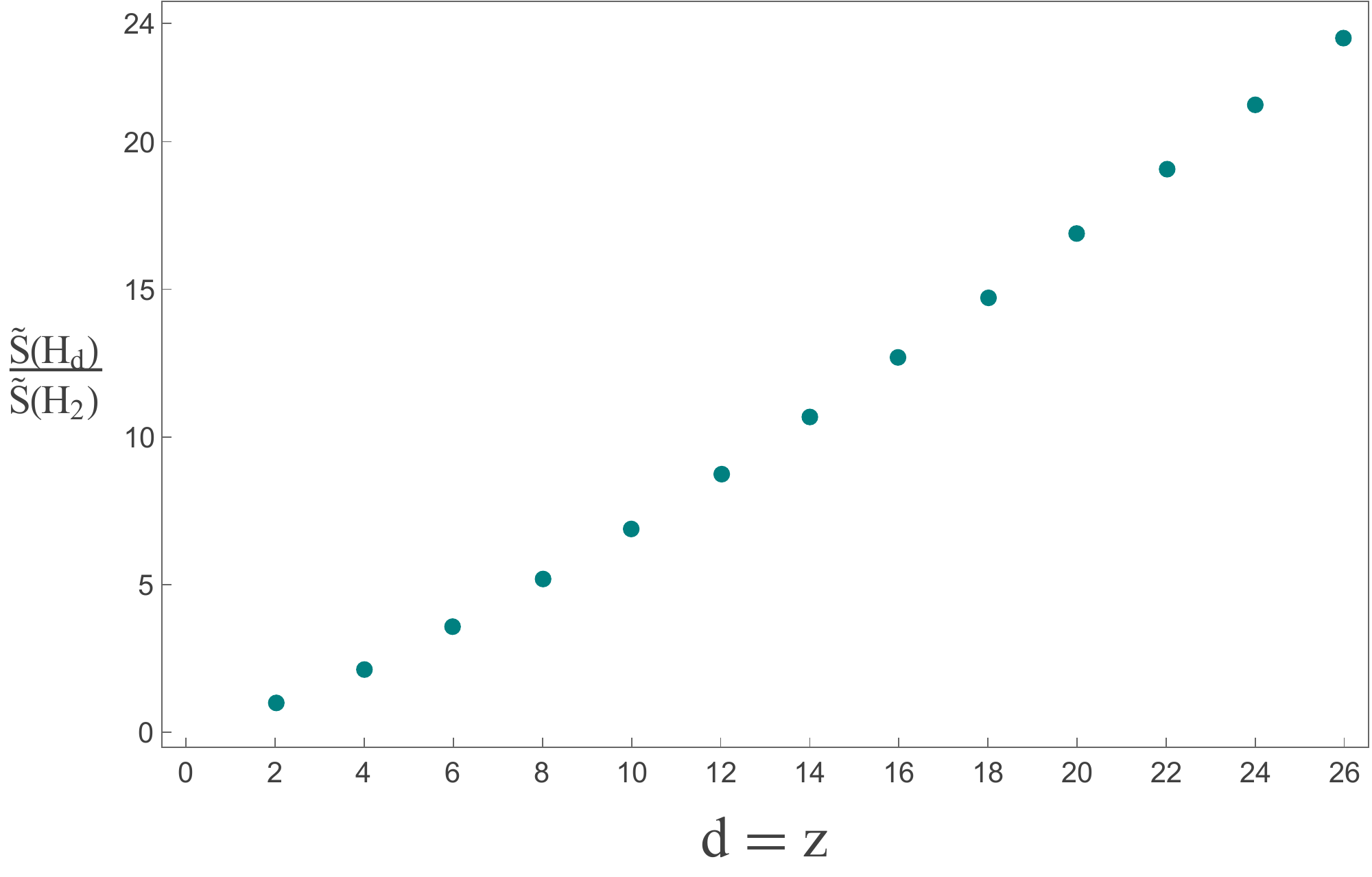}
\caption{The universal finite term \eqref{final-EE-sphere} in the entanglement entropy of GQLM on a hemisphere plotted
against the number of spatial dimension $d$ (which is equal to the critical exponent $z$). We normalise $S[H_d]$ with 
respect to the two-dimensional case, and set $g=R_c=1$.}
\label{fig:EE-hemisphere-2}
\end{center}
\end{figure}

As mentioned above, the log-determinant on the whole sphere is the sum of the log-determinants  
on the hemisphere with Dirichlet and Neumann boundary conditions~\cite{Dowker2011},
\begin{align}
\label{eq:GJMS-logdet-crit-sphere}
\log {\det}' \mathcal{P}_{d,S^d}&=\log \det \mathcal{P}_{d,H^d_{N}}+\log \det \mathcal{P}_{d,H^d_{D}}\,. 
\end{align}
With an eye towards the entropy formula \eqref{eq:entropy-GJMS-sphere}, we express the ratio of determinants as 
\begin{align}
2\log\det \PP_{d,H^d_{D}}-\log{\det}' \PP_{d,S^d}&=-\sum_{n=0}^{d}\left(h^{D}_n(d)-h^{N}_n(d)\right)\zeta'(-n)- f^D(d)+f^N(d)\notag\\
& =\sum_{n=0}^{d}h_n(d)\zeta'(-n)+\log(d-1)! \,,
\end{align}
where, using the properties of the binomial coefficients, one can write $h_n$ in the the following form
\begin{equation}
\label{eq:h-coeff}
h_n(d)=\frac{1}{(d-1)!}\stirling{d}{n+1}+\sum_{j=0}^{d-n}\frac{(-1)^j}{(d-j)!}\binom{d}{j-1}\stirling{d-j+1}{n+1}.
\end{equation}
Putting everything together, we obtain a surprisingly simple expression for the entanglement entropy 
of a hemisphere,
\begin{equation}
\label{final-EE-sphere}
S[H^d]=\frac{1}{2}\sum_{n=0}^{d}h_n(d)\zeta'(-n)+\log\left(\sqrt{4\pi g \mathcal{A}_d(d-1)!}R_c\right)-\frac{1}{2},
\end{equation}
with $h_n(d)$ given above in \eqref{eq:h-coeff}. 
For dimensions $d=2,4,6,$ and $8$, in the critical case $z=d$, the entropy is given explicitly by
\begin{eqnarray}
\label{explicit-ex-EE-sphere}
d=z=2: && ~~
S_{EE}= \log\left(\sqrt{8 \pi g} R_c\right) -{1\over 2}
\\ 
d=z=4: && ~~
S_{EE}= \log\left( 4\sqrt{2 g} \pi R_c \right) -{1\over 2}-{\zeta(3)\over 4\pi^2}
\\ 
d=z=6: &&~~
S_{EE}= \log\left( 16\sqrt{\pi^3 g } R_c\right) -{1\over 2}-{ 15\, \zeta(3)\over 32\pi^2}+{3\, \zeta(5)\over 32 \pi^4 }
\\ 
d=z=8: &&~~
S_{EE}= \log\left( 32\sqrt{3 g } \pi^2 R_c\right) -{1\over 2}-\frac{469\,\zeta(3)}{720\pi^2}+\frac{7\,\zeta(5)}{24\pi^4}-\frac{\zeta(7)}{32\pi^6},
\end{eqnarray}
and more values are plotted in Fig. \ref{fig:EE-hemisphere-2}.
The two-dimensional case agrees with the result presented in \cite{Zhou:2016ykv}. 
Notice that the logarithmic term depends on the product $R_c \sqrt{g}$, which is independent of rescaling of the fields. 
Hence, in the case of a $d$-sphere cut into two $d$-hemispheres, the finite universal terms of the entanglement 
entropy~\eqref{final-EE-sphere} are constant, they only depend on the physical compactification radius $R_c \sqrt{g}$ 
of the target space, which appears in the above expression through the zero modes. 
%


Explicit results can also be obtained for the subcritical case, i.e. when $z < d$. In this case, the entanglement entropy on a hemisphere is simply given by the 
difference of log-determinants on the hemisphere with Dirichlet and Neumann boundary conditions. In \cite{Dowker:2014rva} this difference was shown to be 
equal to a ``boundary free energy’’, initially defined for hemispheres in 4-dimensional CFTs in \cite{Gaiotto:2014gha}. 
Its value for $d=4$ and $z=2$ was calculated in \cite{Klebanov:2011gs, Gaiotto:2014gha, Dowker:2014rva}, while the value for 
$d=6$ and $z=2$ appeared in \cite{Dowker:2014rva}.%
\footnote{We thank Stuart Dowker for bringing these results and references to our attention.}
Here we list a few examples of the entanglement entropy on a hemisphere in the subcritical case with $z$ and $d$ both even integers,
\be
d=4, z=2: \qquad && S_{EE}= -\frac{\zeta (3)}{8 \pi ^2}\,,
\\
\nn
d=6, z=2: \qquad && S_{EE}= \frac{\zeta (3)}{96 \pi ^2}+\frac{\zeta (5)}{32 \pi ^4}\,,
\\ \nn
d=6, z=4: \qquad && S_{EE}= -\frac{5 \zeta (3)}{48 \pi ^2}\,+\frac{\zeta (5)}{16 \pi ^4},
\\ \nn
d=8, z=2: \qquad && S_{EE}= -\frac{\zeta (3)}{720 \pi ^2}-\frac{\zeta (5)}{192 \pi ^4}-\frac{\zeta
	(7)}{128 \pi ^6}\,. 
\ee
The relevant functional determinants (for $z$ and $d$ even integers) were computed originally in \cite{Dowker2011} and are included in Appendix \ref{sec:dowker}.

The result in \eqref{final-EE-sphere} only depends on ``topological data'' represented by the scale invariant compactification 
radius of the target space and not on other geometric features. 
One might object that this is because we initially set the radius of the $d$-sphere to one, 
and thus our computations are insensitive to the geometry. 
Indeed, as mentioned in the Introduction, for smooth entangling cuts in even-dimensional CFTs, the entanglement entropy 
is expected to have a universal term proportional to the logarithm of a characteristic scale of the system with a constant of 
proportionality which depends on the central charge and on the Euler characteristic.
It can be checked that introducing a radius $R$ of the $d$-sphere in our problem
modifies the above results by adding a term proportional to 
\be
\label{log-term-s}
\Delta \chi \log R\,,
\ee
where $\Delta \chi$ is the change in the Euler characteristic due to dividing the $d$-sphere along the entanglement cut. 
For the two-dimensional case this was understood in \cite{Hsu:2008af}. 
Just as for a two-dimensional sphere, the change in the Euler characteristic vanishes for the chosen entanglement
cut (while having a non-smooth entangling surface can introduce further universal logarithmic terms). 
Indeed, on a non-unit sphere all eigenvalues entering our determinants are rescaled, and upon regularising this contributes, 
\be
&& \log \det \PP_{2k, H^d}= -d\, Z_d (0, a_D, k) \log R -Z'_d(0,a_D, k)\,, \\ 
&& \log \det \PP_{2k, S^d}= -d \, (Z_d (0, a_D, k)+Z_d (0, a_N, k)) \log R -Z'_d(0,a_D, k)-Z'_d(0,a_N, k)\,, \nn
\ee
instead of equations \eqref{eq:logdetp2k}, \eqref{eq:logdetp2k-s}.
Including the contribution coming from the normalisation of the zero-mode this would leave us with 
\be
{d\over 2} \left(1+Z_d (0, a_N, k)- Z_d (0, a_D, k)  \right) \log R\,,
\ee
but it is straightforward to check, using \eqref{check-logR-termHD} and \eqref{check-logR-termHN-critical}, that this combination vanishes. 
In fact, Dowker's construction of the determinant for the sphere as sum of determinants on hemispheres with Dirichlet and Neumann 
boundary conditions makes this quite transparent, since the spectral $\zeta$-function in the Neumann case is nothing but the 
Dirichlet one after subtracting the zero mode.%
\footnote{See \cite{Dowker2011,Diaz:2008hy, Bugini:2018def} for related studies of conformal anomalies for GJMS operators on 
spherical manifolds.}
Finally, we should stress that the sub-leading universal terms as \eqref{log-term-s} (which vanish here due to the chosen 
entanglement surface) are those expected in a $d$-dimensional CFT. The quantum field theory we are considering lives on a 
$(d{+}1)$-dimensional manifold, and yet due to the enhanced $d$-dimensional symmetries in the critical $d=z$ case, it 
has entanglement properties typical of $d$-dimensional CFTs. 


\section{Entanglement entropy on cut $d$-torus}
\label{sec:torus}

\begin{figure}
\centering\includegraphics[scale=0.351]{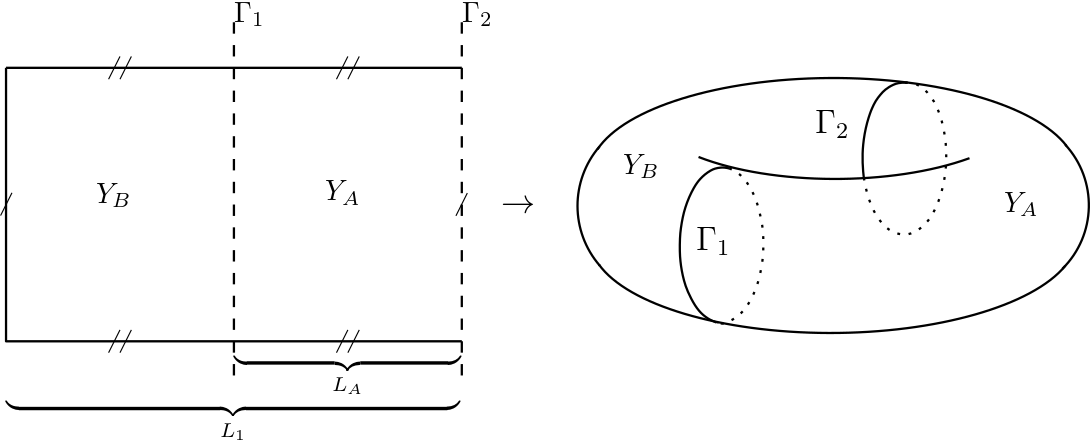}
\caption{The torus is cut into cylinders $Y_A$ and $Y_B$ by the two entangling cuts $\Gamma_1$ and $\Gamma_2$.}
\label{fig:torus}
\end{figure}

We now turn our attention to the sub-leading universal terms in the entanglement entropy 
on a flat $d$-dimensional torus with circumferences $L_1,\ldots,L_d$, 
\be
T^d_{L_1,\ldots,L_d}:\, = \mathbb{R}^d/(L_1\Z\cross\ldots\cross L_d\Z), 
\ee
that is cut into two $d$-cylinders: $Y_B:=[-L_B, 0]\cross T^{d-1}_{L_2,\ldots,L_d}$ and 
$Y_A:=[0,L_A]\cross T^{d-1}_{L_2,\ldots,L_d}$, where our conventions are $L_B>0$ and $L_1= L_A+L_B$. 
The two-dimensional case is shown in Figure~\ref{fig:torus}.
The replica method for the entanglement entropy on the torus was discussed in 
Section \ref{sec:replica}, and it requires us to compute \eqref{EE-torus-done}, 
where the winding sector contribution is given by \eqref{Wn-torus}, with the classical fields 
satisfying the equations of motion and boundary conditions expressed 
in \eqref{initial-bc-torus2-v2} and \eqref{eq:ext-boundary-conditions-appendix}. 
For the $d$-torus, the bulk and boundary terms in the action are given by \eqref{bulk-action-torus} and 
\eqref{boundary-action-torus}, respectively. The operator $\mathcal{P}_{d,T^d}$ in \eqref{operator-torus} 
is simply an integer power of the Laplacian. 
We first compute the quantum contribution to the entanglement 
entropy arising from the partition functions in \eqref{EE-torus-done}, and after that we tackle the 
winding sector contribution. 
All the detailed calculations of functional determinants are relegated to Appendix \ref{sec:Torusdet}, and 
those regarding the winding sector to Appendix \ref{sec:winding-torus}. 
In this section we collect the results and discuss some interesting limits. 

The operator $\mathcal{P}_{d,T^d}$ has a zero mode  
and as result the torus partition function is given by 
\be
Z_{A\cup B} = 2\pi R_c \sqrt{{g \over \pi} A_d} \, \left({\det}' {\left(\frac{g}{\pi} \PP_{d, T^d}\right)}\right)^{-\half},
\ee
where $A_d$ is the area of the $d$-torus.
As was the case for the sphere, the $\frac{g}{\pi}$ factor in the determinant only amounts 
to a rescaling of the torus to which the entanglement entropy is not sensitive, and we can ignore it in our calculations. 
On the $d$-cylinder with Dirichlet boundary conditions, on the other hand, there is no zero mode and we 
can write the (sub-leading terms of) entanglement entropy as
\be
\label{EE-tous-det-expression}
S[A]=\frac{1}{2}\log \left(\frac{\det \PP_{d, Y_{A,D}} \det \PP_{d, Y_{B, D} }}{{\det}' {\PP_{d, T^d}}} \right)
+\log\left(\sqrt{4\pi g \, A_d}R_c\right)-\frac{1}{2} -\overline{W}'(1)\,. 
\ee
The required functional determinants are evaluated in Appendix~\ref{sec:Torusdet}. 

By means of equations \eqref{log-det-Lk-torus} and \eqref{eq:log-det-flat-torus}, we find that the determinant on the full torus is given by
\begin{align}
\label{eq:log-det-flat-torus-GJMS}
\begin{split}
\log {\det}' \Delta_{T^{d}_{L_1,\ldots,L_d}}^{d/2}&=-\frac{d}{2}\zeta'_{T^{d}_{L_1,\ldots,L_d}}(0)\\
&=d\log(L_1)+\frac{d}{2}L_1\ \zeta_{T^{d-1}_{L_2,\ldots,L_d}}(-1/2)-\frac{d}{2}G'(0;L_1,\ldots,L_d)\,,
\end{split}
\end{align}
where $\zeta_{T^{d-1}_{L_2,\ldots,L_d}}(s)$ is the spectral $\zeta$-function on the $(d-1)$-torus. 
The auxiliary function $G$ is defined in Appendix \ref{app:det-L-torus}, as
\begin{align}
& G(s;L_1,L_2,\ldots,L_d):\, = \\ \nonumber
& \frac{2^{3/2-s}L_1^{s+1/2}}{\Gamma(s)\sqrt{\pi}}\sumprime_{\vec{n}_{d-1}\in\mathbb{Z}^{d-1}}
\sum_{n_1=1}^\infty\ \left(\frac{n_1}{\sqrt{\vec{n}_{d-1}^T \,\Xi_{d-1}\, \vec{n}_{d-1}}}\right)^{s-1/2} 
K_{s-1/2}\left(L_1 n_1 \sqrt{\vec{n}_{d-1}^T \,\Xi_{d-1} \, \vec{n}_{d-1}}\right),
\end{align}
where the primed sum indicates the omission of the zero mode, $K_\nu(z)$ is a modified 
Bessel function of the second kind and
$\Xi_{d-1}= \text{diag}\left(\left(2\pi/L_2\right)^2,\ldots,\left(2\pi/L_d\right)^2\right)$ is a diagonal matrix.
We have explicit expressions both for the spectral $\zeta$-function on the torus 
in \eqref{eq:closed-torus-zeta-function} and its derivative evaluated at $s=0$
in \eqref{eq:closed-torus-zeta-prime}, but at this stage we find it more convenient to use the above 
expression, and only insert explicit formulae at the end, after some cancellations. 

For $d$-cylinders with Dirichlet boundary conditions, using \eqref{log-det-Lk-cyl} and \eqref{eq:log-det-cut-torus},  
we obtain
\begin{align}
\label{eq:log-det-cut-torus-GJMS}
\log \det \Delta^{d/2}_{[0, L]\cross T^{d-1}_{L_2,\ldots,L_d}}
=&\>\frac{d}{2}\log(2 L)+\frac{d}{4}\zeta'_{T^{d-1}_{L_2,\ldots,L_d}}(0) \nonumber\\
&+\frac{d}{2}L \zeta_{T^{d-1}_{L_2,\ldots,L_d}}(-1/2)-\frac{d}{4}G'(0;2L,\ldots,L_d),
\end{align}
where $L=L_A$ for $Y_A$ and $L=L_1-L_A$ for $Y_B$. 
%
%
We can rewrite the difference between the log-determinants as
\begin{align}
\label{det-again-v2}
&\log \det \Delta^{d/2}_{[0,L_1-L_A]\cross T^{d-1}_{L_2,\ldots,L_d}} 
+\log \det \Delta^{d/2}_{[0, L_A]\cross T^{d-1}_{L_2,\ldots,L_d}}-\log {\det}' \Delta_{T^{d}_{L_1,\ldots,L_d}}^{d/2} \nonumber \\
&\quad=\frac{d}{2}\log(4 u(1-u))+\frac{d}{2}\zeta'_{T^{d-1}_{L_2,\ldots,L_d}}(0)-\frac{d}{4}G'(0;2 L_A,\ldots,L_d)+\nonumber\\
&\quad\quad-\frac{d}{4} G'(0;2(L_1-L_A),\ldots,L_d)+\frac{d}{2} G'(0;L_1,\ldots,L_d),
\end{align}
where the parameter $u=L_A/L_1$ characterises the relative size of the two $d$-cylinders. 

The explicit expression for $\zeta'_{T^{d-1}_{L_2,\ldots,L_d}}(0)$ is given by \eqref{eq:closed-torus-zeta-prime} 
with the replacement $d \to d{-}1$ and a relabelling of the sides $L_i$. 
As discussed in Appendix~\ref{sec:Torusdet}, despite its appearance the above expression 
is rather convenient to handle, thanks to the fast convergence of the modified Bessel functions 
contained in the auxiliary function $G$. 
The derivative of the function $G$ with respect to $s$, evaluated at $s=0$, is given by
\be
\label{def-Gprime}
G'(0, L, L_2, \dots, L_d)
&=& \sqrt{{8 L\over \pi}}
\sumprime_{\vec{n}_{d-1}\in\mathbb{Z}^{d-1}}\sum_{n_1=1}^\infty
\frac{(\vec{n}_{d-1}^T \,\Xi_{d-1}\, \vec{n}_{d-1})^{1/4}}{\sqrt{n_1}}
K_{-1/2}\big(L \,n_1 \sqrt{\vec{n}_{d-1}^T \,\Xi_{d-1} \, \vec{n}_{d-1}}\big) 
\nn \\  &=& 
2 \sumprime_{\vec{n}_{d-1}\in\mathbb{Z}^{d-1}}\sum_{n_1=1}^\infty\  
{\exp\left({-L\, n_1 \sqrt{\vec{n}_{d-1}^T \,\Xi_{d-1} \, \vec{n}_{d-1}}}\right)\over n_1}\,,
\ee
where we have used the explicit expression \eqref{bessel-half} for the modified Bessel function $K_{-\half}$.

As an explicit example of the above result, the determinant ratio for $z=d=2$ is explicitly given by
\begin{align}
\label{det-difference-torus-d2}
&\log \det \Delta_{[0,L_1-L_A]\cross S^1_{L_2}} 
+\log \det \Delta_{[0, L_A]\cross S^2_{L_2}}-\log \det \Delta_{T^{2}_{L_1, L_2}}
\nonumber\\
&\quad = \log\left( 2 u |\tau_1| \eta^2(2 u \tau_1)\right)
+ \log\left( 2 (1-u) |\tau_1| \eta^2(2 (1-u) \tau_1)\right) -\log\left(L_1^2 \eta^4(\tau_1)\right)
\nonumber\\
&\quad= -2\log L_1+\log\left(  4 u (1-u) |\tau_1|^2   \right)
+ 2 \log \left( {\eta(2(1-u) \tau_1)\, \eta(2 u \tau_1)\over \eta^2(\tau_1)}\right)\,,
\end{align}
where we used \eqref{det-torus-2d} and \eqref{det-cyl-2d} and introduced the notation 
$\tau_k  = i {L_1\over L_{k+1}}$, for $k=1\,, \dots\,, d{-}1$, 
for the aspect ratios of the general $d$-torus.

For the winding sector, the computations are detailed in Appendix \ref{sec:winding-torus}. The end result, 
given in \eqref{final-Wprime-torus}, is 
\begin{equation}
\label{final-Wprime-torus-2}
-\overline{W}'(1)=\log\sqrt{\Lambda_z}-\frac{1}{2}-\int\limits_{-\infty}^\infty\frac{dk}{\sqrt{\pi}}e^{-k^2}
\log(\sum_{\omega\in\Z} \exp(-\frac{\pi}{\Lambda_z}\omega^2-2i\sqrt{\frac{\pi}{\Lambda_z}}k\,\omega)), 
\end{equation}
with $\Lambda_z$ given by 
\be
\Lambda_z= 
g \, \pi \, R_c^2 \frac{ (-1)^{z/2}z! }{(1-2^z)B_z}\, \left( u^{1-z}+(1-u)^{1-z}\right) {1\over |\tau_1| \dots |\tau_{d-1}|}\,.
\ee
where $B_z$ are the Bernoulli numbers. For instance, in $d=2$ we have 
\be
\Lambda_2 = 4 \pi \, g\, R_c^2 \, {1\over u(1-u)} {1\over |\tau_1|}\,.
\ee

Finally, putting together the contributions from the functional determinants and the winding sector, \eqref{det-again-v2} 
and \eqref{final-Wprime-torus-2} respectively, we get the following (rather long) expression for the entanglement entropy 
\eqref{EE-tous-det-expression},
\begin{align}
\label{EE-final-torus-v2}
S[A]=
& \frac{d}{4}\log(4 u(1-u))+\frac{d}{4}\zeta'_{T^{d-1}_{L_2,\ldots,L_d}}(0)-\frac{d}{8}G'(0;2 L_A,\ldots,L_d)
-\frac{d}{8} G'(0;2L_B,\ldots,L_d)\nonumber \\
&+\frac{d}{4} G'(0;L_1,\ldots,L_d) +\log\left(4\pi g\,R^2_c\right)+\log{\sqrt{A_d}}
-1- \half \log|\tau_1 \dots \tau_{d-1}| \nonumber\\
&+\half \log\left(\frac{ (-1)^{d/2}d! }{4(1-2^d)B_d}\, \left( u^{1-d}+(1-u)^{1-d}\right) \right) \nonumber\\
&-\int\limits_{-\infty}^\infty\frac{dk}{\sqrt{\pi}}e^{-k^2}\log(\sum_{\omega\in\Z} \exp(-\frac{\pi}{\Lambda_d}\omega^2-2i\sqrt{\frac{\pi}{\Lambda_d}}k\,\omega))\,.
\end{align}
It can be verified that the entanglement entropy is symmetric under the transformation $u\to 1-u$ \cite{Chen:2016bk}
as required for a pure state of the full system. 

For the special case of $d=z=2$ we obtain (using \eqref{det-difference-torus-d2} and \eqref{def-lambdad2})
\begin{align}
\label{EE-torus-2d-moreexp}
 S[A_2]  =&-\half \log|\tau_1|+\half\log\left(  4 u (1-u) |\tau_1|^2   \right)
 + \log \left( {\eta(2(1-u) \tau_1)\, \eta(2 u \tau_1)\over \eta^2(\tau_1)}\right) -1 +\log( 4 \pi \, g\, R_c^2)
 \nonumber\\
&- \half \log(u(1-u))- \half\log|\tau_1|
 -\int\limits_{-\infty}^\infty\frac{dk}{\sqrt{\pi}}e^{-k^2}\log(\sum_{\omega\in\Z} 
 \exp(-\frac{\pi}{\Lambda_2}\omega^2-2i\sqrt{\frac{\pi}{\Lambda_2}}k\,\omega))\,
\nonumber \\
 =&\log \left( {\eta(2(1-u) \tau_1)\, \eta(2 u \tau_1)\over \eta^2(\tau_1)}\right) -1 +\log( 8 \pi \, g\, R_c^2)
\nonumber \\ 
 & -\int\limits_{-\infty}^\infty\frac{dk}{\sqrt{\pi}}e^{-k^2}
 \log(\sum_{\omega\in\Z} \exp(-\frac{\pi}{\Lambda_2}\omega^2-2i\sqrt{\frac{\pi}{\Lambda_2}}k\,\omega))\,, 
\end{align}
where $\Lambda_2 = 4 \pi \, g\, R_c^2 \, {1\over u(1-u)} {1\over |\tau_1|}\,$. The final result looks relatively simple
due to some cancellations between the classical and quantum contributions.
To our knowledge, this is the first time that universal finite terms in the entanglement entropy on a torus have been 
obtained in closed form using path integral methods, even for the two-dimensional case. 
They have been computed numerically in \cite{Stephan:2009dy} and by means of a boundary field method in \cite{Oshikawa:2010kv}.

We will now check some interesting limits of our general expressions. 
%

\paragraph{Halved $d$-torus.}

\begin{figure}
\begin{center}
\includegraphics[scale=0.5]{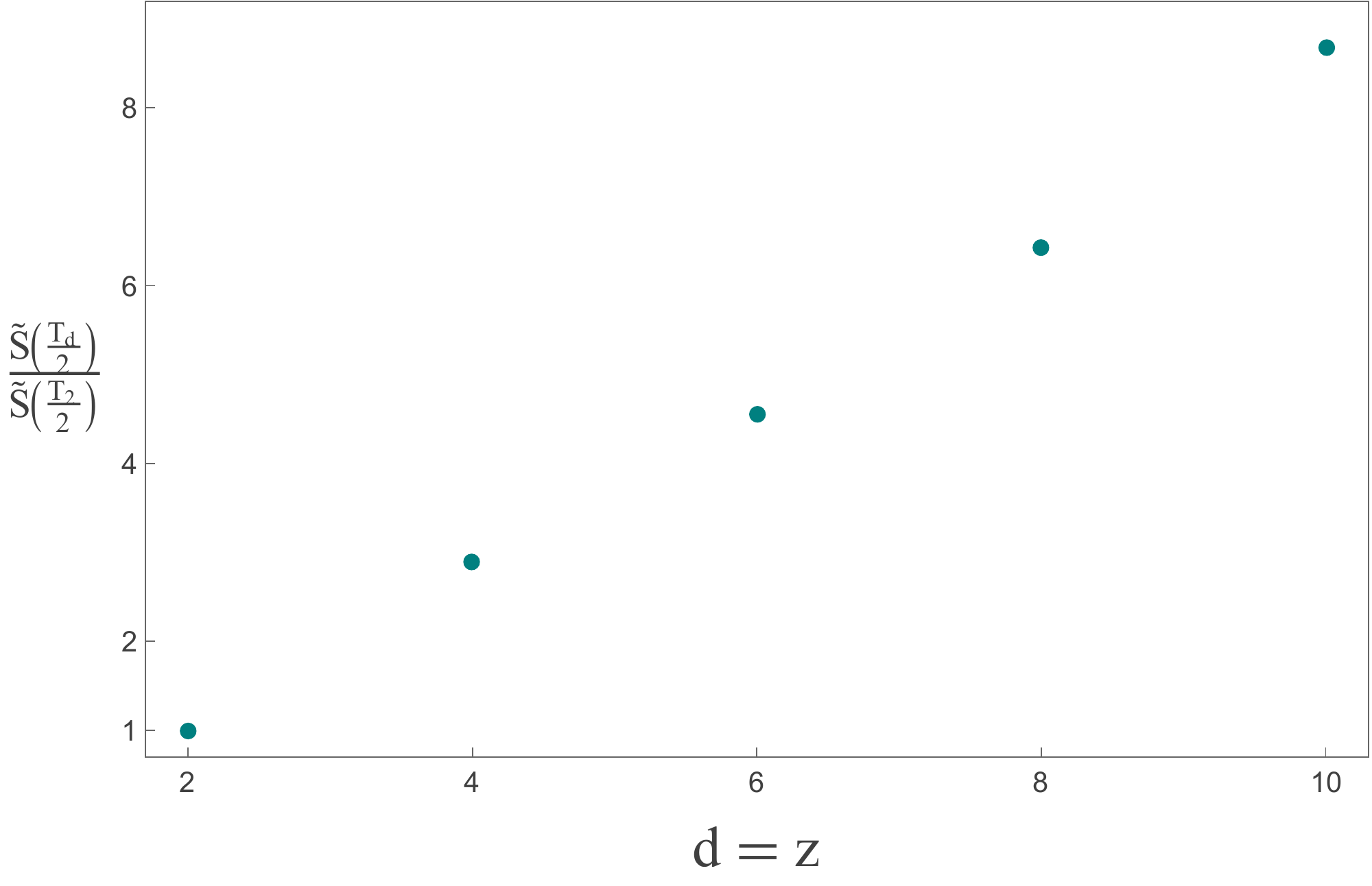}
\caption{We plot the final expression for the universal finite term in the entanglement entropy on a half 
torus \eqref{EE-torus-d-half} against the number of spatial dimension $d$ (which is equal to the critical exponent $z$). 
We normalise $S[T_d/2]$ with respect to the two-dimensional case, and set $g=R_c=L_1= \dots = L_d=2L_A=1$ in the plot.}
\label{fig:EE-torus-2}
\end{center}
\end{figure}
The first simplifying special case that that we consider is when the torus is divided into two equal parts:
\be
L_A=L_B\,, \qquad L_1= 2L_A\,, \qquad u=\half\,.
\ee
The contribution from the functional determinants \eqref{det-again-v2} simplifies tremendously, leaving only a single term,
\begin{equation}
\log \det \Delta^{d/2}_{[0,L_A]\cross T^{d-1}_{L_2,\ldots,L_d}} 
+\log \det \Delta^{d/2}_{[0,L_A]\cross T^{d-1}_{L_2,\ldots,L_d}}
-\log {\det}' \Delta^{d/2}_{T^d_{2L_A,\ldots,L_d}}
=\frac{d}{2}\zeta'_{T^{d-1}_{L_2,\ldots,L_d}}(0)\,, ~~~
\end{equation}
and we obtain for the universal terms in the entanglement entropy
\begin{align}\label{EE-torus-d-half}
\begin{split}
& S[A]=
\frac{d}{4}\zeta'_{T^{d-1}_{L_2,\ldots,L_d}}(0)
+\log\left(4\pi g\,R^2_c\right)+{d\over 2}\log L_1-1-  \log|\tau_1 \dots \tau_{d-1}| \\
&+\half \log\left(\frac{ (-1)^{d/2}\,d! }{4(1-2^{-d})B_d}\, \right)
-\int\limits_{-\infty}^\infty\frac{dk}{\sqrt{\pi}}e^{-k^2}\log(\sum_{\omega\in\Z} 
\exp(-\frac{\pi}{\Lambda_d}\omega^2-2i\sqrt{\frac{\pi}{\Lambda_d}}k\,\omega))\,,
\end{split}
\end{align}
where now
\be
\Lambda_d (u=1/2)= 
g \, \pi \, R_c^2  \frac{  (-1)^{d/2} 2^{d} d!}{\left(1- 2^d\right) B_d}
 \, {1\over |\tau_1| \dots |\tau_{d-1}|}\,.
\ee
In $d=2$ this reduces to 
\begin{align}\label{EE-torus-2-half}
\begin{split}
S[A_2]&= -1 +\log( 8 \pi \, g\, R_c^2)
 -\int\limits_{-\infty}^\infty\frac{dk}{\sqrt{\pi}}e^{-k^2}\log(\sum_{\omega\in\Z} 
 \exp(-\frac{\pi}{\Lambda_2}\omega^2-2i\sqrt{\frac{\pi}{\Lambda_2}}k\,\omega))\,,
\end{split}
\end{align}
since the contributions from the Dedekind-eta functions in \eqref{EE-torus-2d-moreexp} cancel against each other when $u=1/2$. 
Moreover, $\Lambda_2$ is given here by 
\be
\Lambda_2 (u=1/2)= 16 \, g \, \pi \, R_c^2 \,  {1\over |\tau_1|}\,.
\ee

\paragraph{Thin $d$-torus.}

In $d=2$, the infinitely thin torus limit (sometimes called also the long torus limit) amounts to $|\tau_1|\gg 1$ and $u$ fixed. 
It can be helpful to think of this limit as $L_2 \to 0$ while all the other lengths ($L_1, L_A$) are kept fixed. 
In this case, the contribution from the integral in the expression \eqref{EE-torus-2d-moreexp} is exponentially 
suppressed, and moreover, we have the asymptotic behaviour \eqref{dedekind-explarge} for the Dedekind $\eta$ function.
It is then clear that all the contributions from the Dedekind $\eta$-functions vanish in this limit, and we are left with the simple result
\be\label{thin-2torus}
S[A_2]  &= \log( 8 \pi \, g\, R_c^2)-1\,,
\ee
which agrees with \cite{Oshikawa:2010kv}.
In this limit the entanglement entropy for the thin torus is {\it twice} the entanglement entropy for 
the thin cylinder \cite{Oshikawa:2010kv}, since the entropy still carries information about the two boundaries of the torus. 

We can take a look at the same limit for the $d$-torus. 
In the $d$-dimensional case we assume that $L_1, L_A$ are fixed and of order one, while all the other sides are 
approaching zero, that is $L_2, \dots, L_d\to 0$. 
There is an ambiguity in how to take this limit, so as a first step we consider the case 
\be\label{lim-widetorus-1}
L_1, L_A \gg L_2 \gg \dots \gg L_d\,, 
\ee
which can also be written as
\be
1\ll |\tau_1|\ll \dots\ll |\tau_{d-1}|\,. 
\ee
Let us examine how the different terms in \eqref{EE-final-torus-v2} behave when the inequalities in \eqref{lim-widetorus-1} hold. 
First, all the functions $G'(0, L, L_2, \dots, L_d)$ \eqref{def-Gprime} with $L=2L_A, 2(L_1-L_A), L_1$ 
are exponentially suppressed, since all the elements of the matrix $\Xi_{d-1}$ diverge, while $L$ is kept fixed. 
Now consider the term $\zeta'_{T^{d-1}_{L_2,\ldots,L_d}}(0)$ in \eqref{EE-final-torus-v2}. 
This term is defined in \eqref{eq:closed-torus-zeta-prime} with a shift $d\to d{-}1$ and subsequent relabelling of the torus sides. 
With the choice \eqref{lim-widetorus-1} all the Bessel functions contained in \eqref{def-sfunction}, and thus in 
$\zeta'_{T^{d-1}_{L_2,\ldots,L_d}}(0)$, are exponentially suppressed. 
It then follows that the leading piece of $\zeta'_{T^{d-1}_{L_2,\ldots,L_d}}(0)$ in \eqref{eq:closed-torus-zeta-prime} is given 
by the highest term in the sums, that is 
\be
\nn
\zeta'_{T^{d-1}_{L_2,\ldots,L_d}}(0)  &\approx& 4 {(-\pi)^p\over p!} {L_2 \dots L_{2p+1}\over L_{2p+2}^{2p}} \zeta'(-2p) 
+2 {L_2\dots L_{2 p}\over L_{2p+1}^{2p-1}}{(-2\pi)^p\over (2p-1)!!} \zeta(-2p+1) 
\\
&\approx& 4 {(-\pi)^p\over p!} {L_2 \dots L_{2p+1}\over L_{2p+2}^{2p}} \zeta'(-2p) \,,
\ee
where $p=\lceil {d-1\over 2}\rceil -1= \lfloor {d-1\over 2}\rfloor$ for even $d$. 
We can rewrite the term more elegantly as a function of the aspect ratios,%
\footnote{Where we have used the $\zeta$-function identity $\zeta'(-2n)=(-1)^n {(2n)!\over 2 (2\pi)^{2n}}\zeta(2n+1)\,, 
\quad n\in\mathbb{N}$.}
\be
\nn
\zeta'_{T^{d-1}_{L_2,\ldots,L_d}}(0)  &\approx& {2 \,\Gamma\left(p+\half\right)\over \pi^{p
+\half}}\zeta\left(p+\half\right) {|\tau_{2p+1}|^{2p}\over |\tau_1| \dots |\tau_{2p}|}\,.
\ee
Finally, the integral over $k$ in \eqref{EE-final-torus-v2} is also exponentially suppressed and will not contribute to the final expression. 
Then, keeping only the most divergent term according to \eqref{lim-widetorus-1}, we obtain
\be\label{thin-dtorus}
S[A_d]\approx {d \,\Gamma\left(p+\half\right)\over 2\, \pi^{p+\half}}\zeta\left(p+\half\right) {|\tau_{2p+1}|^{2p}\over |\tau_1| \dots |\tau_{2p}|}\,, \qquad p= \left\lfloor {d-1\over 2}\right\rfloor\,. 
\ee
Similar limits were discussed in \cite{Chen:2016bk} for the Renyi entropies of $3{+}1$-dimensional 
relativistic fields theories with various twisted boundary conditions. Except for having the same power-law divergence, 
our results appear not to agree with their findings. The comparison is tricky though, as there are effectively three length 
scales in the $d=4$, and since we are looking at the regularised entanglement entropy we do not have an 
explicit cut-off as in \cite{Chen:2016bk}. 

We should stress that when $d=2$ all the sums in $\zeta'_{T^{d-1}_{L_2,\ldots,L_d}}(0)$ in \eqref{eq:closed-torus-zeta-prime} 
are empty and the only contribution from this term is the logarithm $-2 \log{L_2}$. 
Then the only divergent contributions are coming from the $\log$ terms 
(see {\it e.g.} \eqref{EE-torus-2d-moreexp}) and they cancel, leaving the finite term shown in \eqref{thin-2torus}.

\paragraph{The thin sliced $d$-torus.}

In $d=2$ this limit corresponds to $L_A\to 0$ while all the other length scales involved remain fixed, 
that is $u\to 0$ while $|\tau_1|$ is kept fixed. 
The integral in \eqref{EE-torus-2d-moreexp} can then be evaluated, for instance by means of the Poisson summation formula \eqref{poisson-formula}, and at leading order it gives $\half\log u$. 
Considering only the leading term in the expansion of the Dedekind function \eqref{dedekind-exp}, 
we obtain, for $u\to 0$,
\be\label{thin-sliced-2torus}
S[A_2] &= - {\pi\over 24 |\tau_1| u}+\dots\,,
\ee
which agrees with the entanglement entropy for the infinite long and thin sliced cylinder computed in \cite{Zhou:2016ykv}. 
Indeed, in this limit the torus and the cylinder are indistinguishable at leading order. 

We can proceed with similar arguments in higher dimensions, assuming $u\to 0$ while all the aspect ratios $|\tau_i|$, 
with $i=1, \dots , d-1$ are kept fixed. 
In order to simplify the computation we assume all the aspect ratios to be equal, $|\tau_i|=\sigma$ for all $i=1, \dots, d-1$. 
Then, the leading divergent terms are contained in $G'(0, 2L_A, L_2, \dots, L_d)$ in \eqref{EE-final-torus-v2}, 
and by estimating the $d{-}1$-dimensional sum in $G'(0, 2L_A, L_2, \dots, L_d)$  (cf. \eqref{def-Gprime}) 
with an integral we obtain the following leading behaviour for the entanglement entropy,
\be\label{thin-sliced-dtorus}
S[A_d] \approx {\kappa_d\over u^{d-1} \sigma^{d-1}}\,, 
\ee
where $\kappa_d$ is a numerical coefficient that depends on the number of dimensions $d$. 
Similar behaviour was obtained for the three-dimensional torus in conformal field theories in~\cite{Witczak-Krempa:2016jhc} 
(see also \cite{Casini:2009sr}), and also in \cite{Bueno:2016rma} from a holographic approach.

\paragraph{The wide $d$-torus.}

As our final example, we consider the so-called wide torus limit, that is when the directions transverse to the cut
are very large while $L_A, L_1$ are kept fixed. 
Let us start by considering this limit for $d=2$. This means that $|\tau_1|\to 0$ while $u$ is kept fixed.  
Using the expansion of the Dedekind-eta function \eqref{dedekind-exp}, 
we see that the term containing the logarithm of the ratio of Dedekind-eta functions in 
\eqref{EE-torus-2d-moreexp} produces the leading divergence. 
Hence, from the general expression for the entanglement entropy in $d=2$ \eqref{EE-torus-2d-moreexp}, 
we obtain
\be\label{s-widetorus-anyu-2}
S[A_2] \approx  -{\pi\over 24 u (1-u) |\tau_1| }+{\pi \over 6 |\tau_1|}\,.
\ee
This asymptotic behaviour is also expected for the universal function of the Renyi entropies 
of the two-dimensional torus, cf. \cite{Chen:2016bk} and references therein, and was found in holographic CFTs
in \cite{Chen:2014zea}. 

In higher dimensions we can consider the limit when $u$ is kept fixed, and all the transverse directions are very large compared to $L_1, L_A$, but all the aspect ratios approach zero at the same rate, that is 
$|\tau_i|=\varepsilon$, with $i=1, \dots,d-1$ and $\varepsilon \to 0$. 
In this case, the expressions in \eqref{EE-final-torus-v2}  simplify, and, as in the two-dimensional case, 
the leading divergent contribution is contained in the functions $G'(0, L, L_2, \dots, L_d)$ (cf. \eqref{def-Gprime}), 
where $L$ can be $L_1, 2L_A$ or $2(L_1-L_A)$. 
Using a similar expansion as performed in the thin sliced torus limit, we obtain 
\be\label{s-widetorus-anyu-d}
S[A_d]\approx {f_d(u)\over \varepsilon^{d-1}}\,,
\ee
where $f_d(u)$ is a function symmetric under the exchange $u\to 1-u$. 

In the above discussion, the case $u=\half$ is special for any dimension $d$, since the function $f_d(u=1/2)=0$, 
so that the sub-leading but still diverging terms become important.
Looking directly at \eqref{EE-torus-d-half} and \eqref{EE-torus-2-half}, 
there is no contribution now coming from $G'(0, L, L_2, \dots, L_d)$, 
and the next divergent term is logarithmic in the aspect ratios $\tau_i$, 
which in the two-dimensional torus is entirely coming from the integral in \eqref{EE-torus-2-half}, 
while in higher dimensions it receives contributions also from the area term $\log\sqrt{A_d}$ and $\zeta'_{T^{d-1}}(0)$. 
In our simplified limit where all the ratios $|\tau_i|$ approach zero at the same rate, we see that 
\be\label{s-widetorus-halfu-2}
S[A_d] \approx \half \log \varepsilon \,, \qquad u=\half\,, ~~ d\ge 2\,.
\ee
This is consistent with the findings of \cite{Chen:2016bk}, where for the $z=2$ free boson field theory in 3+1~dimensions
the universal function of Renyi entropies  $J_n$ satisfies the relation $\lim_{|\tau|\to 0} J_n(u=1/2, |\tau|) |\tau|=0$.%
\footnote{Note that the bosons are not compactified in \cite{Chen:2016bk}.} 
This clearly holds in our case since the universal term of the entanglement entropy has a logarithmic divergence. 
Rather different behaviour was observed in free two-dimensional CFTs  for Renyi entropies \cite{Chen:2016bk} 
and also in holographic CFTs in two and three space dimensions for entanglement entropy \cite{Bueno:2016rma}, 
where also for $u=1/2$ the universal part continues to have a power-law divergence similar to \eqref{s-widetorus-anyu-2} 
and \eqref{s-widetorus-anyu-d}. 
The disagreement was already observed in \cite{Chen:2016bk}.

\section{Discussion}
\label{sec:conclusions}

In this work we have analytically computed the universal finite corrections to the entanglement entropy for 
GQLMs in arbitrary $d{+}1$~dimensions, for even integer $d$, and on either a $d$-sphere cut into two $d$-dimensional 
hemispheres or a $d$-torus cut into two $d$-dimensional cylinders. 
GQLMs are free field theories where the Lifshitz exponent is equal to the number of spatial dimensions, 
and they are described in terms of compactified massless scalars. 
When $d=z=2$ the GQLM reduces to the quantum Lifshitz model \cite{Ardonne:2003wa}, and our findings  
confirm the known results of \cite{Oshikawa:2010kv, Zaletel:2011ir, Zhou:2016ykv}. 
The calculations are performed by means of the replica method. Caution is required when performing the cut
as the massless scalar field in the GQLM is compactified. 
It is useful to discern between the role of the fluctuating fields and the classical modes. 
In essence, the periodical identification mixes with the boundary conditions imposed at the cut on the 
replicated fields~\cite{Stephan:2009dy, Hsu:2010ag, Oshikawa:2010kv, Zaletel:2011ir, Zhou:2016ykv}, and disentangling the winding modes 
from the rest leads to an additional universal sub-leading contribution to the entanglement entropy. 
The fluctuating fields satisfy Dirichlet conditions at the cut (as well as further conditions imposed on their even-power derivatives), 
while the classical fields take care of the periodic identification. 
The contribution from the fluctuating fields comes from the ratio of functional determinants of the relevant operators, 
which we compute via spectral $\zeta$-function methods. 
The classical fields contribute via the zero-mode and via the winding sector summarised in the function $W(n)$.  

For the spherical case, the full analytic expression of the universal finite terms turned out to be a constant, 
depending only on the scale invariant compactification radius, cf. \eqref{final-EE-sphere}. 
This is a consequence of the presence of zero modes, and their normalisation.

For the toroidal case, the story is rather rich. 
The general expression is \eqref{EE-final-torus-v2}, while \eqref{EE-torus-d-half} is valid when we cut the torus by half.
In both cases, the universal term is comprised of a scaling function, which depends on the relevant aspect ratios of the 
subsystems, and a constant term, which contains the ``physical'' compactification radius. 
The last one comes from the zero mode of the partition function of the $d$-torus as well as from the winding sector. 
We considered various limits, such as the thin torus limit, which results in the simple expressions \eqref{thin-2torus} 
and \eqref{thin-dtorus} in two and $d$ dimensions respectively, the thin sliced $d$-torus, cf. \eqref{thin-sliced-2torus} 
and \eqref{thin-sliced-dtorus}, and finally we examined the wide torus limit, cf. \eqref{s-widetorus-anyu-2}, 
\eqref{s-widetorus-anyu-d}, and \eqref{s-widetorus-halfu-2} where the last expression is valid for $u=1/2$. 
Notice that in the toroidal case, where the winding sector is non trivial, it also contributes to the scaling function. 
For example, in the thin torus limit its contribution is decisive in order to cancel divergences and leave a 
finite result, cf. \eqref{thin-2torus} and \eqref{thin-dtorus}. 
Our findings confirm expectations from the study of the (2+1)-dimensional QLM 
\cite{Stephan:2009aa,Oshikawa:2010kv, Zaletel:2011ir, Zhou:2016ykv}, that also for critical non-relativistic 
theories entanglement entropy can encode both local and non-local information of the whole system. 
Moreover, our results give substance to the field theoretic intuition that entanglement entropy should depend 
on the dynamical critical exponent, see also \cite{Gentle:2017ywk, He:2017wla, MohammadiMozaffar:2017nri} 
for analogous results in this direction. 
The specific dependence is rather non-trivial already in the simplest spherical case. 

The next step would be to extend our analysis to even-dimensional spacetime, that is when $d$ is an odd integer. 
Progress has been recently made in {\it e.g.} \cite{Dowker:2018lus} (and references therein), in computing 
determinants of GJMS operators on odd-dimensional spheres, see also \cite{Barvinsky_2017} for a different approach based on heat kernel techniques.
It would also be interesting to broaden our study of GQLMs by examining entanglement entropy for non-smooth 
entangling surfaces. 
In the QLM for non-smooth boundaries, sharp corners source a universal logarithmic contribution (i.e. $\log L_A$), 
with a coefficient that depends on the central charge and the geometry of the surgery \cite{Fradkin:2006mb, Cardy:1988tk}. 
For $d$-dimensional CFTs  and in presence of non-smooth entangling surfaces further UV divergences also appear 
whose coefficients are controlled by the opening angle.
In particular, for a conical singularity, the nearly smooth expansion of the universal corner term (seen as a function 
of the opening angle $\theta$) is simply proportional to the central charge of the given 
CFT \cite{Bueno:2015lza, Bueno:2015rda, Bueno:2015qya, Bueno:2015xda, Bueno:2015ofa, Klebanov:2012yf, 
Elvang:2015jpa, Miao:2015dua}.%
\footnote{See also \cite{Pang:2015lka, Alishahiha:2015goa} for related holographic results.} 
Another relevant direction to pursue is the study of post-quench time evolution of entanglement in these systems, 
see {\it e.g.} \cite{Zhou_2016} for recent results in QLM and \cite{MohammadiMozaffar:2018vmk} for Lifshitz-type scalar theories.%
\footnote{See also the work \cite{Plamadeala_2018} for a study of signatures of chaos in the QLM.}
In this respect, GQLMs can provide an interesting and rich playground where one can answer many questions in a 
full analytic manner. 

\section*{Acknowledgements}

We acknowledge useful discussions with J. Bardarson, P. Di Vecchia, J. S. Dowker, D. Friedan, B. Gouteraux, K. Grosvenor, D. Medina-Rincon, R. Leigh, D. Seminara, W. Sybesma, S. Vandoren, and M. Zaletel.   
This research was supported in part by the Icelandic Research Fund under contracts 163419-053 and 163422-053, 
and by grants from the University of Iceland Research Fund.
%

\appendix

\section{The determinant of GJMS operators on spheres and hemispheres}
\label{sec:dowker}

In this appendix we review the calculation of the determinants of GJMS operators on spherical domains, 
originally performed by Dowker in \cite{Dowker2011}, and rewrite his results in a way we find more transparent. 
Building on his previous work, in particular \cite{Chang1993} and \cite{Dowker1994}, Dowker writes the following 
expression for the spectral $\zeta$-function of a GJMS operator of degree $2k$, which we denote by $P_{2k}$, 
on the hemisphere,
\begin{equation}
\label{eq:spectralzeta}
Z_d(s,a_{D/N},k)=\sum\limits_{\mathbf{m}\in\mathbb{N}^d}
\prod_{j=0}^{k-1}\left((\mathbf{m}\cdot\mathbf{d}+a_{D/N})^2-\alpha_j^2\right)^{-s},
\end{equation}
with $\mathbf{d}=(1,\ldots,1)\in\mathbb{R}^d$, $\alpha_j=j+1/2$, $a_D=(d+1)/2$ for Dirichlet boundary conditions, 
and $a_N=(d-1)/2$ for Neumann boundary conditions. The GJMS operators are well defined for 
$k=1,\ldots,d/2$ and we distinguish between the subcritical case $k<d/2$ and the critical case $k=d/2$. 
Using the above form of the spectral $\zeta$-function the log determinant of $P_{2k}$ on the 
$d$-hemisphere $H^d$ with either Neumann or Dirichlet boundary conditions can be found as
\begin{equation}
\label{eq:logdetp2k}
\log \det P_{2k,H^d_{D/N}} =  -Z_d'(0,a_{D/N},k).
\end{equation}
The expression $ Z_d'(0,a_{D/N},k)$ should be interpreted as $\lim_{s\rightarrow 0}\partial_sZ_d(s,a_{D/N},k)$. 
Given \eqref{eq:logdetp2k}, one can then add the Neumann and the Dirichlet cases to find the log determinant 
of $P_{2k}$ on the whole sphere 
\begin{equation}
\label{eq:logdetp2k-s}
\log\det P_{2k, S^d} = -Z_d'(0,a_{D},k) - Z_d'(0,a_{N},k).
\end{equation}
Our task therefore is to evaluate $ Z_d'(0,a_{D/N},k)$. In order to do so, we first review the definition 
and key properties of the Barnes $\zeta$-function. We then turn to the evaluation of the spectral 
$\zeta$-function of one of the factors in \eqref{eq:spectralzeta} and finally put everything together to get the 
desired log determinants.

\subsection{Definition and properties of the Barnes $\zeta$-function}
\label{sec:Barnes}
The Barnes $\zeta$ function is defined for $s>d$ as 
\begin{equation}
\zeta(s,a\vert \mathbf{d}):\,
= \sum\limits _{\mathbf{m}\in\mathbb{N}^d}\left(a+\mathbf{m}\cdot\mathbf{d}\right)^{-s}
= \sum\limits _{m_1,\ldots,m_d=0}^\infty\left(a+m_1d_1+\ldots m_dd_d\right)^{-s}.
\end{equation}
For the special case of $\mathbf{d}=\mathbf{1}=(1,\ldots,1)$ we use the simplified notation,
\begin{equation}
\label{eq:def-barnes-zeta}
\zeta_d(s,a):\,= \zeta(s,a\vert \mathbf{1}).
\end{equation} 
Analytic continuation of $\zeta_d(s,a)$ is facilitated by the relation (see \cite{Srivastava2012}, page 149), 
\begin{equation}
\label{eq:analcontmultzeta}
\zeta_d(s,a) = \Gamma(1-s)I_d(s,a),
\end{equation}
with $I_d(s,a)$ given by the integral
\begin{equation}
I_d(s,a)=-\frac{1}{2\pi i}\int\limits_\infty^{0^+}\frac{(-z)^{s-1}e^{-az}}{(1-e^{-z})^n}dz.
\end{equation}
$I_d(s,a)$ is an entire function in $s$. The new definition of $\zeta_d(s,a)$ is analytic for all $s$ except for 
simple poles at $s=1,\ldots,d$ where it has residues
\begin{align}
\label{eq:barnes-zeta-residue}
\begin{split}
\underset{s=k}{\Res}\zeta_d(s,a)
&=\frac{1}{(d-k)!(k-1)!}\lim_{z\rightarrow 0}\frac{d^{d-k}}{dz^{d-k}}\frac{z^d e^{-az}}{(1-e^{-z})^d}\\
&=\frac{(-1)^{d-k}B^{(d)}_{d-k}(a)}{(d-k)!(k-1)!}\\
&=\frac{(-1)^k}{(k-1)!}I_d(k,a),\qquad k=1,\ldots,d
\end{split}
\end{align}
where $B^{(d)}_{l}(a)$ are generalised Bernoulli polynomials. For future convenience we define
\begin{equation}
\label{eq:R-def}
R_d(k,a): = \frac{(-1)^{d-k}B^{(d)}_{d-k}(a)}{(d-k)!(k-1)!},
\end{equation}
such that for $k=1,\ldots,d$ we can simply write $\Res_{s=k}\zeta_d(s,a)=R_d(k,a)$. 
In particular this means that we have the following expansion around $s=0$ for $k=1,\ldots,d$,
\begin{equation}
\zeta_d(k+s,a)=\frac{1}{s}R_d(k,a)+C_d(k,a)+\mathcal{O}(s).
\end{equation}
Calculating $\partial_s s\zeta_d(k+s,a)\vert_{s=0}$ using the analytic continuation \eqref{eq:analcontmultzeta} one finds that
\begin{equation}
C_d(k,a)=R_d'(k,a)-\psi(k)R_d(k,a),
\end{equation}
where the derivative of $R$ should be read as $R'_d(k,a)=\lim_{s\rightarrow 0} \partial_s R_d(k+s,a)$, 
and $\psi$ is the digamma function. For non-negative $k$, $\zeta_d(-k,a)$ can be evaluated using its 
analytic continuation \eqref{eq:analcontmultzeta}. On page 151 of \cite{Srivastava2012} the following 
expression is given
\begin{equation}
\label{eq:barnes-zeta-at-non-integers}
\zeta_d(-k,a)=(-1)^d\frac{k!}{(d+k)!}B^{(d)}_{d+k}(a),
\end{equation}
where again $B^{(d)}_{l}(a)$ are generalised Bernoulli polynomials.

\subsection{A first step towards $Z_d$}
In order to calculate $Z'_d(0,a_{D/N},k)$ with $Z_d$ as in \eqref{eq:spectralzeta}, we need to evaluate the 
derivative at $s=0$ of $\zeta$-functions of the form
\begin{equation}
\label{eq:zetadirichlet}
\zeta_D(s):\, =\sum\limits_{\mathbf{m}\in\mathbb{N}^d}\left((a_D+\mathbf{m}\cdot\mathbf{d})^2-\alpha^2\right)^{-s},
\end{equation}
 with $\alpha=(d-1)/2$ and $a_D=\sum d_i-(d-1)/2$, with $\mathbf{d}\in\mathbb{N}^d$ for Dirichlet boundary conditions, 
 as such functions roughly correspond to the factors of $Z_d$. 
 From now on we set $\mathbf{d}\equiv(1,\ldots,1)$ as this is the only case relevant to our discussion. 
 In particular, this implies that $a_D=(d+1)/2$. For Neumann boundary conditions we have to set $a_N=(d-1)/2$ 
 instead of $a_D$, which makes the term coming from the origin $\mathbf{0}\in\mathbb{N}^d$ ill-defined. 
 For the Neumann case we thus have to omit the origin from the summation,
 \begin{align}
 \label{eq:zetaneumann}
 \begin{split}
 \bar{\zeta}_N(s)&:\,= \zeta_N(s)-(a_N^2-\alpha^2)^{-s}\\
 &=\sum\limits_{\mathbf{m}\in\mathbb{N}^d\setminus\{\mathbf{0}\}} 
 \left((a_N+\mathbf{m}\cdot\mathbf{d})^2-\alpha^2\right)^{-s}.
 \end{split}
 \end{align}
Below, we derive the following two identities
\begin{align}
\label{eq:zeta-prime-dirichlet}
\zeta'_D(0)&=\zeta'_d(0,a_D+\alpha)+\zeta'_d(0,a_D-\alpha) + \sum\limits_{r=1}^{d/2}\frac{\alpha ^{2r}}{r}H_1(r)R_d(2r,a_D)\,,\\
\label{eq:zeta-prime-neumann}
\bar{\zeta}'_N(0)&=\zeta'_d(0,a_N+\alpha)-\log\rho_d-\log(a_N+\alpha)+ \sum\limits_{r=1}^{d/2}\frac{\alpha ^{2r}}{r}H_1(r)R_d(2r,a_N).
\end{align}
Here the $R_d$ are residues of the Barnes $\zeta$-function given in \eqref{eq:barnes-zeta-residue}, $\log\rho_d$ is a 
$\Gamma$-modular form as described in equations \eqref{eq:multgammadef} and \eqref{eq:neumanzetazero} below, 
and $H_n(r)$ is a harmonic function defined in terms of harmonic numbers $H_r$ via, 
\begin{equation}
\label{eq:hn-def}
H_n(r):\, = \frac{H_{r-1}}{2n}-H_{2r-1}.
\end{equation}

\subsubsection{Dirichlet boundary conditions}
We now proceed with the derivation of \eqref{eq:zeta-prime-dirichlet}. 
Since we will only be dealing with Dirichlet boundary conditions throughout this section, 
we will omit the $D$-index in $a_D$ and only reintroduce it when we reach the final result. 
The first step in the evaluation is to perform a binomial expansion of $\zeta_D(s)$. 
The expansion converges because $(a+\mathbf{m}\cdot\mathbf{d})^2\geq\alpha^2$ for $\mathbf{d}=\mathbf{1}$
in both the Dirichlet and Neumann cases, with equality only arising for the zero mode of the Neumann case, 
which is omitted anyway. We thus get
\begin{align}
\label{eq:zeta0calc1}
\begin{split}
\zeta_D(s)&=\sum\limits_{\mathbf{m}\in\mathbb{N}^d}
\sum\limits_{r=0}^\infty\frac{(s)_r}{r!}\alpha^{2r} \left(a+\mathbf{m}\cdot\mathbf{d}\right)^{-(2s+2r)}\\
&=\sum\limits_{r=0}^\infty\frac{(s)_r}{r!}\alpha^{2r}
\sum\limits_{\mathbf{m}\in\mathbb{N}^d} \left(a+\mathbf{m}\cdot\mathbf{d}\right)^{-(2s+2r)}\\
&=\zeta_d(2s,a)+\sum\limits_{r=1}^\infty\frac{s f(s)}{r!}\alpha^{2r}\zeta_d(2s+2r,a),
\end{split}
\end{align}
where we used the definition of the Barnes $\zeta$-function for $\mathbf{d}=\mathbf{1}$ given in \eqref{eq:def-barnes-zeta} 
and rewrote the Pochhammer symbol for $r>0$ as $(s)_r=s(s+1)\cdots(s+r-1)=\,: sf(s)$ to make the behaviour around $s=0$ 
more transparent. We also note that $f(0)=(r-1)!$. 
Considering that $\zeta_d(2r)$ has simple poles at $r=1,\ldots,d/2$ and converges for higher $r$, 
we can immediately write down the following expression at $s=0$
\begin{equation}
\label{eq:zeta0}
\zeta_D(0)=\zeta_d(0,a)+\frac{1}{2}\sum\limits_{r=1}^{d/2}\frac{\alpha^{2r}}{r}R_d(2r,a)
\end{equation}
with $R_d(k,a)$ defined in equation \eqref{eq:barnes-zeta-residue}.

The next step is to evaluate $\zeta_D'(0)$. In order to do this, we first take a partial derivative with respect to $s$ at 
general values of $s$ and then let $s\rightarrow 0$, 
\begin{align}
\begin{split}
\partial_s\zeta_D(s)&=\partial_s\zeta_d(2s,a)
+\sum\limits_{r=1}^\infty\frac{\alpha^{2r}}{r!}\Bigg(f'(s)s\zeta_d(2s+2r,a)+f(s)\zeta_d(2s+2r,a)\\
&\quad\quad\quad\quad\quad\quad+s f(s)\partial_s\zeta_d(2s+2r,a)\Bigg).
\end{split}
\end{align}
We note that $f'(0)=(r-1)!H_{r-1}$, where $H_{r-1}$ is a harmonic number, and remind the reader that for $r=1,\ldots,d/2$ and small $s$ 
\begin{equation*}
\zeta_d(2s+2r,a)=\frac{R_d(2r,a)}{2s}+C_d(2r,a)+\mathcal{O}(s),
\end{equation*}
 and similarly  
\begin{equation*}
\zeta'_d(2s+2r,a)=-\frac{R_d(2r,a)}{2s^2}+\mathcal{O}(s).
\end{equation*}
Thus at $s=0$ we can write
\begin{align}
\left[f'(s)s\zeta_d(2s+2r,a)\right]_{s=0}&=(r-1)!\frac{H_{r-1}}{2}R_d(2r,a)\,,\\
\left[f(s)\zeta_d(2s+2r,a)+s f(s)\partial_s\zeta_d(2s+2r,a)\right]_{s=0}&=(r-1)!C_d(2r,a)\,.
\end{align}
 Putting these expressions together we find
\begin{align}
\label{eq:zetaprime1}
\begin{split}
\zeta_D'(0)&=2\zeta'_d(0,a)+\sum\limits_{r=1}^{d/2}\frac{\alpha^{2r}}{r}\frac{H_{r-1}}{2}R_d(2r,a)
+\sum\limits_{r=1}^{d/2}\frac{\alpha^{2r}}{r}C_d(2r,a) +\sum\limits_{r=d/2+1}^\infty\frac{\alpha^{2r}}{r}\zeta_d(2r,a).
\end{split}
\end{align}
There is no contribution to $\zeta_d'$ after $r=d/2$, since all such terms in the sum get set to $0$ by the $s$ factor.

The next task is to compute the remaining infinite series in \eqref{eq:zetaprime1}. 
In order to do this we first note that $\zeta_d(s,a)$ admits the following integral representation for $s>d$
\begin{equation}
\zeta_d(s,a)=\frac{1}{\Gamma(s)}\int\limits_0^\infty dt\ \frac{e^{-at}\ t^{s-1}}{(1-e^{-t})^d}.
\end{equation}
Using this we can rewrite
\begin{align}
\label{eq:inftsum}
\sum\limits_{r=u+1}^\infty \frac{\alpha^{2r}}{r} \zeta_d(2r,a)
&= 2\int\limits_0^\infty dt \frac{t^{-1}\ e^{-at}}{(1-e^{-t})^d}
\sum\limits_{r=d/2+1}^\infty\frac{(\alpha t)^{2r}}{(2r)!} \notag\\
&=\int\limits_0^\infty dt \frac{t^{-1}\ e^{-at}}{(1-e^{-t})^d}\left(\underbrace{2\cosh(\alpha t)}_{=e^{\alpha t}
+e^{-\alpha t}} - 2\sum\limits_{r=0}^{d/2}\frac{(\alpha t)^{2r}}{(2r)!} \right) \notag\\
&=\lim_{\sigma\rightarrow 0}\Bigg(\int\limits_0^\infty dt \frac{t^{\sigma-1}\ 
e^{-(a+\alpha)t}}{(1-e^{-t})^d} +\int\limits_0^\infty dt \frac{t^{\sigma-1}\ e^{-(a-\alpha)t}}{(1-e^{-t})^d}\notag\\
&\qquad\qquad-2\sum\limits_{r=0}^{d/2}\frac{(\alpha )^{2r}}{(2r)!} 
\int\limits_0^\infty dt \frac{t^{2r+\sigma-1}\ e^{-at}}{(1-e^{-t})^d}\Bigg)\notag\\
\begin{split}
&=\lim_{\sigma\rightarrow 0}\Big(\Gamma(\sigma)\left(\zeta_d(\sigma,a+\alpha)+\zeta_d(\sigma,a-\alpha)-2\zeta_d(\sigma,a)\right)\\
&\qquad\qquad -2\sum\limits_{r=1}^{d/2}\frac{\alpha ^{2r}}{(2r)!}\Gamma(2r+\sigma)\zeta_d(2r+\sigma,a)\Big),
\end{split}
\end{align}
where we introduced a convergence parameter $\sigma$ in order for the integral representation to make sense.
For the $\sigma\rightarrow 0$ limit we will need to use the analytic continuation of $\zeta_d$. 
We now take a closer look at the different parts of the expression above around $\sigma=0$. 
The $\zeta_d$ functions on the first line of \eqref{eq:inftsum} are convergent at $\sigma=0$ while the Gamma
function has a simple pole.  Expanding order by order in $\sigma$ gives
\begin{align}
\label{eq:limpart1}
\begin{split}
\Gamma(\sigma)&\left(\zeta_d(\sigma,a+\alpha)+\zeta_d(\sigma,a-\alpha)-2\zeta_d(\sigma,a)\right)=\\
&=\frac{1}{\sigma}\left(\zeta_d(0,a+\alpha)+\zeta_d(0,a-\alpha)-2\zeta_d(0,a)\right)+\\
&\quad +\zeta'_d(0,a+\alpha)+\zeta'_d(0,a-\alpha)-2\zeta'_d(0,a)-\\
&-\gamma\left(\zeta_d(0,a+\alpha)+\zeta_d(0,a-\alpha)-2\zeta_d(0,a)\right) +\mathcal{O}(\sigma).
\end{split}
\end{align}
Carrying out the corresponding expansion on the second line of \eqref{eq:inftsum} gives
\begin{align}
\label{eq:limpart2}
\begin{split}
-2\sum\limits_{r=1}^{d/2}\frac{\alpha ^{2r}}{(2r)!}&\left(\Gamma(2r)+\Gamma(2r)\psi(2r)\sigma+\mathcal{O}(\sigma^2)\right)\left(\frac{1}{\sigma}R_d(2r,a)+C_d(2r,a)+\mathcal{O}(\sigma)\right)\\
&=-\frac{1}{\sigma}\sum\limits_{r=1}^{d/2}\frac{\alpha ^{2r}}{r}R_d(2r,a)-\sum\limits_{r=1}^{d/2}\frac{\alpha ^{2r}}{r}\psi(2r)R_d(2r,a)-\sum\limits_{r=1}^{d/2}\frac{\alpha ^{2r}}{r}C_d(2r,a)+\mathcal{O}(\sigma).
\end{split}
\end{align}
To get a finite end result we need the pole to vanish when we add \eqref{eq:limpart1} and \eqref{eq:limpart2}.%
\footnote{More rigorously one would first prove the convergence of the infinite sum \eqref{eq:inftsum}, which after rewriting implies the vanishing of the pole and in turn \eqref{eq:identityofbarnesfcts}.}
This is equivalent to the condition
\begin{equation}
\label{eq:identityofbarnesfcts}
\zeta_d(0,a+\alpha)+\zeta_d(0,a-\alpha)=2\zeta_d(0,a)+\sum\limits_{r=1}^{d/2}\frac{\alpha ^{2r}}{r}R_d(2r,a).
\end{equation}
We can then take the limit and write 
\begin{align}
\begin{split}
\eqref{eq:inftsum}&=-\gamma\left(\zeta_d(0,a+\alpha)+\zeta_d(0,a-\alpha)-2\zeta_d(0,a)\right)+\\
&\quad +\zeta'_d(0,a+\alpha)+\zeta'_d(0,a-\alpha)-2\zeta'_d(0,a)\\
&\quad-\sum\limits_{r=1}^{d/2}\frac{\alpha ^{2r}}{r}\psi(2r)R_d(2r,a)-\sum\limits_{r=1}^{d/2}\frac{\alpha ^{2r}}{r}C_d(2r,a)\\
&=\zeta'_d(0,a+\alpha)+\zeta'_d(0,a-\alpha)-2\zeta'_d(0,a)-\\
&\quad-\gamma \sum\limits_{r=1}^{d/2}\frac{\alpha ^{2r}}{r}R_d(2r,a)-\sum\limits_{r=1}^{d/2}\frac{\alpha ^{2r}}{r}\psi(2r)R_d(2r,a)-\sum\limits_{r=1}^{d/2}\frac{\alpha ^{2r}}{r}C_d(2r,a)\\
&=\zeta'_d(0,a+\alpha)+\zeta'_d(0,a-\alpha)-2\zeta'_d(0,a)-\\
&\quad-\sum\limits_{r=1}^{d/2}\frac{\alpha ^{2r}}{r}H_{2r-1}R_d(2r,a)-\sum\limits_{r=1}^{d/2}\frac{\alpha ^{2r}}{r}C_d(2r,a),
\end{split}
\end{align}
where $H_{2r-1}:\, =\psi(2r)+\gamma $ is a harmonic number. For clarity and later convenience we write down the following identity resulting from the above discussion in the case of Dirichlet boundary conditions explicitly
\begin{align}
\label{eq:sumidentity1}
\begin{split}
\partial_s \sum\limits_{r=1}^\infty\frac{(s)_r}{r!}\alpha^{2r}\zeta_d(2s+2r,a)\Big\vert_{s=0}&=\zeta'_d(0,a+\alpha)+\zeta'(0,a-\alpha)-2\zeta'_d(0,a)+\\
&\quad+ \sum\limits_{r=1}^{d/2}\frac{\alpha ^{2r}}{r}\left(\frac{H_{r-1}}{2}-H_{2r-1}\right)R_d(2r,a).
\end{split}
\end{align}
 Putting everything back together into \eqref{eq:zetaprime1} we can write for the complete $\zeta$-function in the Dirichlet case
\begin{align}
\begin{split}
\zeta'_D(0):\, =\zeta'(0)&=\zeta'_d(0,a_D+\alpha)+\zeta'_d(0,a_D-\alpha)+ \sum\limits_{r=1}^{d/2}\frac{\alpha ^{2r}}{r}\left(\frac{H_{r-1}}{2}-H_{2r-1}\right)R_d(2r,a_D)\\
&=\zeta'_d(0,a_D+\alpha)+\zeta'_d(0,a_D-\alpha)+ \sum\limits_{r=1}^{d/2}\frac{\alpha ^{2r}}{r}H_1(r)R_d(2r,a_D),
\end{split}
\end{align}
where $H_1(r)$ is a special case of
\begin{equation}
H_n(r):\,= \frac{H_{r-1}}{2n}-H_{2r-1}\,.
\end{equation}
defined for any integer $n\geq 1$. Higher values of $n$ will be relevant when carrying out the corresponding analysis for a zeta function 
of the form $\zeta_d(2ns+2r,a)$ instead of $\zeta_d(2s+2r,a)$.

\subsubsection{Neumann boundary conditions}

Calculating the Neumann case is equivalent to setting $a_N-\alpha=\varepsilon$ and letting $\varepsilon$ go to $0$.
In this limit Barnes \cite{Barnes1904} calculated that 
\begin{equation}
\label{eq:neumanzetazero}
\zeta_d'(0,\varepsilon)=-\log\varepsilon-\log \rho_d +\mathcal{O}(\varepsilon),
\end{equation}
where $\rho_d:\ =\rho_d(\mathbf{1})$ is called a $\Gamma$-modular form and 
obeys the following identity involving the multiple Gamma function
\begin{equation}
\label{eq:multgammadef}
\frac{\Gamma_d(a)}{\rho_d}=\frac{\Gamma_{d+1}(a)}{\Gamma_{d+1}(a+1)}.
\end{equation}
The multiple Gamma function $\Gamma_d(a\vert \mathbf{d})$ is defined
\begin{equation}
\frac{\Gamma_d(a\vert \mathbf{d})}{\rho_d(\mathbf{d})}:\,= e^{\zeta'_d(0,a\vert \mathbf{d})}\,,
\end{equation}
and we write $\Gamma_d(a):\ = \Gamma_d(a\vert \mathbf{1})$.
Equipped with these tools we can compute the derivative of the $\zeta$-function in the Neumann case \eqref{eq:zetaneumann},
\begin{align}
\begin{split}
\bar{\zeta}'_N(0)&:\, = \zeta_N'(0)- \partial_s(a_N^2-\alpha^2)^{-s}\big\vert_{s=0}\\
&=\zeta_d'(0,a_N+\alpha)-\log\rho_d-\log(a_N-\alpha)+\log(a_N^2-\alpha^2)+ \sum\limits_{r=1}^{d/2}\frac{\alpha ^{2r}}{r}H_1(r)R_d(2r,a_N)\\
&=\zeta_d'(0,a_N+\alpha)-\log\rho_d+\log(a_N+\alpha)+ \sum\limits_{r=1}^{d/2}\frac{\alpha ^{2r}}{r}H_1(r)R_d(2r,a_N),\\
\end{split}
\end{align}
where $\zeta_N$ is just $\zeta_D$ but with $a_D$ replaced by $a_N$.
We can also write the following Neumann version of the identity \eqref{eq:sumidentity1}
\begin{align}
\label{eq:sumidentity2}
\begin{split}
\partial_s\Bigg( \sum\limits_{r=1}^\infty\frac{(s)_r}{r!}\alpha^{2r}\zeta_d(2s+2r,a_N)&-(a_N^2-\alpha^2)^{-s}\Bigg)\Bigg\vert_{s=0}=\\
&=\zeta'_d(0,a_N+\alpha)-\log\rho_d+\log(a_N+\alpha)-2\zeta'_d(0,a_N)\\
&\quad+\sum\limits_{r=1}^{d/2}\frac{\alpha^{2r}}{r}H_1(r)R_d(2r,a_N).
\end{split}
\end{align}

\subsection{Determinant of GJMS operators}
\label{sec:GJMS-det}

We are now prepared to tackle the calculation of the log-determinants. 

In section \ref{sec:det-hem-D}, for the case of Dirichlet boundary conditions on the hemisphere, we obtain
\begin{equation}
\label{eq:logdet-p2k-D}
\log \det P_{2k,H^d_{D}}=-\zeta'_{d+1}(0,d/2-k+1)+\zeta'_{d+1}(0,d/2+k+1)-M(d,k),
\end{equation}
where the function $M(d,k)$ is defined via equations \eqref{eq:mone}, \eqref{eq:mtwo}, and \eqref{eq:total-m-def}. 
Equation \eqref{eq:logdet-p2k-D} is valid for $k=1,\ldots,d/2$. 

In section \ref{sec:det-hem-N}, we discuss the Neumann case, which is a little more involved, 
as the zero mode makes the log-determinant ill-defined in the critical case $k=d/2$. 
For the subcritical case $k=1,\ldots,d/2-1$ we get
\begin{equation}
\label{eq:logdet-p2k-N-subcrit}
\log \det P_{2k,H^d_{N}}=-\zeta'_{d+1}(0,d/2-k)+\zeta'_{d+1}(0,d/2+k)-M(d,k),
\end{equation}
while for the critical case we get the slightly more complicated result
\begin{equation}
\label{eq:logdet-p2k-N-crit}
\log \det P_{d,H^d_{N}}=-\log(d-1)!+\log \rho_{d+1}+\zeta'_{d+1}(0,d)-M(d),
\end{equation}
where $M(d)=M(d,d/2)$ and $\rho_{d+1}$ is a $\Gamma$-modular form as defined in \eqref{eq:multgammadef}.

In section \ref{sec:det-sphere-ND} we assemble these results to construct the determinant of GJMS operators on a sphere, 
summarised in \eqref{eq:spectralprimesphere}.

\subsubsection{Determinant on the hemisphere with Dirichlet boundary conditions}
\label{sec:det-hem-D}

We will now generalise the results of the last section.
Our starting point is the spectral $\zeta$-function defined in \eqref{eq:spectralzeta}. 
As before, we start by doing a binomial expansion for every term in the product,
\begin{align}
\label{eq:spectralzetacalc1}
\begin{split}
Z_d(s,a,k)&=\sum\limits_{\mathbf{m}\in\mathbb{N}^d}\prod_{j=0}^{k-1}
\sum\limits_{r_j=0}^\infty\frac{(s)_{r_j}}{r_j!}\frac{\alpha_j^{2r_j}}{(\mathbf{m}\cdot\mathbf{d}+a)^{2s+2r_j}}\\
&=\sum\limits_{\mathbf{r}\in\mathbb{N}^k}^\infty\left(\prod_{j=0}^{k-1}\frac{(s)_{r_j}}{r_j!}\alpha_j^{2r_j}\right)
\sum\limits_{\mathbf{m}\in\mathbb{N}^d}\frac{1}{(\mathbf{m}\cdot\mathbf{d}+a)^{2ks+2\sum_{j=0}^{k-1}r_j}}\\
&=\sum\limits_{\mathbf{r}\in\mathbb{N}^k}^\infty
\left(\prod_{j=0}^{k-1}\frac{(s)_{r_j}}{r_j!}\alpha_j^{2r_j}\right)\zeta_d\left(2ks+2\mathbf{r}\cdot\mathbf{d},a\right)\\
&=\underbrace{\zeta_d(2ks,a)}_{\mathbf{r}=\mathbf{0}}+\sum\limits_{i=0}^{k-1}
\underbrace{\sum\limits_{r_i=1}^\infty \frac{(s)_{r_i}}{r_i!}\alpha_i^{2r_i}\zeta_d(2ks+2r_i,a)}_{\mathbf{r}=(0,\ldots,0,r_i,0,\ldots,0)}+\\
&\quad+\sum\limits_{i=1}^{k-1}\sum\limits_{j=0}^{i-1}\underbrace{\sum\limits_{r_i,r_j=1}^\infty\frac{(s)_{r_i}}{r_i!}
\frac{(s)_{r_j}}{r_j!}\alpha_i^{2r_i}\alpha_j^{2r_j} \zeta_d(2ks+2r_i+2r_j,a)}_{\mathbf{r}=(0,\ldots,0,r_i,0,\ldots,0,r_j,0,\ldots,0)}+R(s).\\
\end{split}
\end{align}
We remind the reader that $\alpha_j= j+\half$, $\mathbf{d}=\mathbf{1}$, and 
$a_D=(d+1)/2$ for Dirichlet boundary conditions. 
For simplicity we write $a$ instead of $a_D$ throughout this section.

The remaining sums, denoted by $R(s)$ above, will vanish for $Z_d$ and $Z'_d$ around $s=0$. 
This is easy to see, as $(s)_r=\mathcal{O}(s)$ while the $\zeta_d$ functions will contribute with simple poles and
$\zeta'$ with double poles. When more than two Pochhammer symbols are present, they overcome the poles and 
give zero in the limit. We can make further simplifications in some of the sums above by noting that the $r_i$ are 
dummy indices and rewrite \eqref{eq:spectralzetacalc1} as
\begin{align}
\label{eq:spectralzetacalc2}
\begin{split}
Z_d(s,a,k)&=\zeta_d(ds,a)+\sum\limits_{r=1}^\infty\frac{(s)_r}{r!}\left(\sum\limits_{j=0}^{k-1}\alpha_j^{2r}\right)\zeta_d(2ks+2r,a)\\
&\quad+\sum\limits_{r,r'=1}^\infty\frac{(s)_{r}}{r!}\frac{(s)_{r'}}{r'!}\left(\sum\limits_{i=1}^{k-1}
\sum\limits_{j=0}^{i-1}\alpha_i^{2r}\alpha_j^{2r'}\right)\zeta_d(2ks+2r+2r',a)+R(s).
\end{split}
\end{align}
The double sum vanishes at $s=0$ as $\zeta_d$ only contributes a pole proportional to $1/s$
while the Pochhammer symbols each contribute a factor of $s$. 
The rest of the expression looks just like the $\zeta$-functions in the previous section, 
so the result follows directly from \eqref{eq:zeta0} and \eqref{eq:zeta0calc1},
\begin{equation}
\label{check-logR-termHD-Juan}
Z_d(0,a,k)=\zeta_d(0,a)+\frac{1}{2k}\sum\limits_{r=1}^{d/2}\frac{1}{r}
\left(\sum\limits_{j=0}^{k-1}\alpha_j^{2r}\right)R_d(2r,a)\,.
\end{equation}
Using equation \eqref{eq:identityofbarnesfcts} we can rewrite this as
\begin{equation}
Z_d(0,a,k)=\frac{1}{2k}\sum\limits_{j=0}^{k-1}\left(\zeta_d(0,a+\alpha_j)+\zeta_d(0,a-\alpha_j)\right).
\end{equation}
By virtue of \eqref{eq:barnes-zeta-at-non-integers} evaluated at $k=0$, this can in turn be written in terms of 
generalised Bernoulli polynomials \cite{Dowker2011},
\be
\label{check-logR-termHD}
Z_d(0, a, k)= {1\over 2 k\, d!} \sum_{j=0}^{k-1} \left(B_d^{(d)}(d/2+j+1)+B_d^{(d)}(d/2-j)\right).
\ee
Evaluating the derivative at $s=0$ is also relatively painless now. For the first two terms the result 
follows directly from the discussion in the previous sections, in particular from equation \eqref{eq:sumidentity1} 
in the Dirichlet and \eqref{eq:sumidentity2} in the Neumann case. In the Dirichlet case we thus have
\begin{align}
\label{eq:mone}
\begin{split}
\partial_s\Bigg(\zeta_d(2ks,a)&+\sum\limits_{r=1}^\infty\frac{(s)_r}{r!}
\left(\sum\limits_{j=0}^{k-1}\alpha_j^{2r}\right)\zeta_d(2ks+2r,a)\Bigg)\Bigg\vert_{s=0}=\\
&=\sum\limits_{j=0}^{k-1}\left(\zeta'_d(0,a+\alpha_j)+\zeta'_d(0,a-\alpha_j)\right)
+\sum\limits_{r=1}^{d/2}\frac{1}{r}\left(\sum\limits_{j=0}^{k-1}\alpha_j^{2r}\right)H_k(r)R_d(2r,a)\\
&=\,: \sum\limits_{j=0}^{k-1}\left(\zeta'_d(0,a+\alpha_j)+\zeta'_d(0,a-\alpha_j)\right)+M_1(d,a,k),
\end{split}
\end{align}
where we defined $M_1$ in the last line to simplify the notation. 
To calculate the contribution from the double sum in \eqref{eq:spectralzetacalc2} 
we remind the reader that $\zeta_d(s+n,a)=\frac{1}{s}R_d(n,a)+\mathcal{O}(s^0)$ 
and $\zeta'_d(s+n,a)=\frac{-1}{s^2}R_d(n,a)+\mathcal{O}(s)$ for $n=1,\ldots,d$. 
Since $(s)_r=\,: sf(s)=\mathcal{O}(s)$ with $f(0)=(r-1)!$, this means that the only 
terms in the sum that will survive are those for which $2r+2r'\leq d$. 
Also noting that $\partial_s (s)_r\big\vert_{s=0}= (r-1)!$ we get the following result
\begin{align}
\label{eq:mtwo}
\begin{split}
\partial_s \sum\limits_{r,r'=1}^\infty & \frac{(s)_{r}}{r!}\frac{(s)_{r'}}{r'!}
\underbrace{\left(\sum\limits_{i=0}^{k-1}\sum\limits_{j=0}^{i-1}\alpha_i^{2r}
\alpha_j^{2r'}\right)}_{= \, :\,  A(r,r')}\zeta_d(2ks+2r+2r',a)\Bigg\vert_{s=0}\\
&=\sum_{r=1}^{d/2}\sum_{r'=1}^{d/2-r}\frac{1}{r}\frac{1}{r'} A(r,r')
\left[s^2\partial_s\zeta_d(2ks+2r+2r',a)\right]_{s=0}\\
&\qquad\qquad 
+ 2 \sum_{r=1}^{d/2}\sum_{r'=1}^{d/2-r}\frac{1}{r}\frac{1}{r'} A(r,r')\left[s\zeta_d(2ks+2r+2r',a)\right]_{s=0}\\
&=\sum_{r=1}^{d/2}\sum_{r'=1}^{d/2-r}\frac{1}{r}\frac{1}{r'} A(r,r')\left(-\frac{1}{2k}R_d(2r+2r',a)+\frac{2}{2k}R_d(2r+2r',a)\right)\\
&=\frac{1}{2k}\sum_{r=1}^{d/2}\sum_{r'=1}^{d/2-r}\frac{1}{r}\frac{1}{r'} A(r,r')R_d(2r+2r',a)=\,: M_2(d,a,k).
\end{split}
\end{align}
Putting \eqref{eq:mone} and \eqref{eq:mtwo} together we get the result 
\begin{align}
\label{eq:spectralprime}
\begin{split}
Z'_d(0,a,k)&=\sum\limits_{j=0}^{k-1}\left(\zeta'_d(0,a+\alpha_j)+\zeta'_d(0,a-\alpha_j)\right)+M(d,a,k)\\
&=\log(\frac{1}{\rho_d^{2k}}\prod_{j=0}^{k-1}\Gamma_d(a+\alpha_j)\Gamma_d(a-\alpha_j))+M(d,a,k),
\end{split}
\end{align}
where in the last line we used the definition of the multiple Gamma function \eqref{eq:multgammadef} 
to rewrite the result, and where we defined
\begin{equation}
\label{eq:total-m-def}
M(d,k):\,= M_1(d,a,k)+M_2(d,a,k).
\end{equation}
We omitted $a$ in $M(d,k)$, because the property $B_n^{(\alpha)}(x)=(-1)^n B^{(\alpha)}_n(\alpha-x)$ \cite{Srivastava2012} of generalised Bernoulli polynomials implies, according to \eqref{eq:R-def}, that $R_d(2r,a_N)=R_d(2r,a_D)$ for even dimension, which in turn implies that $M(d,k)$ does not depend on whether we chose $a_D$ or $a_N$.
It is possible to further simplify the result by using \eqref{eq:multgammadef} 
and noting that $\alpha_j+n=\alpha_{j+n}$. This induces a telescope-like cancellation in the product,
\begin{align}
\label{eq:spectralprimethree}
\begin{split}
\log(\frac{1}{\rho_d^{2k}}\prod_{j=0}^{k-1}\Gamma_d(a+\alpha_j)\Gamma_d(a-\alpha_j))&
=\log(\prod_{j=0}^{k-1}\frac{\Gamma_d(a+\alpha_j)}{\rho_d}\frac{\Gamma_d(a-\alpha_j)}{\rho_d})\\
&=\log(\prod_{j=0}^{k-1}\frac{\Gamma_{d+1}(a+\alpha_j)}{\Gamma_{d+1}(a+\alpha_{j+1})}
\frac{\Gamma_{d+1}(a-\alpha_j)}{\Gamma_{d+1}(a-\alpha_{j-1})})\\
&=\log(\frac{\Gamma_{d+1}(a+\alpha_0)}{\Gamma_{d+1}(a+\alpha_{k})}
\frac{\Gamma_{d+1}(a-\alpha_{k-1})}{\Gamma_{d+1}(a-\alpha_{-1})})\\
&=\log(\frac{\Gamma_{d+1}(d/2-k+1)}{\Gamma_{d+1}(d/2+k+1)}).
\end{split}
\end{align}
Putting everything together we get the following expression for the log-determinant
\begin{align}
\label{eq:logdet-d}
\begin{split}
\log \det P_{2k,H^d_{D}}&=-Z'_d(0,a_D,k)\\
&=-\log(\frac{\Gamma_{d+1}(d/2-k+1)}{\Gamma_{d+1}(d/2+k+1)})-M(d,k)\\
&=-\zeta'_{d+1}(0,d/2-k+1)+\zeta'_{d+1}(0,d/2+k+1)-M(d,k),\\
\end{split}
\end{align}
which is valid for $k=1,\ldots,d/2$, {\it i.e.} in the critical as well as subcritical case.

\subsubsection{Determinant on the hemisphere with Neumann boundary conditions}
\label{sec:det-hem-N}

In the subcritical case there is no zero mode and we can continue from 
\eqref{eq:spectralprimethree}, inserting the appropriate $a$ for Neumann boundary conditions, {\it i.e.} $a_N=(d-1)/2$. 
Then the expression for the spectral $\zeta$-function is as in 
\eqref{check-logR-termHD}, thanks to the above mentioned property of the generalised Bernoulli polynomials \cite{Dowker2011}, 
and the subcritical expression for the determinant is
\begin{align}
\label{eq:logdet-n-subcrit}
\begin{split}
\log \det P_{2k,H^d_{N}}&=-Z'_d(0,a_N,k)\\
&=-\log(\frac{\Gamma_{d+1}(d/2-k)}{\Gamma_{d+1}(d/2+k)})-M(d,k)\\
&=-\zeta'_{d+1}(0,d/2-k)+\zeta'_{d+1}(0,d/2+k)-M(d,k).\\
\end{split}
\end{align}
In the critical case $k=d/2$ the calculation is more subtle. In order to get a well-defined expression, we must subtract the 
zero mode contribution from the sum, 
\begin{align}
\label{eq:spectralzetaneumann}
\begin{split}
\bar{Z}_d(s,a_N,k)&=\sum\limits_{\mathbf{m}\in\mathbb{N}^d\setminus\{0\}}
\prod_{j=0}^{k-1}\left((\mathbf{m}\cdot\mathbf{d}+a_N)^2-\alpha_j^2\right)^{-s}\\
&=Z_d(s,a_N,k)-\prod_{j=0}^{k-1}\left(a_N^2-\alpha_j^2\right)^{-s}.
\end{split}
\end{align}
At $s=0$ we have for the critical case $k=d/2$ \cite{Dowker2011}
\be
\label{check-logR-termHN-critical}
\bar Z_d(0, a_N, k)= {1\over 2 k\, d!} \sum_{j=0}^{k-1} \left(B_d^{(d)}(d/2-j-1)+B_d^{(d)}(d/2+j)\right) -1\,.
\ee
We can now use \eqref{eq:spectralprime} as well as \eqref{eq:neumanzetazero}, to write down the following expression, 
\begin{align}
\label{eq:spectralbarprime}
\begin{split}
\bar{Z}'(0,a_N,k)&=\lim_{\varepsilon\rightarrow 0} 
\sum_{j=0}^{k-2}\left(\zeta'_d(0,a_N+\alpha_j)+\zeta'_d(0,a_N-\alpha_j)\right) +\zeta'_d(0,a_N+\alpha_{k-1})\\
&\quad\quad -\log(\varepsilon)-\log(\rho_d)+M(d,k)+\underbrace{\log(a_N^2-\alpha_{k-1}^2)}_{=\log(a_N+\alpha_{k-1})
+\log(\varepsilon)}+\sum_{j=0}^{k-2}\log(a_N^2-\alpha_{j}^2)+\mathcal{O}(\varepsilon),
\end{split}
\end{align}
where we denote $\varepsilon=a_N-\alpha_{k-1}$. The logarithmic divergence cancels and we obtain
\begin{align}
\label{eq:spectralbarprime2}
\begin{split}
\bar{Z}'(0,a_N,k)&=\sum_{j=0}^{k-2}\left(\zeta'_d(0,a_N+\alpha_j)+\zeta'_d(0,a_N-\alpha_j)+ \log(a_N^2-\alpha_{j}^2)\right)\\
&\quad +\zeta'_d(0,a_N+\alpha_{k-1}) -\log(\rho_d)+\log(d-1) +M(d,k)\\
&=\sum_{j=0}^{k-2}\left(\zeta'_d(0,a_N+\alpha_j)+\zeta'_d(0,a_N-\alpha_j)\right) +\zeta'_d(0,d-1) -\log(\rho_d)+\log((d-1)!)+M(d,k)\\
&=\log(\frac{(d-1)!}{\rho_d^{2k}}\Gamma_d(d-1)\prod_{j=0}^{k-2}\Gamma_d(a_N+\alpha_j)\Gamma(a_N-\alpha_j)) +M(d,k)
\end{split}
\end{align}
where we used the fact that $a_N+\alpha_{k-1}=d-1$ and $\sum_{j=0}^{k-2}\log(a_N^2-\alpha_j^2)=\log((d-2)!)$ for 
Neumann boundary conditions. As before, we can use identities involving the multiple Gamma function to induce 
cancellations. This leads us to the final result in the critical case,
\begin{align}
\label{eq:logdet-n-crit}
\begin{split}
\log \det P_{d,H^d_{N}}&=-\bar{Z}'_d(0,a_N,d/2)\\
&=-\log(\frac{(d-1)!}{\rho_d}\frac{\Gamma_{d+1}(1)}{\Gamma_{d+1}(d)})-M(d)\\
&=-\log(\frac{(d-1)!}{\Gamma_{d+1}(d)})-M(d)\\
&=-\log((d-1)!)+\log \rho_{d+1}+\zeta'_{d+1}(0,d)-M(d),\\
\end{split}
\end{align}
where we used $\rho_d=\Gamma_{d+1}(1)$, and the abbreviated notation $M(d,d/2)=\,: M(d)$. 
Since the critical case is the most relevant to this paper, we give the explicit form of $M(d)$,
\begin{align}
\label{eq:m-explicit}
\begin{split}
M(d)&=\sum\limits_{r,l=1}^{d/2}\frac{1}{r (d-2r)! (2r-1)!}\left(l+\frac{1}{2}\right)^{2r}
\left(\frac{H_{r-1}}{d}-H_{2r-1}\right)B_{d-2r}^{(d)}\left(\frac{d+1}{2}\right)\\
&+\frac{1}{d}\sum_{r,l=1}^{d/2}\sum_{t=1}^{d/2-r}\sum_{m=1}^{l}\frac{1}{rt (d{-}2r{-}2t)! (2r{+}2t{-}1)!}
\left(l{+}\frac{1}{2}\right)^{2r}\left(m{+}\frac{1}{2}\right)^{2t}B_{d{-}2r{-}2t}^{(d)}\left(\frac{d{+}1}{2}\right).
\end{split}
\end{align}
As mentioned in the main body, our result differs by Dowker's result due to the different sign in front of the $\log (d-1)!$ term
because we are considering the functional determinant of the GJMS operators on the sphere with the zero mode removed. 

\subsubsection{Determinant on the sphere}
\label{sec:det-sphere-ND}

The log-determinant on the sphere is obtained by adding the log-determinants on the hemisphere for 
Dirichlet and Neumann boundary conditions. In the critical case,
the spectral $\zeta$-function for the sphere at $s=0$ is given by \eqref{check-logR-termHD} and 
\eqref{check-logR-termHN-critical}, \cite{Dowker2011}, so that
\be
\label{check-logR-termS-crit}
Z_d(0, k)= {1\over k\, d!} \sum_{j=0}^{k-1} \left( B_d^{(d)}(d/2+j+1)+B_d^{(d)}(d/2+j)\right)-1\,, 
\ee
and the log-determinant on the sphere is thus given by
\begin{align}
\label{eq:spectralprimesphere}
\begin{split}
\log\det P_{d, S^d}&=\log\det P_{d, H^d_D}+\log\det P_{d, H^d_N}\\
&=-\log((d-1)!\frac{\Gamma_{d+1}(1)}{\Gamma_{d+1}(d)\Gamma_{d+1}(d+1)})-2M(d)\\
&=-\zeta'_{d+1}(0,1)+\zeta'_{d+1}(0,d+1)+\zeta'_{d+1}(0,d)\\
&\qquad-\log((d-1)!)+\log \rho_{d+1}-2M(d).
\end{split}
\end{align}
In the subcritical case we instead have 
\be
\label{check-logR-termS-subcrit}
Z_d(0, k)= {1\over k\, d!} \sum_{j=0}^{k-1} \left( B_d^{(d)}(d/2+j+1)+B_d^{(d)}(d/2+j)\right)\,, 
\ee
and the functional determinant reads
\begin{align}
\label{eq:spectralprimespheresubcrit}
\begin{split}
\log\det P_{2k, S^d}&=\log\det P_{2k, H^d_D}+\log\det P_{2k, H^d_N}\\
&=-\log(\frac{\Gamma_{d+1}(d/2-k)}{\Gamma_{d+1}(d/2+k)}\frac{\Gamma_{d+1}(d/2-k+1)}{\Gamma_{d+1}(d/2+k+1)})-2M(d, k)\\
&=-\zeta'_{d+1}(0,d/2-k)+\zeta'_{d+1}(0,d/2+k)+\\
&\qquad-\zeta'_{d+1}(0,d/2-k+1)+\zeta'_{d+1}(0,d/2+k+1)-2M(d, k).\\
\end{split}
\end{align}

\section{An alternative form for determinants of GJMS operators on spheres and hemispheres}
\label{sec:alternative-dowker}

In this appendix we will show how to rewrite Dowker's expressions \eqref{eq:logdet-d} 
and \eqref{eq:logdet-n-crit} for the log-determinants of GJMS operators on hemispheres as
\begin{align}
\log \det P_{2k,H^d_{D}}&=-\sum_{n=0}^{d}h^{D}_n(d,k)\zeta'(-n)- f^D(d,k),\notag\\
\log \det P_{2k,H^d_{N}}&=-\sum_{n=0}^{d}h^{N}_n(d,k)\zeta'(-n)- f^N(d,k),
\end{align}
where $\zeta$ is the Riemann $\zeta$-function, and $h_n$ and $f$ are functions that we derive and that depend 
on the boundary conditions, the dimension, and the degree of the GJMS-operator. The spherical case then follows 
directly as the sum of Dirichlet and Neumann hemispherical results,
\begin{equation}
\log \det P_{2k,S^d}=-\sum_{n=0}^{d}\left(h^{D}_n(d,k)+h^{N}_n(d,k)\right)\zeta'(-n)- f^D(d,k)-f^N(d,k).
\end{equation}

\subsection{Rewriting $\zeta_d$ in terms of the Riemann $\zeta$-function}
In \cite{Adamchik2003}, Adamchik gives a closed form of the Barnes $\zeta$-function in terms of a series of Riemann $\zeta$-functions. 
His calculation is summarized by equations (14), (17), and (23) in the reference, which in our slightly different notation 
and after a bit of rearranging reads
\begin{align}
\label{eq:Adamchikscalc1}
\begin{split}
\zeta_d'(0,z)&=(-1)^{d+1}\log G_d(z)+\sum_{k=0}^{d-1}(-1)^{k}\binom{z}{k}R_{d-k},
\end{split}
\end{align}
where we note that the $G_d(z)$ are multiple Gamma functions with a different normalization compared to the one used by Dowker. 
We have the following explicit closed forms for all the parts of the above expression,
\begin{align}
\label{eq:Adamchikscalc2}
\begin{split}
\log G_d(z)&=\frac{(-1)^d}{(d-1)!}\sum_{k=0}^{d-1}P_{k,d}(z)\left( \zeta'(-k)-\zeta'(-k,z)\right),\quad \Re(z)>0\\
R_{d-k}&=\frac{1}{(d-k-1)!}\sum_{l=0}^{d-k-1}\stirling{d-k}{l+1}\zeta'(-l).
\end{split}
\end{align}
For our purposes, it is not necessary to know how the polynomials $P_{k,d}(z)$ are defined, it suffices to know that they 
satisfy \cite{Adamchik2003}
\begin{equation}
\label{eq:Pkdpoly}
\sum_{k=0}^{d-1}P_{k,d}(z)n^k=(n-z+1)_d = \prod_{k=1}^{d-1}(n+k-z).
\end{equation}
We are only interested in the special case of the above formula, for which $z$ is a positive integer. 
For $z=1$ it is clear that $\log G_d(1)=0$. For $z>1$ we can use the fact that
\begin{equation}
\zeta(s,z)=\zeta(s)-\sum_{n=1}^{z-1}\frac{1}{n^s}
\end{equation}
to simplify 
\begin{equation}
\zeta'(-k)-\zeta'(-k,z)=-\sum_{n=1}^{z-1}n^k\log n.
\end{equation}
We can now define the following quantity
\begin{align}
\label{eq:Addef}
\begin{split}
A(d,z)&:\, = (-1)^{d+1}\log G_d(z)\\
&=\frac{1}{(d-1)!}\sum_{k=0}^{d-1}P_{k,d}(z)\sum_{n=1}^{z-1}n^k\log n\\
&=\frac{1}{(d-1)!}\sum_{n=1}^{z-1}\log(n)(n-z+1)_d\,,\quad z>1\,.
\end{split}
\end{align}
It is further clear from $\zeta(s)=\zeta(s,1)$ that $A(d,1)=0$. 
Another very useful simplification comes from rewriting the $R_{d-k}$ sum in \eqref{eq:Adamchikscalc1} as
\begin{align}
\begin{split}
\sum_{k=0}^{d-1}(-1)^k\binom{z}{k}R_{d-k}&=\sum_{k=0}^{d-1}
\binom{z}{k}\frac{(-1)^{k}}{(d-k-1)!}\sum_{l=0}^{d-k-1}\stirling{d-k}{l+1}\zeta'(-l)\\
&=\sum_{l=0}^{d-1}\left(\sum_{k=0}^{d-l-1}\frac{(-1)^{k}}{(d-k-1)!}\binom{z}{k}\stirling{d-k}{l+1}\right)\zeta'(-l)\\
&=\, :  \sum_{l=0}^{d-1} D_{l}(d,z)\zeta'(-l),
\end{split}
\end{align}
where we define $D_{k}(d,z)$ in the last line as 
\begin{equation}
\label{eq:Dkddef}
D_{k}(d,z):\, = \sum_{j=0}^{d-k-1}\frac{(-1)^j}{(d-j-1)!}\binom{z}{j}\stirling{d-j}{k+1}.
\end{equation}
With these definitions it possible to write down the $\zeta'_d(0,z)$ in the following very simple way
\begin{equation}
\label{eq:barneszetaprimezero}
\zeta'_d(0,z)=A(d,z)+\sum_{k=0}^{d-1} D_{k}(d,z)\zeta'(-k),\quad z\in\mathbb{N}^+.
\end{equation}

\subsection{Determinants in terms of Riemann $\zeta$-functions}
\label{sec:det-hem-alt}
\paragraph{Dirichlet boundary conditions on the hemisphere.} We can now use the above technology to rewrite the log-determinant \eqref{eq:logdet-d} on the hemisphere 
for Dirichlet boundary conditions,
\begin{align}
\label{eq:det-hem-d-riem}
\begin{split}
\log \det P_{2k,H^d_{D}}&=-\zeta'_{d+1}(0,d/2-k+1)+\zeta'_{d+1}(0,d/2+k+1)-M(d)\\
&=-A(d+1,d/2-k+1)+A(d+1,d/2+k+1)-M(d)-\\
&\quad-\sum_{n=0}^{d}\left(D_{n}(d+1,d/2-k+1)-D_{n}(d+1,d/2+k+1)\right)\zeta'(-n)\\
&=\, :-\sum_{n=0}^{d}h^{D}_n(d,k)\zeta'(-n)- f^D(d,k).
\end{split}
\end{align}
Here we have defined two functions
\begin{align}
h^{D}_n(d,k)&:\, = D_{n}(d+1,d/2-k+1)-D_{n}(d+1,d/2+k+1),\\
f^D(d,k)&:\, =  A(d+1,d/2-k+1)-A(d+1,d/2+k+1)+M(d).
\end{align}
Since the critical case $k=d/2$ is of particular interest to us, we give the explicit form of $h^D_n$ and $f^D$ in that case,
\begin{align}
h^D_n(d)&:\, = h^D_n (d,d/2)\notag\\
\label{eq:det-hem-dir-coeff-h-crit-appendix}
&=-\frac{1}{(d-1)!}\stirling{d}{n+1}+\frac{1}{d!}\stirling{d+1}{n+1}-\sum_{j=0}^{d-n}\frac{(-1)^j}{(d-j)!}\binom{d+1}{j}\stirling{d-j+1}{n+1},\\
f^D(d)&:\, = f^D(d,d/2) \notag\\
\label{eq:det-hem-dir-coeff-f-crit-appendix}
&= -\frac{1}{d!}\sum_{j=1}^{d-1}\log(j)(j-d)_{d+1}+M(d).
\end{align}

\subsubsection*{Neumann boundary conditions on the hemisphere}

Writing the subcritical, i.e. $k=1,\ldots,d/2-1$, determinant \eqref{eq:logdet-n-subcrit} for Neumann boundary conditions is straightforward,
\begin{align}
\log \det P_{2k,H^d_{N}}&=-\zeta'_{d+1}(0,d/2-k)+\zeta'_{d+1}(0,d/2+k)-M(d)\notag\\
&=-A(d+1,d/2-k)+A(d+1,d/2+k)-M(d)-\notag\\
&\quad-\sum_{n=0}^{d}\left(D_{n}(d+1,d/2-k)-D_{n}(d+1,d/2+k)\right)\zeta'(-n)\notag\\
\label{eq:det-hem-n-riem-subcrit}
&=\, : -\sum_{n=0}^{d}h^{N}_n(d,k)\zeta'(-n)- f^N(d,k),
\end{align}
with the functions $h^N_n(d,k)$ and $f^N(d,k)$ defined in the last line as
\begin{align}
h^{N}_n(d,k)&:\, = D_{n}(d+1,d/2-k)-D_{n}(d+1,d/2+k),\\
f^N(d,k)&:\, =  A(d+1,d/2-k)-A(d+1,d/2+k)+M(d).
\end{align}
The critical case \eqref{eq:logdet-n-crit} is slightly harder, as we also need to rewrite $\log \rho_{d+1}$. 
In order to do this, we use the following formula from \cite{Adamchik2003},
\begin{equation}
\log \rho_d = -\frac{1}{(d-1)!}\sum_{n=0}^{d-1}\stirling{d}{n+1}\zeta'(-n).
\end{equation}
Reminding the reader that $\stirling{d}{d+1}=0$ we can thus write
\begin{align}
\label{eq:det-hem-n-riem}
\begin{split}
\log \det P_{d,H^d_{N}}&=-\log (d-1)!+\log \rho_{d+1}+\zeta'_{d+1}(0,d)-M(d)\\
&=-\sum_{n=0}^d\left(\frac{1}{d!}\stirling{d+1}{n+1}-D_{n}(d+1,d)\right)\zeta'(-n)-\log (d-1)!+A(d+1,d)-M(d)\\
&=\, :  -\sum_{n=0}^{d}h^{N}_n(d)\zeta'(-n)- f^N(d)
\end{split}
\end{align}
where we define $h^N_n(d)$ and $f^N(d)$ in the last line. 
In analogy to the Dirichlet case we can define $h^N_n(d,d/2):\, =  h^N_n(d)$ and 
$f^N(d,d/2):\, =  f^N(d)$, as with this definition equation \eqref{eq:det-hem-n-riem-subcrit} 
becomes valid for for all $k$. Again we write these functions explicitly for the critical case,
\begin{align}
\label{eq:det-hem-dir-coeff-h-crit-appendix}
h^N_n(d)&=\frac{1}{d!}\stirling{d+1}{n+1}-\sum_{j=0}^{d-n}\frac{(-1)^j}{(d-j)!}\binom{d}{j}\stirling{d-j+1}{n+1},\\
\label{eq:det-hem-dir-coeff-f-crit-appendix}
f^N(d) &=\log (d-1)!-\frac{1}{d!}\sum_{j=1}^{d-1}\log(j)(j-d)_{d+1}+M(d).
\end{align}

\section{Functional determinants on the flat $d$-torus}
\label{sec:Torusdet}

In this appendix we calculate functional determinants of the Laplacian on the flat torus (Appendix \ref{app:det-L-torus})
and on a cylinder obtained by cutting the torus along a cycle and imposing Dirichlet boundary conditions along the cut
(Appendix \ref{app:det-cut-torus}). 
After that, in Appendix \ref{app:torus-final-det}, we derive the functional determinant of $k$-powers of the Laplacian 
both on the flat torus as well as on the cylinder.

\subsection{Determinant of the Laplacian on the flat torus}
\label{app:det-L-torus}

The $d$-dimensional flat torus can be defined as the hyperinterval with side lengths given by $L_i$ and opposing sides identified,
 \be
 T^d_{L_1,\ldots,L_d}:\,=\mathbb{R}^d/(L_1\mathbb{Z}\cross\ldots\cross L_d\mathbb{Z}).
 \ee 
Since the manifold is flat, the Laplace-Beltrami operator is just the standard Laplacian on Euclidean space. 

We compute the functional determinant for the Laplacian on the $d$-torus via a spectral $\zeta$-function method. 
The eigenvalues of the Laplace operator $-\p_a\p^a$, are given by
\begin{equation}
\label{eq:torus-ev}
\lambda_{n_1,\ldots,n_d}=\left(\frac{2\pi n_1}{L_1}\right)^2+\cdots+\left(\frac{2\pi n_d}{L_d}\right)^2\,, 
\qquad n_1,\ldots,n_d\in\mathbb{Z}\,.
\end{equation}
As usual, the zero mode, with $n_1= n_2=\dots = n_d=0$, needs to be removed in the computation of the spectral $\zeta$-function.
In order to calculate the determinant of the Laplacian, we need to evaluate
\begin{align}
\label{eq:torus-zeta-function}
\begin{split}
\zeta_{T^d_{L_1,\ldots,L_d}}(s)&:\, = \sumprime_{\vec{n}\in\mathbb{Z}^d} \left(\left(\frac{2\pi n_1}{L_1}\right)^2+\cdots
+\left(\frac{2\pi n_d}{L_d}\right)^2\right)^{-s}\\
&=\sumprime_{\vec{n}\in\mathbb{Z}^d} \left(\vec{n}_d^T\, \Xi \, \vec{n}_d\right)^{-s},
\end{split}
\end{align}
where $\vec{n}_d:\,= (n_1,\ldots,n_d)$ is a $d$-vector, $\Xi:\, = \text{diag}\left(\left(2\pi/L_1\right)^2,\ldots,\left(2\pi/L_d\right)^2\right)$
a $d\times d$ matrix, and the prime on the sum denotes the omission of the zero mode.

In section 2.2 of \cite{Elizalde1998} the analytic continuation of the sum \eqref{eq:torus-zeta-function} is evaluated recursively. In our notation the result is given by
\begin{multline}
\label{eq:torus-zeta-function-elizalde}
\zeta_{T^d_{L_1,\ldots,L_d}}(s)=2\left(\frac{L_1}{2\pi}\right)^{2s}\zeta(2s)
+L_1\frac{\Gamma(s-1/2)}{2\sqrt{\pi}\Gamma(s)}\zeta_{T^{d-1}_{L_2,\ldots,L_d}}(s-1/2)+G(s;L_1,\ldots,L_d) \,,
\end{multline}
where we define the function
\begin{align}
\label{eq:help-function-G}
\begin{split}
& G(s;L_1,L_2,\ldots,L_d):\, = \\
& \frac{2^{3/2-s}L_1^{s+1/2}}{\Gamma(s)\sqrt{\pi}}\sumprime_{\vec{n}_{d-1}\in\mathbb{Z}^{d-1}}
\sum_{n_1=1}^\infty\ \left(\frac{n_1}{\sqrt{\vec{n}_{d-1}^T \,\Xi_{d-1}\, \vec{n}_{d-1}}}\right)^{s-1/2} 
K_{s-1/2}\left(L_1 n_1 \sqrt{\vec{n}_{d-1}^T \,\Xi_{d-1} \, \vec{n}_{d-1}}\right),
\end{split}
\end{align}
with $K_\nu(z)$ being the modified Bessel function of the second kind, $\vec n_{d-1}= (n_2,\ldots,n_d)$, 
and $\Xi_{d-1}= \text{diag}\left(\left(2\pi/L_2\right)^2,\ldots,\left(2\pi/L_d\right)^2\right)$.

We can directly evaluate the log-determinant on the flat torus, as the analytic continuation 
\eqref{eq:torus-zeta-function-elizalde} has only a pole at $s=\frac{d}{2}$ in the whole complex plane, 
\begin{align}
\label{eq:log-det-flat-torus}
\begin{split}
\log \det \mathcal{P}_{2,T^d} &= \log \det \Delta_{T^d_{L_1,\ldots,L_d}} =-\zeta'_{T^d_{L_1,\ldots,L_d}}(0)\\
&=2\log(L_1)+L_1\ \zeta_{T^{d-1}_{L_2,\ldots,L_d}}(-1/2)-G'(0;L_1,\ldots,L_d)\,.
\end{split}
\end{align}
This expression is recursive, and thus not in a closed form yet, however, it turns out to be useful in the 
evaluation of the functional determinant contribution to the entanglement entropy in Section \ref{sec:torus}. 

We now solve the recursion directly and provide a closed form for the $\zeta$-function. 
Our calculation of the closed form is slightly more direct than the one carried out in section 4.2.3 of \cite{Elizalde2012}
but the end result is the same.  
In order to make the calculation more transparent, let us for a while rewrite \eqref{eq:torus-zeta-function-elizalde} as
\begin{equation}
\zeta_{d}(s)=h_1(s)+f_1(s)\zeta_{d-1}\left(s-\frac{1}{2}\right)+g_d(s),
\end{equation}
where the functions above summarize the information in the recursion:
\begin{subequations}
\begin{align}
\zeta_{d-k}(s)&:\, =\zeta_{T^{d-k}_{L_{k+1},\ldots,L_d}}(s)\,,\\
h_k(s)&:\, = 2 \left( \frac{L_k}{2\pi} \right)^{2s}\zeta(2s)\,,\\
f_k(s)&:\, = L_k \frac{\Gamma(s-1/2)}{2\sqrt{\pi}\Gamma(s)}\,,\\
g_{d-k}(s)& :\, = G(s;L_{k+1},\ldots,L_d)\,.
\end{align}
\end{subequations}
In this notation it is not difficult to see, that after $k$ steps we get 
\begin{align}
\label{eq:recursion-k-steps}
\begin{split}
\zeta_{d}(s)&= h_1(s)+\sum_{j=2}^k h_j\left(s-\frac{j-1}{2}\right)\prod_{i=1}^{j-1}f_i\left(s-\frac{i-1}{2}\right)
+g_{d}(s)+\\
&+\sum_{j=2}^k g_{d-j+1}\left(s-\frac{j-1}{2}\right)\prod_{i=1}^{j-1}f_i\left(s-\frac{i-1}{2}\right)
+\zeta_{d-k}\left(s-\frac{k}{2}\right)\prod_{j=1}^{k}f_j\left(s-\frac{j-1}{2}\right)\,.
\end{split}
\end{align}
The anchor for the recursion is 
\begin{align}
\begin{split}
\zeta_1(s) :\, =  \sumprime_{n\in\Z}\left(\frac{2\pi n}{L_d}\right)^{-2s}
= 2\left(\frac{L_d}{2\pi}\right)^{2s}\zeta(2s) =\, : \, h_d(s)\,.\label{def-anchor}
\end{split}
\end{align}
We note that the products of  $f_k$ functions in \eqref{eq:recursion-k-steps} simplify to 
\begin{equation}
\prod_{i=1}^{j-1}f_i\left(s-\frac{i-1}{2}\right)= 
L_1L_2\cdots L_{j-1}\left(\frac{1}{2\sqrt{\pi}}\right)^{j-1}\frac{\Gamma\left(s-\frac{j-1}{2}\right)}{\Gamma(s)},
\end{equation}
hence, performing the recursion for $k=d-1$ steps and reinserting the definitions then gives us the following expression 
\begin{align}
\label{eq:recursion-(d-1)-steps}
\begin{split}
\zeta_{T^{d}_{L_{1},\ldots,L_d}}(s)&=  \frac{2}{\sqrt{\pi}}\left(\frac{1}{2\pi}\right)^{2s}
\sum_{j=1}^{d}\pi^\frac{j}{2}\frac{\Gamma\left(s-\frac{j-1}{2}\right)}{\Gamma(s)}\zeta(2s-j+1) L_1\cdots L_{j-1} L_j^{2s-j+1}+\\
&\quad+\frac{1}{\Gamma(s)}\sum_{j=1}^{d-1}L_1\cdots L_{j-1}
\left(\frac{1}{2\sqrt{\pi}}\right)^{j-1}\Gamma\left(s-\frac{j-1}{2}\right)G\left(s-\frac{j-1}{2};L_{j},\ldots,L_d\right),
\end{split}
\end{align} 
with the understanding that for $j=1$ the product $L_1 \dots L_{j-1}$ is simply 1. 
If we in addition insert the definition of $G$ we can rewrite the complete expression as
\begin{align}
\label{eq:closed-torus-zeta-function}
\begin{split}
\zeta_{T^{d}_{L_{1},\ldots,L_d}}(s) &=  \frac{2}{\sqrt{\pi}}\left(\frac{1}{2\pi}\right)^{2s}
\sum_{j=1}^{d}\pi^\frac{j}{2}\frac{\Gamma\left(s-\frac{j-1}{2}\right)}{\Gamma(s)}\zeta(2s-j+1) L_1\cdots L_{j-1} L_j^{2s-j+1}+\\
&+\frac{2^{2-s}}{\Gamma(s)}\sum_{j=1}^{d-1}L_1\cdots L_{j-1}\frac{L_j^{s-\frac{j-2}{2}}}{(2\pi)^{j/2}}
\sumprime_{\vec{n}\in\mathbb{Z}^{d-j}}\sum_{n_j=1}^\infty\ \left(\frac{n_j}{\sqrt{\vec{n}_{d-j}^T\, \Xi_{d-j}\, \vec{n}_{d-j}}}\right)^{s-j/2} \times \\ 
&  \times K_{s-j/2}\left(L_j n_j\sqrt{\vec{n}_{d-j}^T \, \Xi_{d-j}\, \vec{n}_{d-j}}\right)\,,
\end{split}
\end{align}
where in this notation $\Xi_{d-j}$ is $(d-j)\times (d-j)$-matrix obtained from $\Xi$ by removing the first $j$ columns and $j$ rows.
From this expression we can now directly calculate $\zeta'_{T^{d}_{L_{1},\ldots,L_d}}(0)$, as needed for the determinant. 
Calculating the derivative of the second sum at $s=0$ is simple, since 
\be
{1\over \Gamma(\varepsilon)} = \varepsilon+\dots  \,, \qquad {d\over ds}{1\over \Gamma(s)}_{|s=\varepsilon}
= 1+\dots \,, \qquad \varepsilon <<1\,, 
\ee
so the only term that survives is the one where the derivative hits the $\Gamma$-function, 
and we effectively set $s=0$ everywhere and the $\Gamma$-function to $1$. 
For the first sum the situation is a bit trickier, as there are two $\Gamma$-functions involved, 
whose poles cancel only for even $j$, meaning that we have to separate even and odd $j$ for the calculation. 
Let us first take a look at the even $j$ part of the first sum, that is $j=2\ell$, we have
\begin{multline}
\frac{d}{ds}\frac{2}{\sqrt{\pi}}\left(\frac{1}{2\pi}\right)^{2s}
\sum_{\ell=1}^{\lfloor d/2 \rfloor}\pi^\ell\frac{\Gamma\left(s-\frac{2\ell-1}{2}\right)}{\Gamma(s)}\zeta(2s-2\ell+1) 
L_1\cdots L_{2\ell-1} L_{2\ell}^{2s-2\ell+1}\Big\vert_{s=0}=\\
=2\sum_{\ell=1}^{\lfloor d/2 \rfloor}\pi^{\ell-\frac{1}{2}} \frac{L_1\cdots L_{2\ell-1}}{L_{2\ell}^{2\ell-1}}
\Gamma\left(\frac{1}{2}-\ell\right)\zeta(1-2\ell).
\end{multline}
When $j$ is odd, that is $j=2\ell-1$, we get 
\begin{multline}
 \frac{d}{ds}\frac{2}{\sqrt{\pi}}\left(\frac{1}{2\pi}\right)^{2s}\sum_{\ell=1}^{\lceil d/2 \rceil}\pi^{\ell-1/2}
 \frac{\Gamma\left(s-\ell+1\right)}{\Gamma(s)}\zeta(2s-2\ell+2) L_1\cdots L_{2\ell-2} L_{2\ell-1}^{2s-2\ell+2}\Big\vert_{s=0}=\\
 =-2\log(L_1)+ 4\sum_{\ell=1}^{\lceil d/2 \rceil-1} \frac{ L_1\cdots L_{2\ell}}{L_{2\ell+1}^{2\ell}}\frac{(-\pi)^\ell}{\ell!}\zeta'(-2\ell).
\end{multline}
Before putting everything together, we note that the sums over the modified Bessel functions converge 
exponentially at $s=0$~\cite{Elizalde1998}. We can thus introduce the following notation for their limits
\begin{equation}
\label{def-sfunction}
S_{d-j}(L_{j},\ldots,L_d):\, =\frac{1}{(2\pi)^{j/2}}\sumprime_{\vec{n}\in\mathbb{Z}^{d-j}}
\sum_{n_j=1}^\infty\ \left(\frac{\sqrt{\vec{n}_{d-j}^T \Xi_{d-j}\, \vec{n}_{d-j}}}{n_j}\right)^{\frac{j}{2}} 
K_{-j/2}\left(L_j n_j\sqrt{\vec{n}_{d-j}^T \, \Xi_{d-j}\, \vec{n}_{d-j}}\right),
\end{equation}
where $S_{d-j}$ is convergent for $d>j>0$. 
Finally, we obtain 
\begin{multline}
\label{eq:closed-torus-zeta-prime}
\zeta'_{T^{d}_{L_{1},\ldots,L_d}}(0)=-2\log(L_1)
+4\sum_{j=1}^{\lceil d/2 \rceil-1} \frac{ L_1\cdots L_{2j}}{L_{2j+1}^{2j}}\frac{(-\pi)^j}{j!}\zeta'(-2j)+\\
+2\sum_{j=1}^{\lfloor d/2 \rfloor}\frac{L_1\cdots L_{2j-1}}{L_{2j}^{2j-1}}\frac{(-2\pi)^j}{(2j-1)!!}
\zeta(-(2j-1))+4\sum_{j=1}^{d-1}\frac{L_1\cdots L_{j-1}}{L_j^{\frac{j}{2}-1}}S_{d-j}(L_{j+1},\ldots,L_d).
\end{multline}
The expression above gives us the functional determinant for a $d$-dimensional torus, and despite its
intimidating appearance, it is rather straightforward to handle, since the modified Bessel functions hidden 
in $S$ converge rapidly to zero as the integers $n_i$ increase. 

It is instructive to evaluate the $d=2$ case. First of all, we note that the sum on the first line of \eqref{eq:closed-torus-zeta-prime} is 
empty for $d=2$. The remaining two sums in \eqref{eq:closed-torus-zeta-prime} consist only of the $j=1$ term, giving us
\begin{align}
\zeta'_{T^2_{L_1,L_2}}(0)&=-2\log L_1+\frac{\pi}{3}\frac{L_1}{L_2}+4\sqrt{L_1}S_1(L_2)
 = \nn\\
 &=
-2\log L_1+\frac{\pi}{3}\frac{L_1}{L_2}+ 2 \sumprime_{n_2\in\Z} \sum_{n_1=1}^\infty {e^{-2\pi n_1 |n_2| {L_1\over L_2}}\over n_1}\,. \nn
\end{align}
Using more standard conventions, see {\it e.g.} \cite{DiFrancesco1997}, introducing the modular parameter $\tau = i L_1/L_2$, 
as well as defining $q:= e^{2\pi i \tau}$, and then performing the sum over $n_1$, we obtain
\begin{align}
\zeta'_{T^2_{L_1,L_2}}(0)&=-2\log L_1+\frac{\pi}{3}\frac{L_1}{L_2}+ 2 \sumprime_{n_2\in\Z} 
\sum_{n_1=1}^\infty {e^{-2\pi n_1 |n_2| {L_1\over L_2}}\over n_1} \nn\\
&= -2\log L_1 - \frac{ i \pi}{3}\tau - 4\sum_{m=1}^\infty \log \left(1-q^m\right) = - \log(L_1^2\eta^4(\tau)),
\end{align}
where $\eta$ is the Dedekind $\eta$-function, defined as in \eqref{def-dedekind-eta}.
If we denote the area of the $2$-torus by $A$, where $A=L_1 L_2$ in our convention,%
\footnote{We set here the coupling $g=1$ as well as $2\pi R_c=1$, since we are only interested here in checking that our calculations reproduce well known results in literature.}
then taking into account the contribution from the zero-mode, the log-determinant and the partition function on the torus can be written as
\begin{align}
\begin{split}\label{det-torus-2d}
& \log\det \Delta_{T^2} = - \zeta'_{T^2_{L_1,L_2}}(0) = \log(L_1^2\eta^4(\tau)),\\
& Z(\tau) =\sqrt{A}\exp(\frac{1}{2}\zeta'_{T^2_{L_1,L_2}}(0))=\frac{1}{\sqrt{\text{Im}(\tau)}\eta^2(\tau)},
\end{split}
\end{align}
which is a well known result, see {\it e.g.} \cite{DiFrancesco1997}.

\subsection{Determinant of the Laplacian on the cut $d$-torus}
\label{app:det-cut-torus}

We now consider cutting the torus $T^d_{L_1,\ldots,L_d}$ at $x_1=0$ and $x_1=L< L_1$, as shown in figure \ref{fig:torus}.
As discussed in the main body, this gives rise to two subsystems, each of which is a cylinder represented by an interval times a 
$d{-}1$-dimensional torus. 
Imposing Dirichlet boundary conditions \eqref{dirichlet-conds} the eigenvalues of the Laplacian $-\p_a\p^a$ are 
\begin{align}
\label{eq:cut-torus-ev}
\begin{split}
\mu_{m,n_2,\ldots,n_{d}}= \left(\frac{m\pi}{L}\right)^2 + \lambda_{n_2,\ldots,n_{d}},
\quad m\in\mathbb{N}^+,\quad n_2,\ldots,n_d\in\mathbb{Z}\, ,
\end{split}
\end{align}
with $\lambda_{n_2,\ldots,n_d}$ the eigenvalue on $T^{d-1}_{L_2,\ldots,L_d}$. 
Notice that there is no zero mode now. 
The spectral $\zeta$-function on this geometry thus takes the form
\begin{align}
\begin{split}
\zeta_{[0, L]\cross T^{d-1}_{L_2,\ldots,L_d}}(s) &:\, = \sum_\lambda \sum_{m=1}^\infty\left(\lambda+ \left(\frac{m\pi}{L}\right)^2\right)^{-s} \nonumber\\
&=\left(\frac{L}{\pi}\right)^{2s}\zeta(2s)+\sumprime_\lambda \sum_{m=1}^\infty\left(\lambda+ \left(\frac{m\pi}{L}\right)^2\right)^{-s},\nonumber
\end{split}
\end{align}
where we schematically write $\lambda$ for the $(d{-}1)$-dimensional toroidal part of the eigenvalue, and we explicitly separate 
the $(d{-}1)$-dimensional toroidal zero mode from the rest in the last passage. 
We can now evaluate the primed sum by means of the identities \eqref{gamma-rep}, 
\eqref{poisson-formula}, and \eqref{bessel-integral-rep} collected in appendix \ref{app:formulae},
and we obtain
\begin{align}
\sumprime_\lambda \sum_{m=1}^\infty\left(\lambda+ \left(\frac{m\pi}{L}\right)^2\right)^{-s}
=\ &\frac{1}{\Gamma(s)}\sumprime_\lambda\sum_{m=1}^\infty\int\limits_0^\infty dt\ t^{s-1} 
e^{-t\left(\lambda+\left(\frac{m\pi}{L}\right)^2\right)}
\nn\\
 =\ &\frac{1}{\Gamma(s)}\sumprime_\lambda\int\limits_0^\infty dt\ t^{s-1} 
e^{-t\lambda}\sum_{m=1}^\infty e^{-t\left(\frac{m\pi}{L}\right)^2}\nn\\
=\ &\frac{1}{\Gamma(s)}\sumprime_\lambda \int\limits_0^\infty dt\ t^{s-1} e^{-t\lambda}\left(-\frac{1}{2}+\frac{L}{2\sqrt{\pi}}t^{-\frac{1}{2}}
+\frac{L}{\sqrt{\pi}}t^{-\frac{1}{2}}\sum_{m=1}^\infty e^{-\frac{(m L)^2}{t}}\right)\nn\\
=\ &-\frac{1}{2}\zeta_{T^{d-1}_{L_2,\ldots,L_d}}(s)+\frac{L}{2\sqrt{\pi}}\frac{\Gamma(s-1/2)}{\Gamma(s)}
\zeta_{T^{d-1}_{L_2,\ldots,L_d}}(s-1/2)
\nn\\
&\ \ +\frac{2L^{s+\frac{1}{2}}}{\Gamma(s)\sqrt{\pi}}
\sumprime_\lambda\sum_{m=1}^\infty\left(\frac{m}{\sqrt\lambda}\right)^{s-\frac{1}{2}}K_{s-1/2}\left(2L m \sqrt{\lambda}\right)\,. 
\end{align}
Notice that $\lambda$ is nothing but $\vec n_{d-1}\, \Xi_{d-1} \, \vec n_{d-1}$ as defined in Appendix~\ref{app:det-L-torus}. 
Hence, adopting the same notation here, the above term containing the modified Bessel function can be written in terms 
of the function $G$ as defined in \eqref{eq:help-function-G}, and we can finally write the expression for the $\zeta$-function 
on the cut torus as follows,
\begin{align}
\label{eq:cut-torus-zeta-function}
\zeta_{[0,L]\cross T^{d-1}_{L_2,\ldots,L_d}}(s) &=\left(\frac{L}{\pi}\right)^{2s}\zeta(2s)
-\frac{1}{2}\zeta_{T^{d-1}_{L_2,\ldots,L_d}}(s)+ {L\over 2\sqrt \pi} \frac{\Gamma(s-1/2)}{\Gamma(s)}\zeta_{T^{d-1}_{L_2,\ldots,L_d}}(s-1/2)
+\nn\\
&+\frac{1}{2}G(s;2L ,L_2,\ldots,L_d).
\end{align}
Finally, the log-determinant is given by 
\begin{align}
\label{eq:log-det-cut-torus}
\begin{split}
&\log \det \Delta_{[0, L]\cross T^{d-1}_{L_2,\ldots,L_d}} = -\zeta'_{[0, L]\cross T^{d-1}_{L_2,\ldots,L_d}}(0) = \\
&=\log(2 L)+\frac{1}{2}\zeta'_{T^{d-1}_{L_2,\ldots,L_d}}(0) +L\, \zeta_{T^{d-1}_{L_2,\ldots,L_d}}(-1/2) -\frac{1}{2}G'(0;2 L,\ldots,L_d)\,.
\end{split}
\end{align}
Our expression \eqref{eq:log-det-cut-torus} agrees with the results of~\cite{KIRSTEN20061814} obtained by contour integration. 

Let us check that for $d=2$ we obtain the well-known result for the log determinant of the Laplacian on a cylinder. 
In this case $T^{1}_{L_2}$ is nothing but a circle of length $L_2$, and according to \eqref{def-anchor}, we have
\be
\zeta_{T^{1}_{L_2}} (s)= 2 \left({L_2 \over 2\pi}\right)^{2 s} \zeta(2 s)\,. 
\ee
Hence, we obtain 
\be
\log \det \Delta_{[0, L]\cross S_{L_2}^1} &=&- \zeta'_{[0, L]\cross S_{L_2}^1}(0) 
=\log\left(2 {L\over L_2}\right)- {\pi\over 3} {L\over L_2} - 2 \sum_{n_2=1}^\infty\sum_{n_1=1}^\infty {e^{-4 \pi {L\over L_2} n_1 n_2}\over n_1} 
\nn\\
&=& \log\left(2 {L\over L_2}\right)- {\pi\over 3} {L\over L_2} +2 \sum_{n_2=1}^\infty \log\left(1-e^{-4 \pi {L\over L_2}n_2}\right)\,.
\ee
Introducing modular parameters,
\be
\tau= i {L_1 \over L_2}\,, \qquad \text{and}\qquad \alpha :\, = {L\over L_1},
\ee
as well as the Dedekind function as in \eqref{def-dedekind-eta}, we see that we can rewrite the functional determinant on the cylinder as 
\be
\label{det-cyl-2d}
\log \det \Delta_{[0, L]\cross S_{L_2}^1} = \log\left(2 \alpha |\tau| \, \eta^2(2\alpha \tau) \right)\,. 
\ee
This is a well-known result in literature, see {\it e.g.} \cite{DiFrancesco1997, Ginsparg:1988ui}.
It is convenient to leave $\alpha$ general, so that we can easily use the above results for the functional determinants for arbitrary cuts. 

\subsection{Determinant of powers of the Laplacian on the torus}
\label{app:torus-final-det}

When calculating the entanglement entropy of the GQLM on a $d$-torus with $d$ even, the determinants that arise 
are those of even powers of the Laplacian. On the flat $d$-torus geometry the higher-derivative conformal operator 
$\PP_z$ is indeed just the $z/2$-th power of the standard Laplacian, cf. equation \eqref{operator-torus}  in Section \ref{sec:gen-qlm}. 
In order to generalise our previous result, we first make the observation that, since the flat torus as well as the cut torus are compact 
manifolds, the spectrum of $\Delta^k$ is just given by the set of $\lambda^k$, where $\lambda$ is an eigenvalue  of $\Delta$ as in 
\eqref{eq:torus-ev} or \eqref{eq:cut-torus-ev} in the case of the $d$-torus or the cut $d$-torus respectively. 
In particular, the spectral $\zeta$-function corresponding to $\Delta^k$ is given by
\begin{align}
\begin{split}
\zeta_T(s,k):\, = \sumprime_\lambda \left(\lambda^k\right)^{-s} =\zeta_T(ks),
\end{split}
\end{align}
where we write schematically $T$ for either  $T^{d}_{L_1,\ldots,L_d}$ or $[0, L]\cross T^{d-1}_{L_2,\ldots,L_d}$, 
and $\lambda$ for the corresponding eigenvalues \eqref{eq:torus-ev} or \eqref{eq:cut-torus-ev} respectively. 
This leads to the simple result for the determinants~\cite{Elizalde:2004wz}
\begin{subequations}
\begin{align}
\label{log-det-Lk-torus}
\log \det \Delta^k_{T^d_{L_1,\ldots,L_d}}&= k \log \det \Delta_{T^d_{L_1,\ldots,L_d}}\,,\\
\label{log-det-Lk-cyl}
\log \det \Delta^k_{[0, L]\cross T^{d-1}_{L_2,\ldots,L_d}}&=k\log \det \Delta_{[0, L]\cross T^{d-1}_{L_2,\ldots,L_d}},
\end{align}
\end{subequations}
where the critical case is found by setting $k=d/2$.

\section{The winding sector for the $d$-torus}
\label{sec:winding-torus}

The goal of this appendix is to compute the winding sector contribution \eqref{Wn-torus} that originates from 
cutting the $d$-torus, as discussed in Section \ref{sec:replica}. 
In order to do so, we first need to solve the classical equations of motions for the $n-1$ classical fields, \eqref{eom-torus}, 
obeying the boundary conditions \eqref{initial-bc-torus2-v2} as well as \eqref{eq:ext-boundary-conditions-appendix}. 
As discussed in Section \ref{sec:replica}, the $n$-th classical field is reabsorbed into the constrained partition functions 
to create a free one, cf \eqref{int-n-field-torus}, thus here we are only concerned with $n-1$ classical fields. 

To facilitate reading, we list here again equations of motion and explicit conditions which the $n-1$ classical fields have to fulfill. 
The equations of motion are given by
\begin{equation}
\Delta^{z/2}\bar{\phi}^{\rm cl}_i=0\,\quad i=1,\ldots,n-1.
\end{equation}
We solve these equations on the flat torus given by $[-L_B,L_A]\cross[0,L_2]\cross\ldots\cross[0,L_d]$ with the 
ends of each of the intervals identified and place the cuts $\Gamma_1$ at $x_1=0$ and $\Gamma_2$ at 
$x_1=L_A$ (and thus $x_1=-L_B$). 
With this, the boundary conditions along the $x_1$ direction \eqref{initial-bc-torus2-v2} for $i=1,\ldots,n-1$ 
can be rewritten as
\begin{subequations}
\begin{align}
& \bar\phi^{\rm cl}_i|_{\G_1}(x)= \bar{\phi}^{\rm cl}_i(0, y)=0 \,, \label{eq:appendix-bc-cond-torus-1}\\
& \bar\phi^{\rm cl}_i|_{\G_2}(x)= 2\pi R_c\,  \bar\omega_i = \bar{\phi}^{\rm cl}_i(L_A, y)=\bar{\phi}^{\rm cl}_i(-L_B, y)
=2\pi R_c\, \bar{\omega}_i, \label{eq:appendix-bc-cond-torus-2}
\end{align}
\end{subequations}
where we denote  $x = (x_1, y) = (x_1, x_2, \dots x_d)$ and 
$\bar{\omega}_i :\, =  (M_{n-1})_{ij}\omega_j $ 
throughout this section. 
Notice that the above conditions \eqref{eq:appendix-bc-cond-torus-1}-\eqref{eq:appendix-bc-cond-torus-2} 
have to hold for all the coordinates $y$ in the $d{-}1$-dimensional torus, i.e. $y \in [0,L_2]\cross\ldots\cross[0,L_d]$. 
Along the $d{-}1$-toroidal directions we have the periodicity conditions
\begin{equation}
\label{eq:appendix-bc-cond-torus-3}
\bar{\phi}^{\rm cl}_i(x_1, y )=\bar{\phi}^{\rm cl}_i(x_1, y+ \beta)\,, \qquad \beta :=  (L_2,\ldots,L_d)\,. 
\end{equation}
The standard way of solving such a partial differential equation is by separation of variables, 
and here it suffices to separate only the first variable $x_1$  from the remaining orthogonal $d-1$ directions $y$, as {\it e.g.} 
\be
\bar{\phi}^{\rm cl}_i(x_1, y )= f_i(x_1) \, g_i(y)\,,  \qquad i=1, \dots, n-1\,. 
\ee
The boundary condition \eqref{eq:appendix-bc-cond-torus-2} shows that $g_i(y)$ can only be a constant for all 
$i=1, \dots, n-1$, and from now on we set $g_i(y)=1$ and work only with the functions $f_i(x_1)$. 
Hence, the equations of motion and boundary conditions expressed on $f_i$ ($i=1, \dots, n-1$) are simply 
\begin{subequations}
\begin{align}
&\Delta^{z/2}\bar{\phi}^{\rm cl}_i(x,y ) = 0 ~~~\Rightarrow ~~~ \partial_{x_1}^z  f_i(x_1)=0\,, \label{eom-torus-again}\\
& \bar{\phi}^{\rm cl}_i(0, y)=0 ~~~\Rightarrow ~~~ f_i(0)=0\,,  \label{cut1-torus-again}\\
& \bar{\phi}^{\rm cl}_i(L_A,y)=\bar{\phi}^{\rm cl}_i(-L_B,y )=2\pi R_c\, \bar{\omega}_i   ~~~\Rightarrow ~~~ 
f_i(L_A)= f_i(-L_B) = 2\pi R_c\, \bar{\omega}_i\,. 
\label{cut2-torus-again}
\end{align}
\end{subequations}
Notice that the last condition above has to be imposed either at $x_1=L_A$ or at $x_1=-L_B$, or, in other words, 
we solve for classical fields in the region $A$ and in the region $B$ separately and then we glue the solutions at the boundary. 
As we discussed in Section \ref{sec:replica}, these conditions \eqref{cut1-torus-again}-\eqref{cut2-torus-again} are not 
sufficient to specify a solution of the equation of motion \eqref{eom-torus-again}, which is in general a polynomial of 
degree $z-1$, expressed in terms of $z$ coefficients. 
A choice of supplementary boundary conditions are given by \eqref{eq:ext-boundary-conditions-appendix}, 
which now imply 
\begin{subequations}
\begin{align}
& \partial_n\Delta^{k}\bar{\phi}^{\rm cl}_i\big\vert_{\Gamma_1}=0\quad k=0,\ldots,\frac{z}{2}{-}2\,, 
~ \Rightarrow ~ \partial_{x_1}^{2\ell-1}f_i(0)=0\,,\quad \ell=1,\ldots,\frac{z}{2}{-}1,
\\
& \partial_n\Delta^{k}\bar{\phi}^{\rm cl}_i\big\vert_{\Gamma_2}=0\quad k=0,\ldots,\frac{z}{2}{-}2\,, 
~ \Rightarrow ~ \partial_{x_1}^{2\ell-1}f_i(L_A)=\partial_{x_1}^{2\ell-1}f_i(-L_B)=0\,,\quad \ell=1,\ldots,\frac{z}{2}{-}1\,.
\end{align}
\end{subequations}
We then solve the equations respectively in $A$ and $B$, and glue the solutions at the two cuts, that is the whole solution 
is given by
\begin{equation}
f_{i}(x_1)=
 \begin{cases} 
f_{i,A}(x_1),\quad & x_1\in[0,L_A]\,,\\ 
f_{i,B}(x_1),\quad & x_1 \in [-L_B,0]\,,
 \end{cases}
\end{equation}
where 
\begin{equation}
f_{i, A (B)}(x_1)=2\pi R_c \, \bar{\omega}_i \left(a_{z-1}\left(\frac{\vert x_1 \vert}{L_{A (B)}}\right)^{z-1}
+\sum\limits_{k=1}^{z/2-1}a_{2k}\left(\frac{x_1}{L_{A(B)}}\right)^{2k}\right).
\end{equation}
While the explicit value of the coefficients is quite complicated, their $i$ dependence is simple. 
For instance, for the first few values of $z$, the functions $f_{i,A}$ read
\begin{align*}
f_{i,A}(x_1)=
\begin{cases}
2\pi R_c \, \bar{\omega}_i\frac{|x_1|}{L_A},\quad & z=2,\\
2\pi R_c \, \bar{\omega}_i\left( -2\left(\frac{|x_1|}{L_A}\right)^3+3\left(\frac{x_1}{L_A}\right)^2 \right),\quad & z=4,\\
2\pi R_c \, \bar{\omega}_i\left(\left(\frac{|x_1|}{L_A}\right)^5-\frac{5}{2}\left(\frac{x_1}{L_A}\right)^4
+\frac{5}{2}\left(\frac{x_1}{L_A}\right)^2\right),\quad & z=6, \\
2\pi R_c \, \bar{\omega}_i\left(-\frac{4}{17}\left(\frac{|x_1|}{L_A}\right)^7
+\frac{14}{17}\left(\frac{x_1}{L_A}\right)^6-\frac{35}{17}\left(\frac{x_1}{L_A}\right)^4+\frac{42}{17}\left(\frac{x_1}{L_A}\right)^2\right),
\quad & z=8,
\end{cases}
\end{align*}
and the same for the functions $f_{i, B}$ after the replacement $L_A\to L_B$. 
Notice that the functions $f_i$ are continuous everywhere on $A\cup B$, but they are not differentiable at the cuts $\Gamma_1, \Gamma_2$, even though the left and right derivatives exist and are finite. 
This is not unusual, and it is true already in $d=2$, see for example the cylindric case in \cite{Zhou:2016ykv, Zaletel:2011ir}. 

Now we are ready to evaluate the winding sector contribution \eqref{Wn-torus}, that is the function
\begin{equation}\nn
\overline{W}(n)=\sum_{\bar{\phi}^{\rm cl}_{i}} e^{-\sum_{i=1}^{n-1}S[\bar{\phi}^{\rm cl}_{i}]}\,. 
\end{equation}
The action is given by the bulk term $S_0$ \eqref{bulk-action-torus} and the boundary part 
$S_\p$ \eqref{boundary-action-torus}, and it is clear that only the latter will contribute to $\overline{W}(n)$. 
As it turns out, evaluating $S_\p[\bar{\phi}^{\rm cl}_{i}]$ results in the following rather simple closed expression 
(notice that only the highest derivative term contributes to the action \eqref{boundary-action-torus}),%
\footnote{As we discussed the classical fields are not differentiable at the cuts, however left and right derivatives 
exist and they are finite, so here the integrals are evaluated using left and right limits, 
exactly as in $d=2$ dimensions.}
\begin{equation}
S[\bar{\phi}^c_{i}]= g \, \pi^2 R_c^2\, \bar{\omega}_i \cdot \,\bar \omega^i \, \frac{ (-1)^{z/2}z! }{(1-2^z)B_z}\, 
\left(1+ \left({L_A \over  L_B}\right)^{z-1}\right)\, {L_2\cdots L_d \over L_A^{z-1}}\,,
\end{equation}
where the $B_z$ are Bernoulli numbers.
Notice that only the case $z=d$ gives a scale invariant winding sector contribution as expected from a critical theory. 
With this result, we can use the analytic continuation found in appendix F of \cite{Zhou:2016ykv} to write
\begin{align}
\overline{W}(n)&=\sum_{\omega\in\Z^{n-1}}\exp(-\pi\ \Lambda_z(L_A,L_B,L_2,\ldots,L_d)\  \omega^T \cdot T_{n-1}\omega)\\
&= \sqrt{n} \Lambda_z^{-\frac{n-1}{2}} \int\limits_{-\infty}^\infty\frac{dk}{\sqrt{\pi}}e^{-k^2}
\left[\sum_{\omega\in\Z} \exp(-\frac{\pi}{\Lambda_z}\omega^2-2i\sqrt{\frac{\pi}{\Lambda_z}}k\,\omega)\right]^{n-1},
\end{align}
where $T_{n-1}:\, = M^T_{n-1}M_{n-1}$ and
\begin{equation}
\Lambda_z(L_A,L_B,L_2,\ldots,L_d):\, =  g \, \pi \, R_c^2 \frac{ (-1)^{z/2}z! }{(1-2^z)B_z}\ 
\frac{L_A^{z-1}+L_B^{z-1}}{L_A^{z-1}L_B^{z-1}}\ L_2\cdots L_d\,.
\end{equation}
The derivative with respect to $n$ at $1$ is finally given by
\begin{equation}\label{final-Wprime-torus}
-\overline{W}'(1)=\log\sqrt{\Lambda_z}-\frac{1}{2}-\int\limits_{-\infty}^\infty\frac{dk}{\sqrt{\pi}}
e^{-k^2}\log(\sum_{\omega\in\Z} \exp(-\frac{\pi}{\Lambda_z}\omega^2-2i\sqrt{\frac{\pi}{\Lambda_z}}k\,\omega)).
\end{equation}

In order to make comparison in a more transparent way it is useful to rewrite $\Lambda_z$ in terms of the aspect ratios of the $d$-torus. 
Introducing the dimensionless parameters $\tau$, as well as $u$ the parameter controlling the cut, as follows
\be
\label{def:aspect-ratio} 
\tau_k  = i {L_1\over L_{k+1}}\,, \qquad k=1\,, \dots\,, d-1\,, \qquad u={L_A\over L_1}\,,
\ee
and noticing that $L_B=L_1-L_A$, we can write 
\be\label{invariant-lambda}
\Lambda_z= g \, \pi \, R_c^2 \frac{ (-1)^{z/2}z! }{(1-2^z)B_z}\, \left( u^{1-z}+(1-u)^{1-z}\right) {1\over |\tau_1| \dots |\tau_{d-1}|}\,.
\ee
In particular, for $d=2$ and $z=2$, we obtain 
\be\label{def-lambdad2}
\Lambda_2 = 4 \pi \, g\, R_c^2 \, {1\over u(1-u)} {1\over |\tau_1|}\,.
\ee
In the semi-infinite limit, that is when $|\tau_1| >>1$, the integral in $\overline{W}'(1)$ \eqref{final-Wprime-torus} is exponentially 
suppressed, hence at the leading order in $\Lambda_2$ we have 
\be
-\overline{W}'(1) = \half  \log{(4 \pi\, g \,R_c^2)}-\half \log(u(1-u)) -\half \log|\tau_1|-\half +\dots\,,
\ee
and for $u=\half$ this reduces to
\be
-\overline{W}'(1) =\half  \log{(16 \pi\, g \,R_c^2)} -\half \log|\tau_1|-\half +\dots\,. 
\ee

\section{Useful formulae}
\label{app:formulae}

Here we collect some useful formulae and definitions of special functions used in the paper. 

The integral representation of the Gamma function immediately leads to
\be\label{gamma-rep}
\ell^{-s}=\frac{1}{\Gamma (s)}\int\limits_0^\infty dt\ t^{s-1}\ e^{-\ell t}\,. \qquad
\ee
The Poisson summation formula is given by
\be\label{poisson-formula}
&& \sum\limits_{n=1}^\infty e^{-a n^2} =-\frac{1}{2}+\frac{1}{2}\sqrt{\frac{\pi}{a}} +\sqrt{\frac{\pi}{a}}\sum\limits_{n=1}^\infty e^{-\frac{n^2\pi^2}{a}}\,,\\
\nn
&& \sum\limits_{n=-\infty}^\infty e^{-\pi n^2 A+2 \pi n A s}={1\over \sqrt A}e^{\pi A s^2} \sum\limits_{m=-\infty}^\infty e^{-{\pi\over A}m^2-2 i \pi m s}\,.
\ee
The integral representation of a modified Bessel function is
\begin{equation}
\label{bessel-integral-rep}
K_\nu(z)=\frac{1}{2}\int\limits_0^\infty du\ e^{-\frac{z}{2}\left(u+\frac{1}{u}\right)}\ u^{\nu-1}\,,
\end{equation}
and the explicit expression for the special case $\nu=-{1\over 2}$ is 
\be 
\label{bessel-half}
K_{-\half}(x)=\sqrt{{\pi\over 2}}{e^{-x}\over\sqrt x}\,. 
\ee
The Dedekind function is defined as follows
\be\label{def-dedekind-eta}
\eta(\tau ):= q^{1/24} \prod_{n=1}^\infty (1-q^n)\,, \qquad q:=e^{2 i\pi \tau}\,. 
\ee
Its expansion for small imaginary argument is given by
\be\label{dedekind-exp}
\eta(i |\tau|) \approx {e^{-{\pi\over 12 |\tau|}}\over \sqrt{|\tau|}}\,, \qquad \text{as}\qquad  |\tau|\to 0^+\,,
\ee
while for large imaginary argument we have 
\be\label{dedekind-explarge}
\eta(i |\tau|) \sim e^{-{\pi\over 12}|\tau|}\,,  \qquad \text{as}\qquad |\tau|\to \infty\,.
\ee

\bibliographystyle{nb}
\bibliography{RHL_v2}

\end{document}